\documentclass[twocolumn]{aastex63}

\usepackage{graphicx}
\usepackage{apjfonts}
\usepackage{natbib}
\usepackage{xcolor}
\usepackage{lineno}

\newcommand{\vsi}{$v_{\rm Si}$}
\newcommand{\vsimax}{$v_{\rm Si}^{0}$}

\begin{document}

\title{The Foundation Supernova Survey: Photospheric Velocity Correlations in Type Ia Supernovae}

\author[0000-0001-7519-133X]{Kyle G. Dettman}
\correspondingauthor{Kyle G. Dettman}
\email{dettman@physics.rutgers.edu}
\affiliation{Department of Physics and Astronomy, Rutgers, the State University of New Jersey, 136 Frelinghuysen Road, Piscataway, NJ 08854, USA}
    
\author[0000-0001-8738-6011]{Saurabh W. Jha}
\affiliation{Department of Physics and Astronomy, Rutgers, the State University of New Jersey, 136 Frelinghuysen Road, Piscataway, NJ 08854, USA}
    
\author{Mi Dai}
\affiliation{Department of Physics and Astronomy, The Johns Hopkins University, Baltimore, MD 21218, USA}
\affiliation{Department of Physics and Astronomy, Rutgers, the State University of New Jersey, 136 Frelinghuysen Road, Piscataway, NJ 08854, USA}
    
\author{Ryan J. Foley}
\affiliation{Department of Astronomy and Astrophysics, 1156 High Street, University of California, Santa Cruz, CA 95064, USA}
    
\author{Armin Rest}
\affiliation{Space Telescope Science Institute, 3700 San Martin Drive, Baltimore, MD 21218, USA}
    
\author{Daniel M. Scolnic}
\affiliation{Department of Physics, Duke University, Durham, NC 27708, USA}
    
\author{Matthew R. Siebert}
\affiliation{Department of Astronomy and Astrophysics, 1156 High Street, University of California, Santa Cruz, CA 95064, USA}
    
\author{K. C. Chambers}
\affiliation{Institute for Astronomy, 2680 Woodlawn Dr., University of Hawaii, Honolulu, HI 96822, USA}
    
\author{D. A. Coulter}
\affiliation{Department of Astronomy and Astrophysics, 1156 High Street, University of California, Santa Cruz, CA 95064, USA}
    
\author{M. E. Huber}
\affiliation{Institute for Astronomy, 2680 Woodlawn Dr., University of Hawaii, Honolulu, HI 96822, USA}
    
\author{E. Johnson}
\affiliation{Department of Physics and Astronomy, The Johns Hopkins University, Baltimore, MD 21218, USA}
    
\author{D. O. Jones}
\affiliation{Department of Astronomy and Astrophysics, 1156 High Street, University of California, Santa Cruz, CA 95064, USA}
    
\author{C. D. Kilpatrick}
\affiliation{Department of Astronomy and Astrophysics, 1156 High Street, University of California, Santa Cruz, CA 95064, USA}
    
\author{R. P. Kirshner}
\affiliation{Harvard-Smithsonian Center for Astrophysics, 60 Garden St., Cambridge, MA 02138, USA}
    
\author{Y.-C. Pan}
\affiliation{Graduate Institute of Astronomy, National Central University, 300 Jhongda Road, Zhongli, Taoyuan, 32001, Taiwan}
    
\author{A. G. Riess}
\affiliation{Space Telescope Science Institute, 3700 San Martin Drive, Baltimore, MD 21218, USA}
\affiliation{Department of Physics and Astronomy, The Johns Hopkins University, Baltimore, MD 21218, USA}
    
\author{A. S. B. Shultz}
\affiliation{Institute for Astronomy, 2680 Woodlawn Dr., University of Hawaii, Honolulu, HI 96822, USA}

\begin{abstract}

    The ejecta velocities of type-Ia supernovae (SNe~Ia), as measured by the Si II $\lambda 6355$ line, have been shown to correlate with other supernova properties, including color and standardized luminosity. We investigate these results using the Foundation Supernova Survey, with a spectroscopic data release presented here, and photometry analyzed with the SALT2 light-curve fitter. We find that the Foundation data do not show significant evidence for an offset in color between SNe~Ia with high and normal photospheric velocities, with $\Delta c = 0.005 \pm 0.014$. Our SALT2 analysis does show evidence for redder high-velocity SN~Ia in other samples, including objects from the Carnegie Supernova Project, with a combined sample yielding $\Delta c = 0.017 \pm 0.007$. When split on velocity, the Foundation SN~Ia also do not show a significant difference in Hubble diagram residual, $\Delta HR = 0.015 \pm 0.049$ mag. Intriguingly, we find that SN~Ia ejecta velocity information may be gleaned from photometry, particularly in redder optical bands. For high-redshift SN~Ia, these rest-frame red wavelengths will be observed by the Nancy Grace Roman Space Telescope. Our results also confirm previous work that SN~Ia host-galaxy stellar mass is strongly correlated with ejecta velocity: high-velocity SN~Ia are found nearly exclusively in high-stellar-mass hosts. However, host-galaxy properties alone do not explain velocity-dependent differences in supernova colors and luminosities across samples. Measuring and understanding the connection between intrinsic explosion properties and supernova environments, across cosmic time, will be important for precision cosmology with SNe~Ia.
    
\end{abstract}

\section{Introduction}\label{sec: Intro}

Type Ia supernovae (SNe Ia) are important cosmological probes due to their use as precise extragalactic distance indicators. Most famously they have been used to discover the accelerating expansion of the Universe \citep{Riess1998,Perlmutter1999} and, more recently, have been used to constrain the measurement of the local Hubble parameter to $\sim$2.4\% precision \citep{Riess2016}.
	
SNe Ia are not a perfectly homogeneous population \citep{Branch1987}, so their cosmological power comes from their use as empirically standardized candles, correlating their luminosity with other observed properties. The largest of these luminosity corrections come from the SN light-curve shape and color \citep{Phillips1993,Riess1996a}. More recent work suggests that the host-galaxy or local environment can further improve SN Ia standardization \citep[e.g.][]{Wolf2016,Jones2018,Smith20}.
	
Of particular interest is whether the standardization is correcting for \textit{intrinsic} properties of the supernova (e.g., the light-curve shape) or \textit{extrinsic} factors (e.g., interstellar dust along the line of sight). Host-galaxy correlations, for instance, could be either: different kinds of SNe Ia are found preferentially in different environments \citep[e.g.,][]{Hamuy96Host,Hamuy2000}, but interstellar dust can also vary with galaxy type \citep{ScolnicBrout20}.

Of other properties intrinsic to the supernova, the explosion kinetic energy and ejecta mass are among the most fundamental. We can probe these key physical quantities directly through observations by measuring the photospheric velocity $v_{\rm ph} \sim \left(KE/M_{\rm ej}\right)^{1/2}$. Here we focus on the expansion velocity measured through the blueshift of the Si II $\lambda6355$ line, \vsi; this line is the hallmark of SNe Ia near maximum light \citep{Branch1993}. \citet{Benetti2005} and \citet[hereafter W09]{Wang2009} explore correlations of SN Ia luminosity, as well as light-curve shape and color, with \vsi\footnote{\citet{Benetti2005} prefer using the Si velocity \textit{gradient} to distinguish SN~Ia. W09 show that the maximum light velocity is strongly correlated enough with the velocity gradient to be suitable for standardization analysis, with the advantage of not requiring the multiple epochs of spectroscopy needed to measure the gradient.}.
    
W09 used the photospheric velocity \vsi\ to divide SNe~Ia into two subclasses. They found that objects with high photospheric velocity (\vsi\ $< -11,800$ km s$^{-1}$) were redder than their ``normal'' counterparts, and also had a shallower luminosity-color relation (lower $R_V$). \citet[][hereafter FK11]{FK10} followed up this work arguing that the difference in $R_V$ was driven by a handful of highly-reddened high-velocity objects, but that the bulk of the data could best be explained by a ``color offset'' in which high velocity SNe~Ia were redder by $\sim$0.06 mag in $B-V$ compared to normal velocity SNe~Ia. \citet[][hereafter F11]{Foley2011} continued this analysis by better characterizing the time evolution of \vsi\ and deriving a linear correlation between color and maximum light photospheric velocity (rather than just cleaving the sample into high and normal velocity groups).
    
For their analyses, W09 and FK11 used a literature sample of low-redshift ($z < 0.06$) SNe~Ia with heterogeneous photometry and spectroscopy. Here we aim to investigate these previous results with a new sample of SNe~Ia from the Foundation Supernova Survey \citep{Foundation}. The well-calibrated, single-system (Pan-STARRS PS1) photometry \citep{Schlafly_2012,Magnier_2013}, allows for uniform light-curve analysis of a large, low-redshift SN~Ia sample. 
    
W09 and FK11 parameterize supernova color with $B_{\rm max} - V_{\rm max}$,  measured from fits to the \textit{B} and \textit{V} band light curves. Our Foundation photometry, however, was obtained in the \textit{griz} filters, making a direct comparison more difficult. Instead, we choose to use the SALT2 light-curve fitter \citep{Guy2007} and its color parameter \textit{c} for the comparison. As a check, and to see if the choice of filter was consequential, we also analyze data from the Carnegie Supernova Project \citep{Hamuy06}, which observed SNe~Ia in both \textit{BV} and \textit{gri}.
    
In \S~\ref{sec: Data} we describe the data used, sample selection, parameterization, and analysis methods. We compare our approach to previous analyses In \S~\ref{sec: PrevRes}. We present our full analysis and results in \S~\ref{sec: Results}, and discuss implications and conclusions in \S~\ref{sec: Discussion}. Tables of our results and a spectroscopic data release from the Foundation Supernova Survey are given in the Appendix.
    
\section{Data and Methods}\label{sec: Data}
	
The data used in this analysis comes primarily from three sources: the Foundation Supernova Survey \citep{Foundation},  the literature sample used by W09 and FK11 (see references therein), and the Carnegie Supernova Project \citep[CSP; ][]{Folatelli2010}.Our spectroscopic analysis is based on public spectra from the Transient Name Server\footnote{\url{https://www.wis-tns.org/}} and WISeREP\footnote{\url{https://wiserep.weizmann.ac.il/}} \citep{Yaron2012} and the Open Supernova Catalog\footnote{\url{https://sne.space}} \citep{OSC}, and we here also present and release our Foundation spectroscopic data (see Appendix).
	
\subsection{Foundation Data}	
    
The Foundation Supernova Survey \citep{Foundation} is a large (180 objects which pass cuts required to be included in a cosmological analysis), low redshift ($0.015 < z \lesssim 0.08$) survey observed on the Pan-STARRS1 (PS1) \citep{Kaiser2010,Chambers2016} telescope. Foundation is a follow-up survey; we observed SN~Ia targets discovered in untargeted, wide-field surveys,  primarily ASAS-SN \citep{Holoien2017}, the Pan-STARRS Surveys for Transients \citep[PSST][]{Huber2015}, Gaia \citep{Gaia2016}, ATLAS \citep{Tonry2018}, and MASTER \citep{MASTER} among others. Each SN included in the sample was spectroscopically confirmed as a requirement for follow-up; the classification spectra near maximum light are used in our spectroscopic analysis. Cosmological analysis of the Foundation sample is presented by \cite{Jones19}. Some Foundation SNe have redshifts which were measured from the SN itself, we exclude these objects. We present unpublished spectra of Foundation objects in the Appendix (see Figure \ref{fig:team spec}). 

\subsection{W09/FK11 Data}

The W09 dataset comprises ``relatively normal SNe~Ia with good photometry and [\ldots] at least one spectrum within one week after \textit{B} maximum'' from the available literature at the time. This low-redshift sample excluded 91T-like and 91bg-like objects. FK11 use these same data with more restrictive selection cuts, and F11 supplement it with additional spectroscopy. It is important to note that nearly all of these supernovae were discovered in \emph{targeted} surveys, like the Lick Observatory Supernova Search \citep[LOSS; ][]{LOSS}, with relatively narrow-field observations pointed at catalogued galaxies. 
    
\subsection{CSP Data}

The Carnegie Supernova Project \citep{Hamuy06} is a follow-up program that obtained exquisite optical and NIR photometry of a large sample of low-redshift SN~Ia \citep{Contreras10,Folatelli2010,Stritzinger2011,Krisciunas17}. In our analysis we only use the CSP data in optical bands (\textit{BVgri}) and apply selection cuts similar to those used in the CSP cosmological analysis \citep{Burns2018}. Like the W09/FK11 samples, the CSP sample was drawn primarily from supernova searches that targeted catalogued host-galaxies\footnote{The second phase of the Carnegie Supernova Project \citep[CSP-II; ][]{Phillips19} emphasizes SN~Ia found in wide-field untargeted surveys.}. A handful of objects are in both the CSP and W09/FK11 samples.
    
\subsection{Measuring Supernova Photospheric Velocity}

We homogenize measurements of the \ion{Si}{2} velocity by re-analyzing near-maximum-light rest-frame spectra for our sample, using two methods. First, we use astropy \citep{astropy} to model the \ion{Si}{2} line as the sum of a polynomial continuum and Gaussian absorption, and derive \vsi\ from the wavelength of the Gaussian minimum (adopting $\lambda_{\rm rest} = 6355$ \AA\ for \ion{Si}{2}). We use the residuals around this fit to generate 100 perturbed realizations of the original spectrum and refit, and take the standard deviation in the Monte Carlo velocity fits as our  estimate of the velocity uncertainty. Second, we apply the technique used in the kaepora database \citep{Siebert19}, smoothing the spectrum (using variance weighting and a kernel with width of $\sim 300$ km s$^{-1}$), and directly measuring the wavelength of the \ion{Si}{2} line minimum. In most cases, these two methods gave consistent results; occasionally one or the other was preferred based on manual inspection. For example, in cases where the line shape was strongly non-Gaussian, we favored the smoothing method, while in some cases of noisy data, the fitting method provided more reliable results.
    
For cases where we did not have the spectra to perform these measurements, we took previously stated measurements of \vsi\ from CBETs. A number of CSP velocities also came from \citet{Folatelli2013}.
    
The observed \ion{Si}{2} velocity varies as the SN photosphere recedes in to the ejecta over time. To compare objects we correct the measured \vsi\ from the observed phase to \textit{B}-maximum light, using the average evolution found by F11 (equation 5):
\begin{equation}
\centering
v_{\rm Si}^{0}=\frac{ v_{\rm Si} \; + \; 0.2850 \, t}{1 \; - \; 0.0322 \, t},
\end{equation}
where \vsimax\ is the \ion{Si}{2} inferred velocity at phase $t=0$, \vsi\ is the measured rest-frame \ion{Si}{2} velocity (both in units of 1000 km s$^{-1}$), and $t$ is the rest-frame phase in days. We restrict this phase correction to spectra with $-6 < t < +10$ days; this also helps to ensure that \vsi\ reflects the photospheric velocity rather than a separate high-velocity feature \citep{Marion2013}. Objects without spectra in this phase range are excluded in our final sample. F11 estimate that this velocity evolution correction has an uncertainty of $\pm 220$ km s$^{-1}$, so we add $0.22$ in quadrature to the uncertainty for \vsimax. In most cases, this dominates the measurement uncertainty. Our derived velocities are given in Tables \ref{table: data with cuts}, \ref{table: csp data with cuts}, and \ref{table: w09 data with cuts} and the results from our two different fitting methods are given in Table \ref{table: velocity fits}. For notational simplicity, hereafter we refer to maximum light \ion{Si}{2} velocity \vsimax\ simply as \vsi; the correction to zero phase is implied.
    
\subsection{SALT2 Light-Curve Fitting and Sample Selection \label{sec:SALT2}}

For consistency, all photometry for each supernova in our sample was refit using the SALT2 model \citep{Guy2007,Betoule2014} with two different fitting programs, sncosmo \citep{sncosmo} and SNANA \citep{SNANA}. These two environments have slightly different implementations of the SALT2 model; we use them as a consistency check. We exclude SNe from our sample for which the fitters give discrepant results in the derived SALT2 parameters. In particular, we require the sncosmo and SNANA fits to agree in the time of \textit{B}-maximum $t_0$ to within 5 days, the SALT2 color parameter $c$ to within 0.05, the SALT2 light-curve shape parameter $x_1$ to within 0.2 and the \textit{B} apparent magnitude at maximum $m_B$ to within 0.05 mag\footnote{SNANA and sncosmo use slightly different zeropoints for their magnitude scales, $m_B = -2.5 \log x_0 + ZP$; we use the median difference across the sample to correct for this.}. The majority of objects show much better agreement than these bounds. The most significant differences between SNANA and sncosmo results come from light curves that begin well after maximum light, so we additionally require our sample SNe to have photometric observations starting no more than 7 days after $t_0$. 
    
We adopt the SNANA SALT2 light curve fits for our primary results, and present the derived SALT2 parameters in Tables \ref{table: data with cuts}, \ref{table: csp data with cuts}, and \ref{table: w09 data with cuts}.

\begin{figure*}
    \centering
    \includegraphics[width=\textwidth]{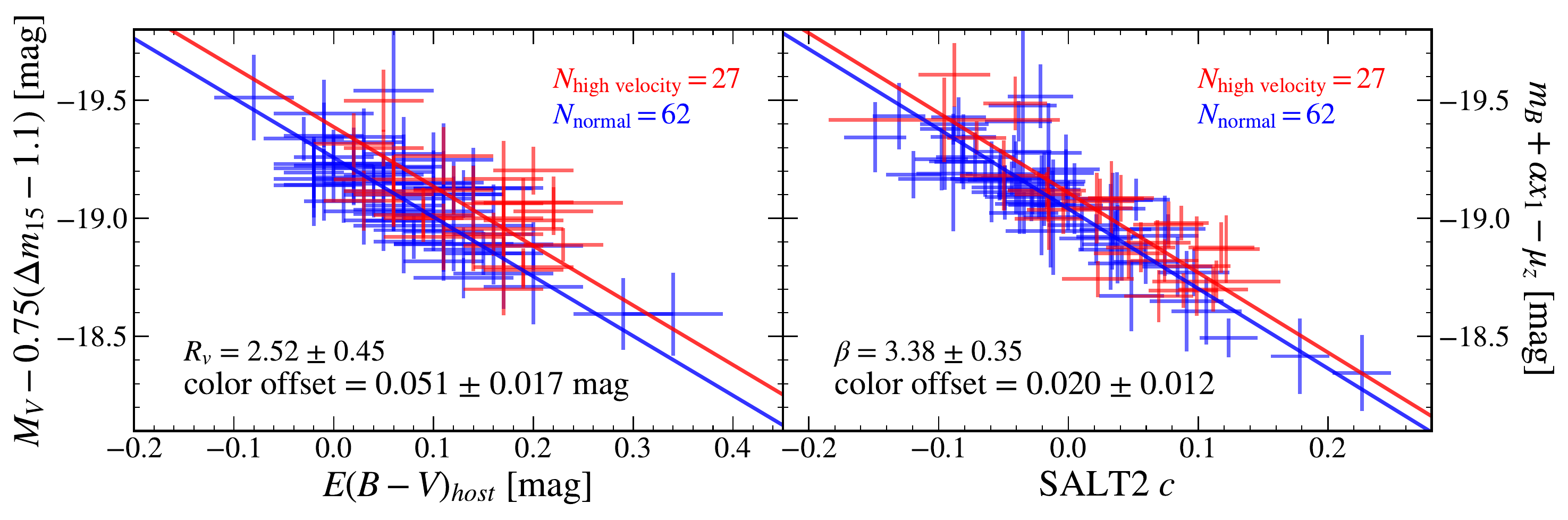}
    \caption{A comparison of the W09 data as analyzed in FK11 (\textbf{left}; see Figure 2 of FK11) and our SALT2 analysis (\textbf{right}). The high-velocity SN~Ia sample comprises objects with \vsi\ $< 11,800$ km s$^{-1}$. Both panels show the same 89 supernovae; these are the subset of the FK11 sample that gave good SALT2 fits. Additionally we use updated redshifts for some objects (with minimal impact on the results). We derive slopes and color-offsets adapting the linmix algorithm \citep{Kelly2007}, as described in the text. The slope and color offset measured in the left panel are consistent with FK11 (who found $R_V \approx 2.50$ and color offset $\approx 0.06$ mag). The SALT2 analysis in the right panel gives a color slope in accord with the expectation $\beta = R_V + 1$. Both panels show a consistent sense for the color offset (high-velocity SN~Ia redder than their normal-velocity counterparts at the same shape-corrected absolute magnitude), but the offset is at higher significance when using $M_V$, $\Delta m_{15}$, and $E(B-V)_{\rm host}$ (left panel; 3.3$\sigma$) compared to SALT2 $m_B$, $x_1$, and $c$ (right panel; 1.8$\sigma$).
    \label{fig:W09 Comp Fits} }
\end{figure*}

\section{Comparison with Previous Results}\label{sec: PrevRes}
    
The main result we aim to test with the Foundation sample is the color offset found by FK11 between high-velocity and normal SN~Ia, following on the work of W09. Both these papers standardize the absolute \textit{V} magnitude of the supernovae as follows (cf. W09 equation 1):
\begin{equation}
M_V = m_V - \mu_z = M^0_V + \alpha (\Delta m_{15} - 1.1) +R_{V} E(B-V)_{\rm host}.
\end{equation}
where $M_V$ and $m_V$ are the absolute and apparent peak \textit{V} magnitudes, and $\mu_z$ is the distance modulus (at redshift $z$). The light-curve shape of each SN is parameterized by $\Delta m_{15}$, the \textit{B}-band magnitude decline in 15 days after maximum light \citep{Phillips1993}, while the SN color $B_{\rm max} - V_{\rm max}$ determines $E(B-V)_{\rm host}$. The coefficients $\alpha$ and $R_V$ apply to all of the SN in the sample, and $M_V^0$ is the fiducial absolute magnitude (i.e., for a supernova with $\Delta m_{15} = 0$ and $E(B-V)_{\rm host} = 0$).

As described above, W09 argue that high-velocity (\vsi\ $< -11,800$ km s$^{-1}$) and normal SN~Ia differ in their inferred $R_V$, though they have consistent $M_V^0$ and $\alpha$. FK11 show that the $R_V$ difference largely disappears if only ``cosmological" SN~Ia (i.e., those with low $E(B-V)_{\rm host} < 0.35$ mag) are considered, but that there is a ``color offset" between the two samples split on \vsi. FK11 analyze this color offset by plotting a shape-corrected absolute magnitude, $M_{\text{shape-corrected}} = M_V - \alpha (\Delta m_{15} - 1.1)$ versus SN color $E(B-V)_{\rm host}$. We recreate such a plot in the left panel of Figure \ref{fig:W09 Comp Fits}, using the values of $M_V$ and $\Delta m_{15}$ from W09, and taking $\alpha = 0.75$ as used in FK11 (consistent with the results of W09)\footnote{Some of the W09/FK11 objects are also in the CSP sample and have slightly updated values of $E(B-V)_{\rm host}$ and $\Delta m_{15}$ presented by \citet{Burns2018}. We have verified that these updates do not significantly change our results.}.
    
Because the Foundation supernova sample was observed in \textit{griz}, we cannot directly replicate the FK11 analysis using $M_V$, $\Delta m_{15}$, and $E(B-V)_{\rm host}$ (from $B_{\rm max} - V_{\rm max}$). Instead, we use SALT2 fits as described in section \ref{sec:SALT2}. To investigate the impact of this change in the analysis, we first compare SALT2 results for the W09/FK11 sample, as shown in the right panel of Figure \ref{fig:W09 Comp Fits}, with the same 89 SNe~Ia (27 high-velocity and 62 normal) in both panels. This is slightly smaller than the sample presented in FK11 (their Figure 2) because we require sufficient photometry for a robust SALT2 fit (and consistent SNANA and sncosmo results, as described in section \ref{sec:SALT2}). In the SALT2 analysis, using the \citet{Tripp1998} standardization, the shape-corrected absolute magnitude is given by
\begin{equation}
\centering
M_{\text{shape-corrected}} = m_{B} + \alpha x_{1} - \mu_z,
\end{equation}
which we regress against the SALT2 color parameter $c$ (with slope $\beta$). For the distance modulus $\mu_z$, we adopt cosmological parameters $H_0 = 70.5$ km s$^{-1}$ Mpc$^{-1}$, $\Omega_{\rm m} = 0.3$, $T_{\rm CMB} = 2.725$ K using astropy's FlatLambdaCDM cosmology object, and only use objects in the Hubble flow ($z > 0.008$), with an assumed redshift uncertainty of 300 km s$^{-1}$ from peculiar velocities.
    
Visually, the data points in Figure \ref{fig:W09 Comp Fits} look similar across both panels, suggesting similar results in either the ($M_V$, $\Delta m_{15}$, $E(B-V)$) analysis or the SALT2 analysis. To test this quantitatively, we encounter the difficulty of linear regression with significant uncertainties in both variables. To deal with this, we adapt the Gaussian mixture model regression of \citet{Kelly2007}, using the Python implementation in the linmix package\footnote{\url{https://github.com/jmeyers314/linmix}}. We use linmix to fit both the high- and normal-velocity samples separately and check that both have compatible slopes (true in all cases). We then join MCMC samples with matching slopes (the only case where a color offset is meaningful) and calculate the offset. This approach allows us to measure the color offset using the linmix method, while marginalizing over the distribution of slope and intrinsic scatter (the latter is consistent with zero for all samples). We validated our method on simulated data with known color offsets.

The results of our fitting method for both the original W09/FK11 data and our SALT2 analysis in the solid lines and annotations on Figure~\ref{fig:W09 Comp Fits}.  Based on the shapes of the normal- and high-velocity distributions, FK11 argue that an intrinsic color offset is the best explanation for the difference, noting for instance that the locus of the bluest high-velocity points remains redder than the bluest normal-velocity objects at all magnitudes. Our results in the left panel are consistent with FK11 both in terms of slope and color offset. In our SALT2 analysis (right panel), we find a slope consistent with the expectation $\beta \approx R_B = R_V + 1$, but the inferred color offset ($0.021 \pm 0.012$) is lower, and at lower significance. 
    
In principle, the SALT2 color parameter $c$ aims to be an analogue of the SN $B - V$ color, and thus we would expect $c$ and $E(B-V)_{\rm{host}}$ to be commensurate, perhaps with a constant offset between them. However, whereas W09/FK11 derive $E(B-V)_{\rm{host}}$ from $B_{\rm max} - V_{\rm max}$, the SALT2 fit for $c$ makes use of the full light curve, in multiple bands. Indeed, if we empirically compare the derived values for these objects in the W09 sample, we find a shallower relation than expected, with $c \approx 0.7 \, E(B-V)_{\rm{host}}$ in the range $-0.3 \leq c \leq 0.3$. In this color range, the SN color can arise from both intrinsic color variations as well as dust reddening, so the scale difference between $c$ and $E(B-V)_{\rm{host}}$ may not be a complete surprise. Such a rescaling of the $x$-axis would bring the color offsets into better accord, with the SALT2 color offset value of 0.020 rescaled to a color offset of $0.021/0.7 = 0.029$ mag in $E(B-V)$ space. This is not a completely satisfactory resolution, however, because such a rescaling would also affect the slope, suggesting $R_B \approx 3.38(0.7) = 2.37$, at odds with $R_B = R_V + 1 \approx 3.6$ expected from the left panel of Fig.~\ref{fig:W09 Comp Fits}.
    
While the SALT2-based analysis does not completely recreate the FK11 $B-V$ color offset seen in the W09 data, we nonetheless proceed with the analysis using SALT2, primarily because it is the light-curve fitter of choice in most supernova cosmology analyses today. Thus, for cosmological applications of SN~Ia, it behooves us to understand any differences related to explosion velocity in the SALT2 context. We discuss this issue further in \S~\ref{sec: Discussion}.
 
\begin{figure*}[ht]
    \centering
    \includegraphics[width=\textwidth]{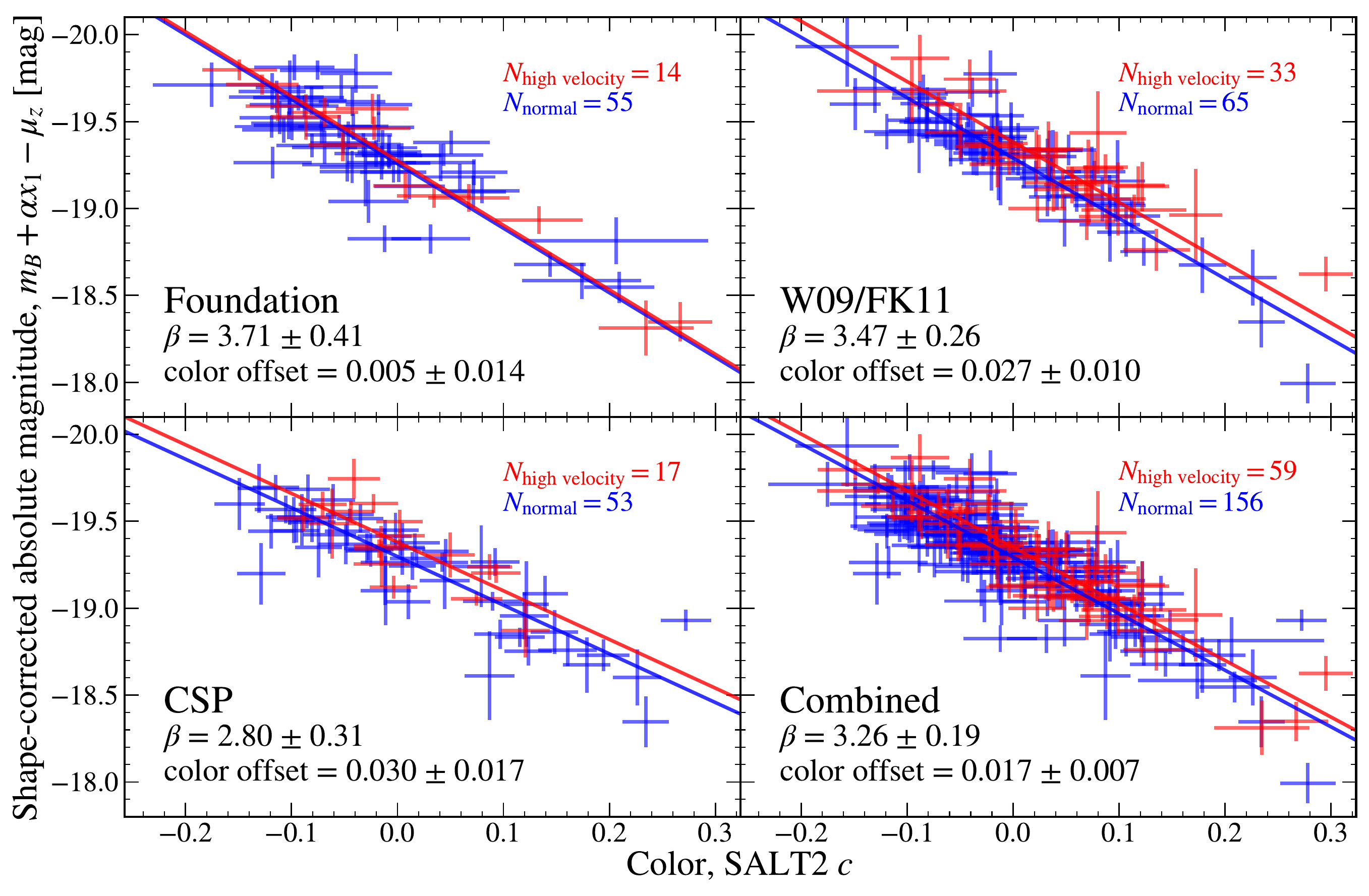}
    \caption{The three data sets (Foundation, \textbf{top left}; W09/FK11 \textbf{top right}; CSP, \textbf{bottom left}) used in this analysis as well as a combined sample (\textbf{bottom right}) of all three, using SALT2 fits. High-velocity objects are denoted by red points and normal-velocity objects are denoted by blue points. In the top right corner of each subplot is the number of normal- and high-velocity objects. In the bottom left corner of each subplot are the fit slope $\beta$ and the fit color offset. The W09/FK11 sample in the upper right panel differs slightly from Fig.~\ref{fig:W09 Comp Fits} because we used our own cuts from the larger sample. The Foundation sample does not show evidence of a color offset, unlike the CSP and W09/FK11 datasets.  \label{fig:fourplot}}
\end{figure*}

\section{Results}\label{sec: Results}

\subsection{Color Offsets for High-Velocity Supernovae \label{sec:coloroffsets}\label{sec:Res-DirComp}\label{sec: combSamp}}

We show the SALT2 shape-corrected absolute magnitude versus color for the Foundation sample in the upper left panel of Figure~\ref{fig:fourplot}. We have 69 Foundation objects (14 high velocity with \vsi\ $< -11800$ km s$^{-1}$ and 55 normal velocity) that have good velocity measurements, good SALT2 light curve fits (as described in section \ref{sec:SALT2}), and that are appropriate for a cosmological sample \citep[e.g.,][]{Jones19}. Specifically, we require $z > 0.008$, $|c| \leq 0.3$, and $|x_1| \leq 3$ (see Tables \ref{table: data with cuts} and \ref{table: velocity fits}). 

Surprisingly, in the Foundation sample we find no evidence of a color offset between the high-velocity and normal-velocity SNe~Ia, with our best fit offset value of $0.005 \pm 0.014$. In the upper right panel of Figure~\ref{fig:fourplot} we show the SALT2 analysis of the W09/FK11 sample for comparison\footnote{The W09/FK11 sample in the upper right panel of Figure \ref{fig:fourplot} is slightly different than the right panel of Figure \ref{fig:W09 Comp Fits}. In Figure \ref{fig:fourplot} we use the SALT2-based sample cuts described in this section, whereas in Figure \ref{fig:W09 Comp Fits} we adopt the $\Delta m_{15}$ and $E(B-V)_{host}$ cuts used by FK11.}. While the Foundation data show a slope consistent with the W09/FK11 data, the Foundation sample has a lower fraction of high-velocity SNe~Ia and a lower color offset.
    
We consider several explanations for this discrepancy in the next sections. Potential culprits could be small number statistics, the SALT2 vs.~$E(B-V)_{host}$ analysis choices (though we have already shown there is a small color offset in W09/FK11 even when using SALT2; Figures \ref{fig:W09 Comp Fits} and \ref{fig:fourplot}), the filters used for the photometry: typically \textit{gri} for Foundation and \textit{BVRI} for W09/FK11, or systematic differences in the kinds of supernovae that comprise either of the samples.
    
To gain further insight into this question, we also fit the CSP sample in the same manner, shown in the lower left panel of Figure~\ref{fig:fourplot}. The CSP objects have a color offset of $0.030 \pm 0.017$, larger than even in the W09/FK11 data, but at lower significance. Curiously the CSP sample also has a relatively low fraction of high-velocity objects. Moreover the shape-corrected magnitude vs.\ color relation has a shallower slope $\beta$ for the CSP objects compared to either W09/FK11 or Foundation. Making a slightly stricter color cut $c < 0.25$ would remove one object from the CSP sample, the slight outlier SN~2007ba, and give a slope more consistent with the other samples.
    
Though the samples show differences that we explore further below, they are marginally consistent with each other. Combining them yields the lower right panel in Figure~\ref{fig:fourplot}, with a fit slope $\beta = 3.26 \pm 0.19$ and a $>2\sigma$ measurement of the color offset, $\Delta c = 0.17 \pm 0.007$.

\begin{figure*}
    \centering
    \includegraphics[width=\textwidth]{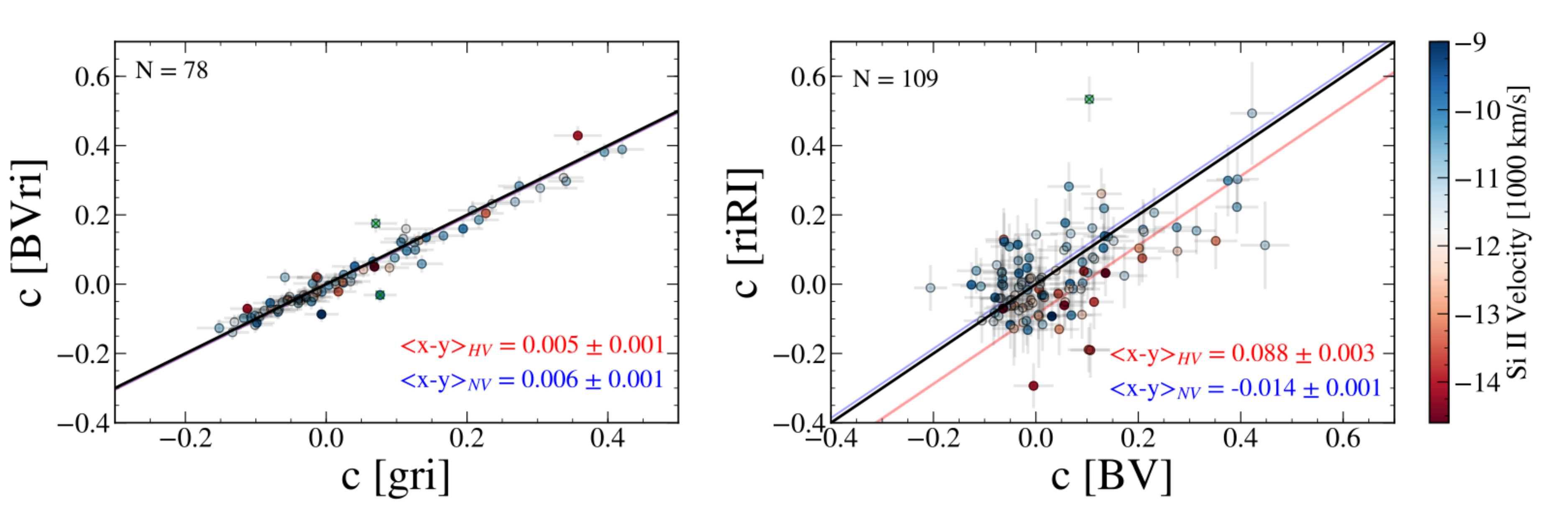}
    \caption{\textbf{Left:} A comparison of the SALT2 color parameter \textit{c} applied to the CSP SN~Ia sample, fitting only the \textit{BVri} photometry versus fitting only the \textit{gri} photometry. Points are color-coded by \vsi\ as shown in the colorbar on the right. Other than a few outliers, the recovered $c$ values are nearly identical (the solid black line indicates exact identity). There is no systematic difference between high-velocity and normal-velocity objects. \textbf{Right:} Same as the left panel, except showing SALT2 $c$ inferred from only two bands, either \textit{ri} or \textit{RI} versus \textit{BV}, again for the CSP sample. Using just two photometric bands results in a noisier measurement (and especially so deriving $c$ from just the two redder bands), but there does seem to be separation in the inferred SALT2 color as a function of velocity. The red and blue lines show the median results for the high-velocity and normal-velocity objects, respectively. This result is intriguing because it suggests that velocity information is encoded in the photometry.
    \label{fig:ri_BV} \label{fig:csp_bvri-gri} }
\end{figure*}

\subsection{Photometric Comparisons \label{sec: Res-PhotComp} \label{sec: color comp}}
    
The W09 and FK11 results are based on the $B_{\rm max} - V_{\rm max}$ color, but the Foundation photometry covers this wavelength range with a single filter, \textit{g}. If the differences between high-velocity and normal SNe~Ia that lead to a color offset are localized to this region of the spectrum, observing it with a single band may suppress the effect, potentially explaining our results. 
    
For a direct comparison between the \textit{g} and \textit{BV} bands we turned to the CSP sample, which (wisely!) observed objects in all of these filters, both to guard against the possibility that the relatively wide \textit{g}-band was leaving important spectral information unresolved and to connect to prior SN~Ia samples. We compare SALT2 fits on CSP objects using just \textit{BVri} photometry versus SALT2 fits of the same objects using just \textit{gri} photometry in the left panel of Figure~\ref{fig:csp_bvri-gri}, adopting the wavelength-extended SALT2 model from \citep{Hounsell2018}\footnote{This is an ``extended'' version of SALT2 in which the model SED covers redder passbands, and is only used in this comparison. For all of the other SALT2 analyses in this paper, we use the SALT2 model described by \citet{Betoule2014}, for consistency with cosmological applications.} The SALT2 color parameter shows a tight, nearly one-to-one (solid line) relation, with a handful of outliers and a Pearson-r coefficient of 0.98. If $3\sigma$ outliers are removed, the scatter in $c_{gri}-c_{BVri}$ is 0.026. In particular, we do not see any systematic difference in the inferred $c$ that depends on velocity. The results for the other SALT2 parameters ($m_B$ and $x_1$) are similar, meaning that \emph{differences between BV and g are not likely to explain the lack of a high-velocity color offset seen in the Foundation sample.}
   
Curiously, we find a velocity-dependent color effect if we compare SALT2 fits using only redder bands (\textit{r} or \textit{R} and \textit{i} or \textit{I}) with bluer bands, as shown in the right panel of Figure \ref{fig:ri_BV}. Because the SALT2 color parameter $c$ is nearly a proxy for the supernova $B-V$ color at maximum, estimating it from redder bands only is more uncertain, and sensitive to the SALT2 SED model (including variation with $x_1$). The highly significant shift seen, a difference in $c(BV) - c(RI)$ of nearly 0.1 between the high-velocity and normal-velocity objects, suggests that SALT2 fits to these redder bands can potentially constrain the SN~Ia velocity \emph{from photometric data alone}. When bluer data are added, however, SALT2 clearly gives information at those wavelengths higher weight in the estimate of $c$; there is little difference in the derived SALT2 color parameter $c$ between \textit{BV} or \textit{BVri} observations.
    
\begin{figure*}
    \centering
    \includegraphics[width=\textwidth]{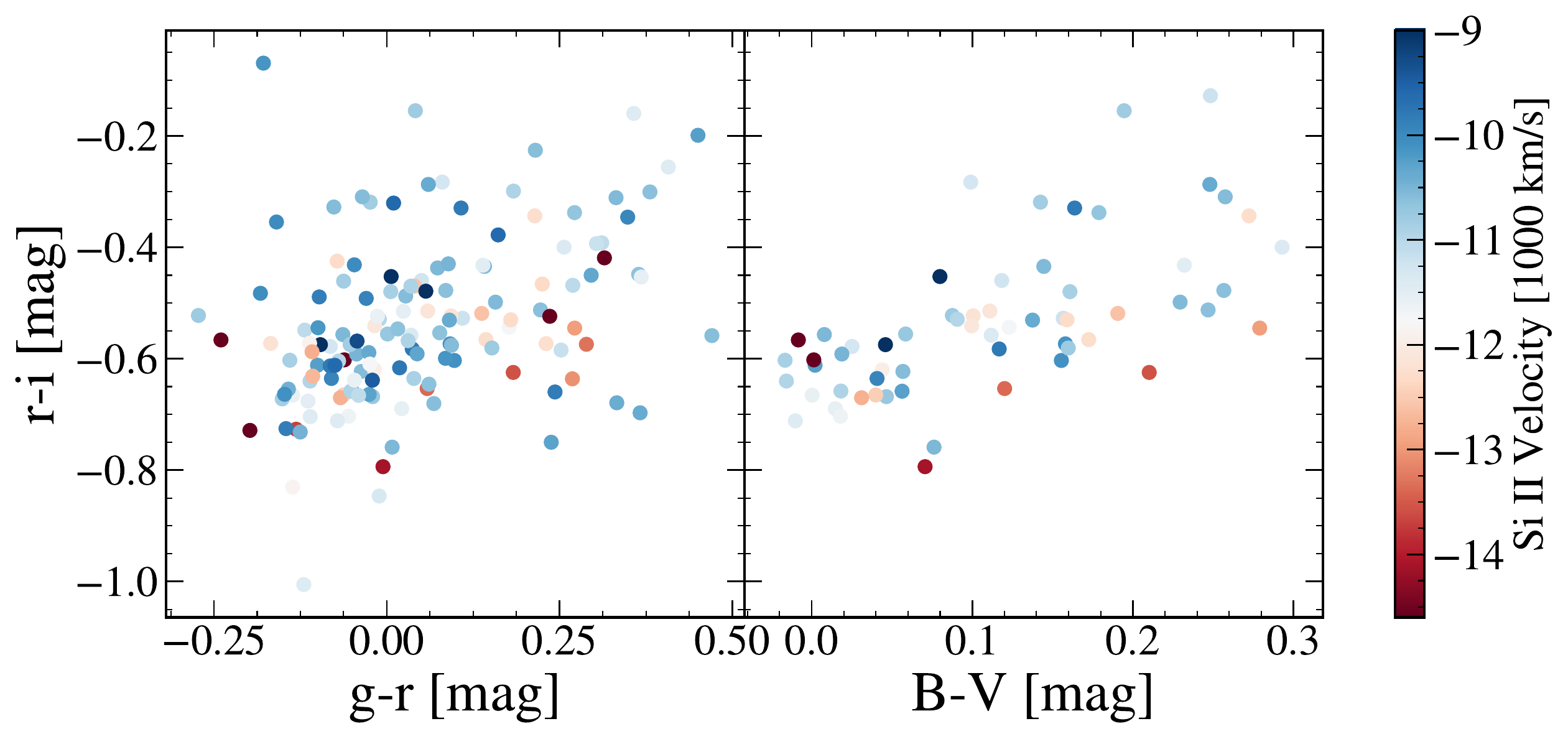}
    \caption{Color-color plots of a subset of SNe in our combined sample. The magnitudes were taken at the time of \textit{B}-band maximum using a Gaussian Process fit to the light curve with $K$-corrections and Milky-Way extinction corrections applied. The colors assigned to each point are as in Fig.~\ref{fig:ri_BV}. \textbf{Left:} Data include CSP and Foundation objects with well-measured \textit{gri} light curves. \textbf{Right:} Data from CSP and W09/FK11, using \textit{BVri} photometry. In the \textit{B}$-$\textit{V} color here we see a similar separation as seen in the right panel of Fig.~\ref{fig:ri_BV}, with the higher velocity points having, on average, bluer \textit{r}$-$\textit{i} colors. In the \textit{g}$-$\textit{r} space though, this trend is less apparent.}
    \label{fig:colDiff}
\end{figure*}

We investigate whether this velocity-dependent SALT2 color effect can be seen directly in SN~Ia colors in Figure \ref{fig:colDiff}, where we show maximum light \textit{r}$-$\textit{i} versus \textit{g}$-$\textit{r} and \textit{B}$-$\textit{V}, color-coded by \vsi. Here, we see some small effect in the \textit{B}$-$\textit{V} vs \textit{r}$-$\textit{i} like we do in the case of SALT2 (Figure \ref{fig:ri_BV}), with the high-velocity points having bluer \textit{r}$-$\textit{i} colors. This effect is not as strong in \textit{g}$-$\textit{r}. Neither has as strong of a relation as the SALT2-fit colors do, suggesting that SALT2 model may be enhancing this effect.

\begin{figure}
    \centering
    \includegraphics[width=0.5\textwidth]{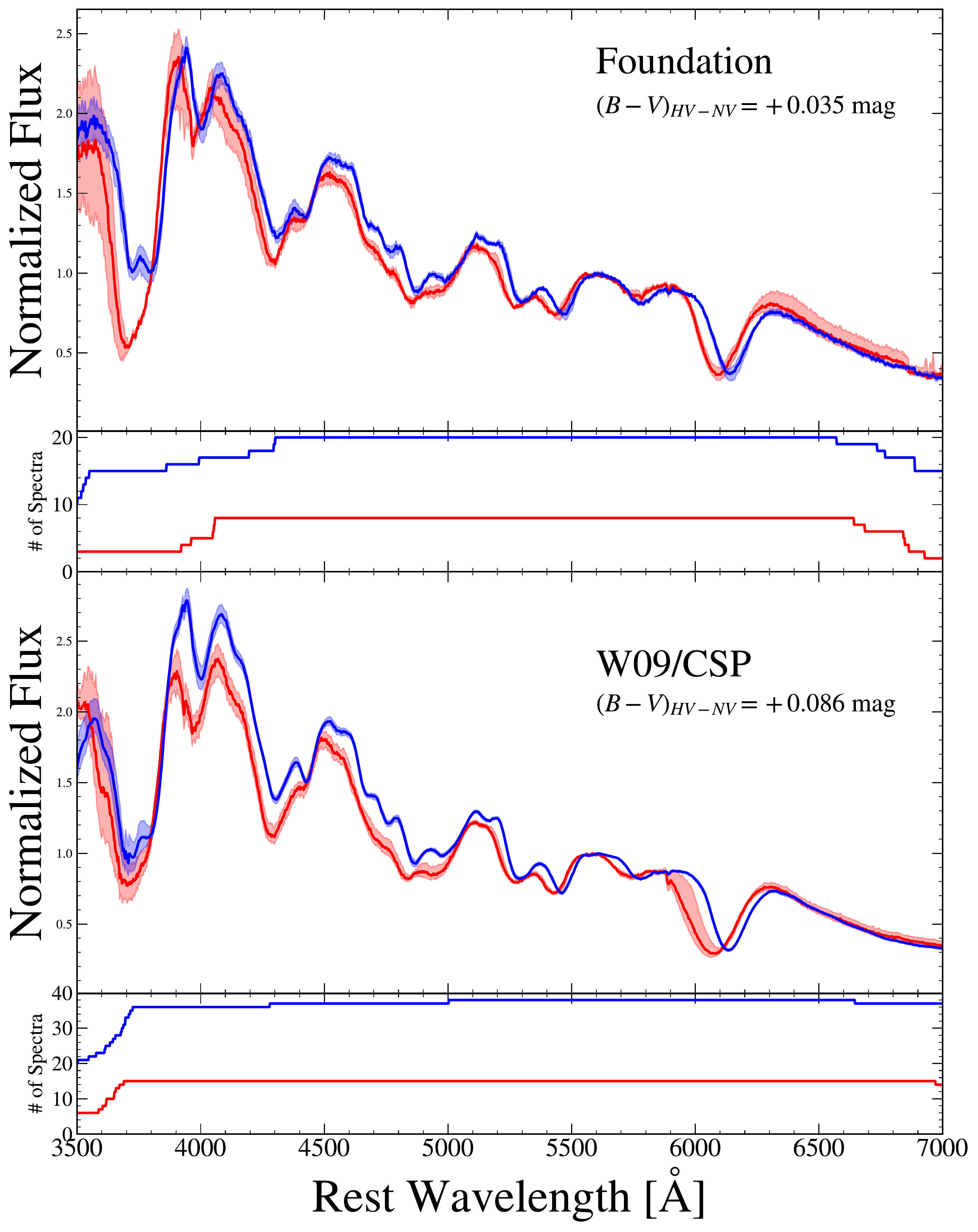}
    \caption{Composite spectra made using kaepora \citep{Siebert19} of Foundation and W09/CSP objects. Red lines are for high-velocity SNe and blue are for normal velocity. The shaded regions are derived from bootstrap resampling of the constituent spectra as described by \citet{Siebert19}. The spectra were normalized to their median continuum between 5500--5700 \AA. The number of spectra that contribute at each wavelength is shown in the bottom panels. In the upper right-hand corner, the difference in the \textit{B}$-$\textit{V} color between the high-velocity and normal-velocity subsets. Like for the color offset, the Foundation sample shows a smaller color difference in the composite spectra. \label{fig:speccompare} }
\end{figure}

\subsection{Spectral Comparisons}\label{sec: Res-SpecComp}

Any relationship between SN~Ia intrinsic color and line velocity should also be evident in spectroscopy. FK11 and \citet{Mandel2014} hypothesize that the redder intrinsic colors of high-velocity SN~Ia may come from broadened, high-opacity lines blocking flux at shorter wavelengths, particularly in the \textit{B} band. Our Foundation observations in \textit{g} band, essentially spanning \textit{B} and \textit{V}, may thus be less sensitive to such an effect. However, as demonstrated earlier (e.g., Figure \ref{fig:csp_bvri-gri} left) we do not find any significant bias in the inferred SALT2 $c$ if we fit with either \textit{BV} or \textit{g}.

To further investigate our results we created composite spectra using the kaepora database \citep{Siebert19}\footnote{https://github.com/msiebert1/kaepora} along with our own near-maximum-light spectra of Foundation objects (see Appendix). The spectra used in this analysis from kaepora come from the following; \citet{Folatelli2013,Blondin2012,Salvo2001,Benetti2004,Kotak2005,Garavini2007,Silverman2012,Stanishev2007,WangX2009,Pereira2013,Mazzali2014}. We created composite spectra using kaepora and the techniques developed in \citet{Siebert19}, though between the host-galaxy corrections being applied and the compositing process, each spectrum was normalized to the median flux between 5500 \AA\ and 5700 \AA. Composites were made for the combined W09/FK11 and CSP sample and for the Foundation data, split between high-velocity and normal velocity SNe.
    
As can be seen in Figure \ref{fig:speccompare}, as expected, the spectral features in the high-velocity composites are clearly broader and blue-shifted. Significantly, the normal-velocity spectra have a bluer continuum than their high-velocity counterparts. This effect is more pronounced in the W09/FK11 and CSP data, with a difference in the \textit{B}$-$\textit{V} color of the high-velocity and normal velocity SNe of $0.086$ mag. In the Foundation sample, this difference is only $0.035$ mag. Each of the four composites were made up of spectra chosen such that the phase of the composite would be close to 0 days and each composite has a phase, over the whole spectral range, which is consistent with 0. The median $x_1$ for each are slightly different, as the Foundation sample has a higher mean $x_1$ than the previous samples, but because $x_1$ and $c$ are not strongly correlated, a difference in $x_1$ cannot be a major factor in explaining the color difference in the composite spectra.
    
\begin{figure*}
    \centering
    \includegraphics[width=\textwidth]{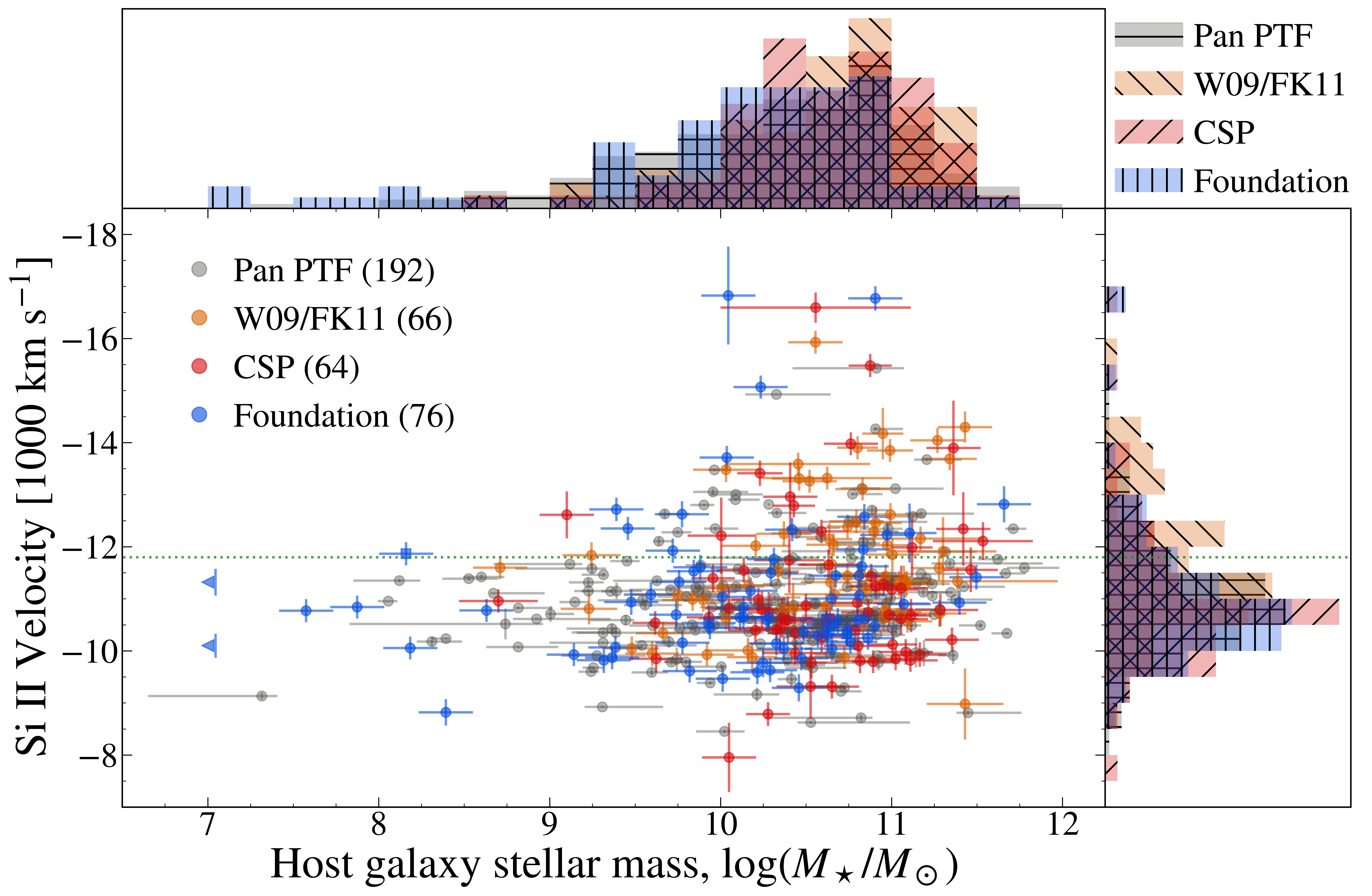}
    \caption{Host-galaxy stellar mass vs.~measured Si II velocity at maximum (\vsi) for each object in our sample (colored and labeled by subsample) along with a large comparison sample (gray) from 
    \citet{Pan2015} and \citet{Pan2020}. The Foundation (blue) 
    points marked with a square did not have previously-reported host masses and were determined using SDSS photometry and \citet{Taylor2011}, eq.~8. Foundation (blue) points marked with a triangle are extremely low stellar-mass host-galaxies, for which we only estimate an upper limit, $\log(M_*/M_\odot) \leq 7.0$. Uncertainties shown for the host-galaxy stellar masses in our sample were taken as the standard deviation of masses from the literature if there was more than one measurement for a given galaxy. If only one measurement was available, we assigned an uncertainty equal to the median uncertainty of the other hosts. The dotted line shows our boundary between high- and normal-velocity SNe. Adjacent histograms show the relative distribution of masses or velocities for each subsample. There is a strong indication that higher-velocity SN~Ia inhabit higher-stellar-mass host galaxies, and nearly all SN~Ia in low-mass host galaxies have normal explosion velocity.}
    \label{fig:velHostmass}
\end{figure*}

\subsection{Host-Galaxy Mass}\label{sec: mass}

One of the important differences between the Foundation SN sample and the CSP and W09 samples used here comes from the SN discovery surveys used in the follow-up. Foundation used discoveries from untargeted, all-sky SN surveys, while the CSP and W09 SN were primarily from SN surveys that targeted and searched catalogued galaxies. Targeted surveys preferentially observe more luminous, higher-mass host galaxies, and so SN with low-mass host galaxies are underrepresented.  Besides affecting the distribution of host masses in the sample, this bias can propagate to SN properties. For example, \citet{Jones2018} found that targeted surveys produced a lower host-galaxy mass magnitude step (in light-curve and color corrected SN luminosity) than untargeted surveys.
    
There is a connection between host-galaxy properties and SN~Ia velocities. \citet{FoleyMass2012} found that the Ca II velocity decreased as host-galaxy stellar mass increased (albeit for a narrow range of masses, and for high-redshift SN). In contrast \citet{Pan2015} and \citet{Pan2020} used a large sample of low-redshift SN~Ia from PTF and found that Si II velocity increased with increasing host-galaxy stellar mass. Moreover, galaxies with stellar mass $M_\star \lesssim 10^{9.5} M_\odot$ did not host high-velocity SN~Ia. We seek to explore this result in the Foundation data and combined samples.

\begin{figure*}
    \centering
    \includegraphics[width=\textwidth]{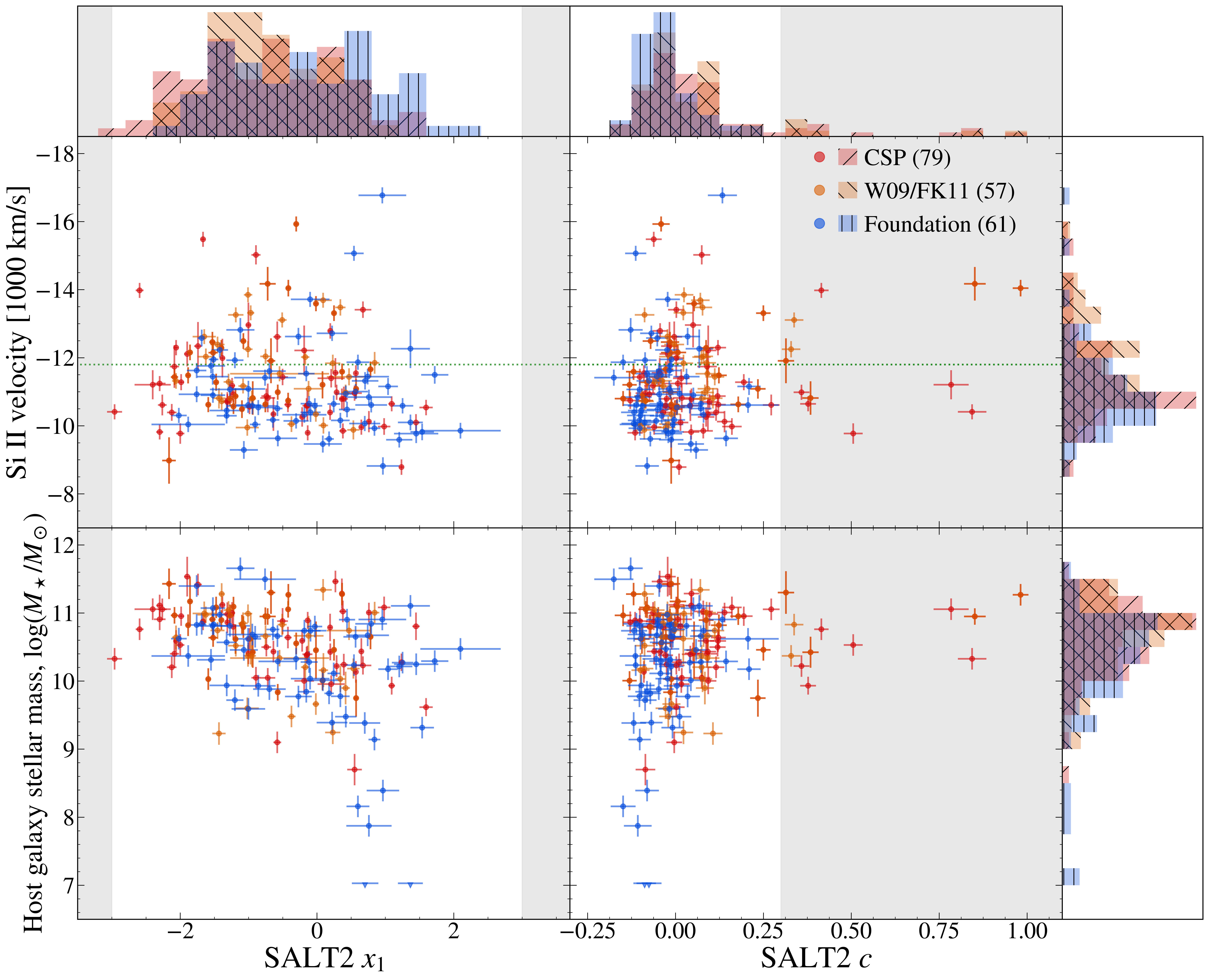}
    \caption{Relations between \vsi\ and host-galaxy stellar mass vs.~ SALT2 light-curve fit parameters $x1$ and $c$. The points are colored by subsample as in Fig.~\ref{fig:velHostmass}. The grey regions show ranges typically excluded by cuts on $x_1$ and $c$ in cosmological applications. The green dotted line denotes the demarcation between high- and normal-velocity; downward facing triangles show upper limits on host-galaxy stellar mass. SN~Ia in the lowest-mass host galaxies show a remarkably narrow range in light-curve properties $x_1$ and $c$ \citep[e.g.,][]{ScolnicBrout20}.}
    \label{fig:x1_c}
\end{figure*}

\subsubsection{Host-Galaxy Stellar Mass Data}\label{sec: mass data}

Because of the importance of the host-galaxy stellar mass as an additional parameter for SN~Ia luminosity standardization, many authors have worked to derive host-galaxy stellar mass estimates for SN~Ia samples, typically from galaxy photometry. We compiled mass measurements for our objects from \citet{Neill2009},  \citet{Sullivan2010}, \citet{Smith2012}, \citet{Chang2015}, \citet{Campbell2016}, \citet{Wolf2016}, \citet{Uddin17}, \citet{Burns2018}, and \citet{Jones2018}. We did not try to homogenize the reported masses (for example, to enforce a consistent stellar initial mass function), and rather have taken them ``as is''. Many objects in our sample have more than mass measurement, and show good agreement across sources. There appears to be systematic differences in the mass estimates based on the method used (for example UV versus optical/NIR photometry), but not at a level significant to our results. 
    
To ensure consistency with published cosmological results, for the Foundation objects we take host-galaxy stellar masses only from \citet{Jones2018}. For the other samples, to combine the mass measurements we took the median for each object as the point estimate, and used the standard deviation of the measurements of the object as an estimate for the mass uncertainty. For host-galaxies with only a single stellar mass measurement, we assigned an uncertainty equal to the median uncertainty of the multiply-measured objects, $\pm 0.16$ dex, i.e., in $\log(M_\star/M_\odot)$.
    
For the stellar mass estimates from \citet{Burns2018}, we followed the prescription in the text using the tabulated \textit{K}-band data, with $\log(M_{\star}/M_{\odot}) = -0.4(K-\mu)+1.04$. Masses for host-galaxies without \textit{K}-band photometry were taken from \citet{Neill2009} or \citet{Chang2015}. One object, CSS160129, did not have a tabulated host-galaxy stellar mass, so we used SDSS photometry and eq.~8 from \citet{Taylor2011} to estimate it at $\log(M_\star/M_\odot) = 8.16$. Two SN, PS15bwh and ATLAS16agv, had host-galaxies too faint for reliable photometry. Based on the imaging depth, we treat their host-galaxies as having a stellar mass of $\log(M_{\star}/M_{\odot}) = 7.0$ as an upper limit; this still places them as the lowest mass host-galaxies in our sample.

\subsubsection{The Effect of Host-Galaxy Stellar Mass}
    
In Figure~\ref{fig:velHostmass} we plot the Si II velocity versus host-galaxy stellar mass for objects in each of our SN subsamples. We immediately see the difference between targeted and untargeted SN surveys, with the objects in the Foundation sample extending to lower host-galaxy stellar masses. We also plot the (untargeted) PTF SN~Ia from \citet{Pan2015} and \citet{Pan2020}; our Foundation data confirm their finding that high-velocity SN~Ia are largely absent from low-mass host-galaxies, $\log(M_\star/M_\odot) \lesssim 9.5$. We ran Anderson-Darling tests to confirm what can be discerned by eye: the host-galaxy stellar mass distribution are largely consistent between CSP and W09/FK11 (drawing on targeted SN surveys with mostly high-stellar-mass host-galaxies) and between Foundation and PTF (SN from untargeted surveys). However, the distributions are strongly inconsistent between W09/FK11/CSP and Foundation/PTF. This effect can explain why the fraction of high-velocity SN~Ia is significantly lower for Foundation than in W09/FK11: surveys that target high-mass host-galaxies will find more high-velocity supernovae relative to normal-velocity supernovae. Clearly, sample selection will play an important role in understanding the connection between host-galaxy properties and SN velocity.
    
To further highlight this interconnectedness, in Figure \ref{fig:x1_c} we show relationships between \vsi\ and host-galaxy stellar mass with SN~Ia SALT2 $x_1$ and $c$. We find these light-curve parameters are largely agnostic to velocity, but stronger trends are present relative to the host-galaxy mass. Analogous to the way faster-declining SN~Ia are found preferentially in early-type host-galaxies \citep[e.g.,][]{Hamuy1995,Hamuy96Host,Hamuy96H0,Hamuy2000}, our results are consistent with previous studies showing these fast-decliners are also in higher-mass host-galaxies \citep[e.g.,][]{Neill2009,Uddin17}. Similarly, we see bluer SN~Ia in lower-mass host-galaxies, confirming the results of, e.g., \citet{Sullivan2010}. Finally we note the presence in the Foundation sample of a ``tail'' of objects with blue colors, $x_1 \approx +1$, and normal \vsi, all from low stellar-mass host-galaxies. \citet{ScolnicBrout20} demonstrate that this quite homogeneous population from low-mass host-galaxies also shows low dispersion on the Hubble diagram.
    
Nevertheless, as we will show later (see \S~\ref{sec: cuts} and Figure ~\ref{fig:cdfs}), simply restricting the Foundation sample to a mass range which reflects the W09/FK11 and CSP objects does not recover a significant offset in the Foundation data. Though there are clear correlations between SN velocity and host-galaxy mass, accounting for mass alone does not bring the three samples into agreement.

\begin{figure*}
    \centering
    \includegraphics[width=\textwidth]{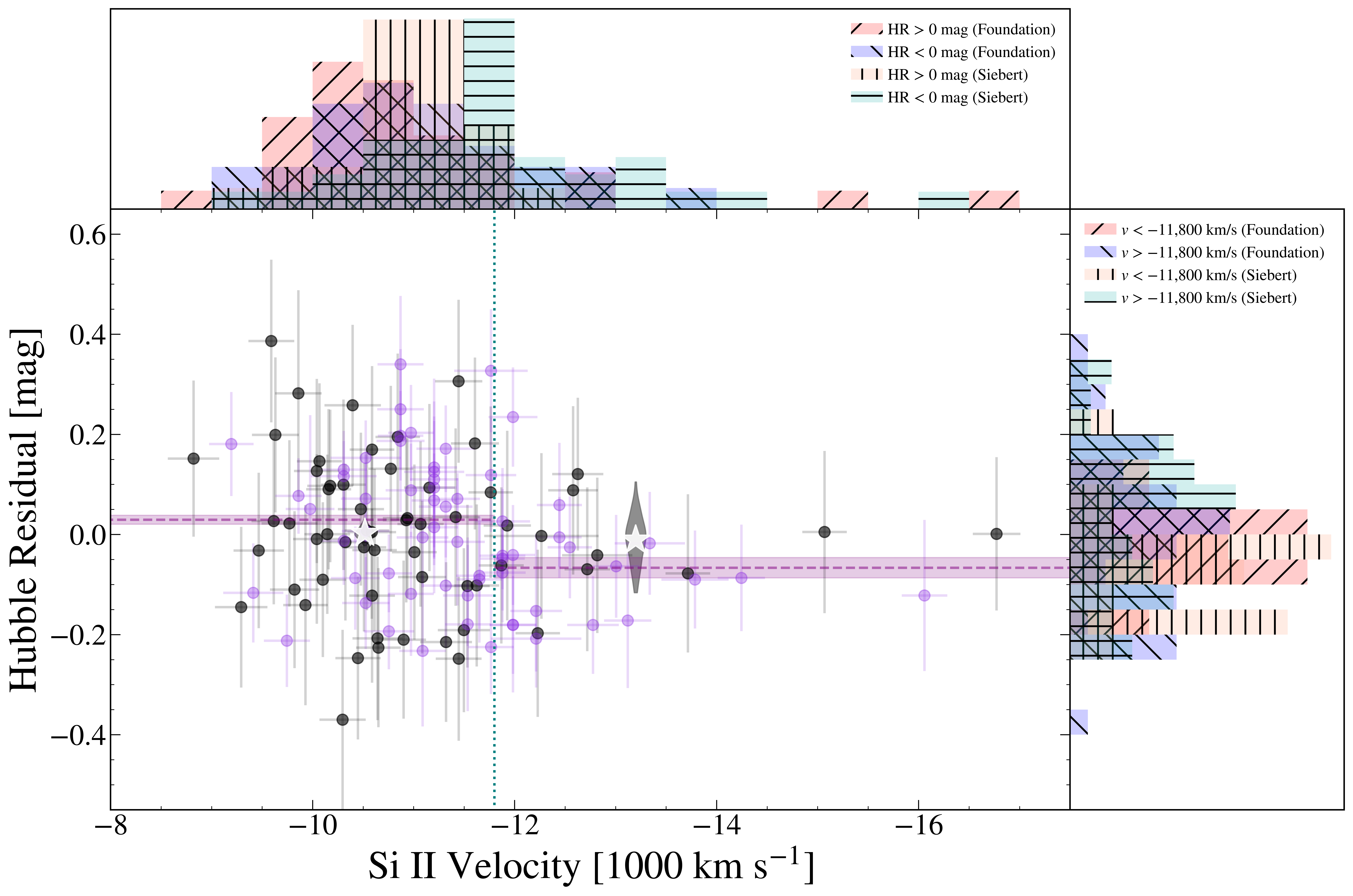}
    \caption{SALT2 Hubble residual vs \vsi\ for objects in the Foundation sample (black) and from \citet{Siebert20} (purple). The gray violin plots show the distribution of the weighted mean for bootstrap resamples of the Foundation data, accounting for uncertainties in both axes; the white stars show the mean of these values. The mean and uncertainty in the mean for the S20 data are shown in the purple dashed line and shaded region. Like in the color offset, the Foundation data show a smaller Hubble-residual difference between normal and high velocity, compared to the previous data.}
    \label{fig:hubbleResiduals_rotated}
\end{figure*}

\subsection{Hubble Residuals}
    
Looking specifically at the question of a relationship between supernova line velocity and luminosity on the Hubble diagram after standardization, i.e. Hubble residual (HR), \citet[hereafter S20]{Siebert20} find a bifurcation in data from W09/FK11 and CSP: SNe with negative HRs (overluminous SNe after standardization) have a systematically higher velocity than SNe with positive HR, with a mean \vsi\ difference of $-0.98 \pm 0.22 \times 10^3$ km s$^{-1}$ in the two samples. Similarly, S20 also find that objects with an expansion velocity above the median value ($-11 \times 10^3$ km s$^{-1}$) have a systematically lower HR, with a $0.091 \pm 0.035$ mag offset in the sample split about the median velocity. We re-examine this finding with the Foundation sample and light-curve fits from \citet{Jones2018}, using a similar method as S20 for bootstrap resampling to account for uncertainties. Our approach differs slightly from S20. As we have done throughout this paper, we use velocities at maximum light, whereas S20 used velocities measured at $+4$ days in their strongest result. For consistency with our measurements, we shift the S20 velocities to maximum light; S20 report largely similar results with maximum light velocities.
    
We find that the Foundation objects with a negative HR have a mean \vsi\ of $-11.07 \pm 0.17 \times 10^3$ km s$^{-1}$, while the Foundation objects with a positive HR have mean velocity of $-10.91 \pm 0.16 \times 10^3$ km s$^{-1}$. Though the Foundation result is in the same direction as what S20 find, the velocity difference in the Foundation subsamples of $-0.16 \pm 0.23 \times 10^{3}$ km s$^{-1}$ is not significant. In Fig.~\ref{fig:hubbleResiduals_rotated} we examine Hubble residuals versus \vsi\ for the Foundation and S20 samples (compare to S20 Figure 8). In our analysis we divide the samples at $-11,800$ km s$^{-1}$, corresponding to the high-velocity and normal-velocity split we have used throughout this paper, again differing slightly from S20 who divided the sample at the median velocity of $-11,000$ km s$^{-1}$. With this split, the S20 data still clearly show an HR offset (at higher significance, in fact) of $0.090 \pm 0.025$ mag, with high-velocity objects showing a more negative HR.  For the Foundation SNe, we find that the objects in the higher velocity bin have a weighted average HR of $-0.012 \pm 0.048$ mag and the objects in the lower velocity bin have a weighted average HR of $0.003 \pm 0.011$ mag, corresponding to an offset of $0.015 \pm 0.049$ mag. 
    
Again the Foundation results are consistent with those of S20, and in the same sense (high-velocity SN with negative HR), but the offset is not significant in the Foundation sample by itself. There is a clear difference in the velocity distribution between the samples, with Foundation showing a much smaller fraction of high-velocity SN~Ia. As discussed in \S~\ref{sec: mass}, the Foundation sample is based on wide-field untargeted surveys, yielding a large number of low-mass host-galaxies that are largely absent in the S20 sample. Nevertheless, we do not believe that host-galaxy differences alone drive the sample differences; Figure \ref{fig:velHostmass} shows that the paucity of high-velocity SN~Ia relative to other samples persists even among higher-mass host-galaxies. 
    
Understanding the origin of these sample differences, whether they are related to selection effects or not, is of clear importance. \citet{Pierel2020} show that, if uncorrected, the HR offset between velocity subsamples found by S20 could be a leading systematic uncertainty in future supernova surveys aiming at precision cosmology, such as from the Nancy Grace Roman Space Telescope \citep{Hounsell2018}. If, on the other hand, supernova cosmology samples are more similar to the Foundation objects, the effect may be muted.
    
\subsection{Color Offsets in Subsamples}\label{sec: cuts}
    
\begin{figure*}
    \centering
    \includegraphics[width=\textwidth]{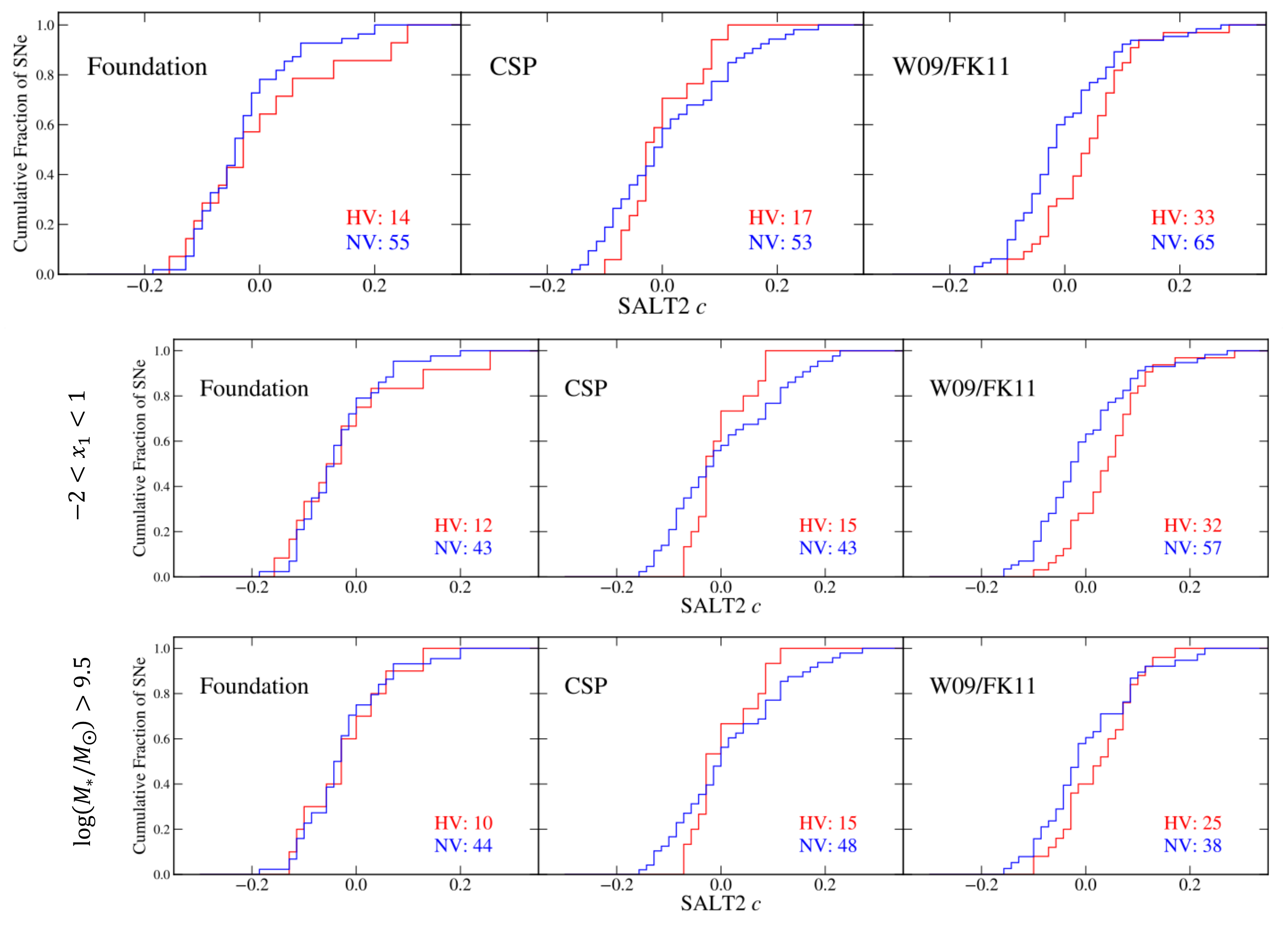}
    \caption{Cumulative distribution functions (CDFs) of the three samples' SALT2 c parameters. \textbf{Top:} CDFs using standard cosmological light-curve cuts. \textbf{Middle:} CDFs restricting SALT2 light-curve shape parameter $-2 < x_1 < +1$. \textbf{Bottom:} CDFs restricted to objects with measured host-galaxy masses $\log(M_{*}/M_{\odot}) > 9.5$.}
    \label{fig:cdfs}
\end{figure*}

Our results so far have shown strong connections between \vsi\ and both light-curve and environmental parameters. Here we circle back to one of the driving questions of our analysis: how do these affect the potential color offset between high-velocity and normal SN~Ia? In Fig.~\ref{fig:cdfs} we show cumulative distribution functions as a function of color for the three samples we analyze, split on velocity. FK11 used such an analysis to support the hypothesis of a color offset; indeed in the upper right panel we see a similar result to what they found: the color distribution of high-velocity objects has a similar shape to the normal objects, just shifted to a redder color. 
    
For the CSP objects, however, the shape of the CDF is different: a redder color offset for high-velocity SN~Ia is only clearly seen in the bluer part of the high-velocity distribution and there are more redder normal-velocity objects than in W09/FK11. The Foundation sample CDF looks different than either CSP or W09/FK11, with nearly identical colors among bluer objects, but with an indication of redder high-velocity objects at the reddest colors. The CSP and Foundation samples suggest that associating the differences between high-velocity and normal SN~Ia primarily with a color offset may not be warranted. A more fine-grained approach, taking into account light-curve properties, host-galaxy environment, and sample selection effects may be necessary.
    
FK11 found the color offset was most significant using a strict cut on light curve decline rate, $1.0 < \Delta m_{15} < 1.5$ mag, aiming particularly to exclude fast-decliners. Translating this to a cut on SALT2 $x_1$, we would need to restrict our sample to either $-2.1 < x_1 < +0.6$ \citep{Guy2007} or $-2.6 < x_1 < +0.2$ \citep{Siebert19}, depending on the transformation used. Unfortunately these cuts would eat into the bulk of the ``normal'' SN~Ia population on the slow-declining side. The SALT2 model is constructed so that the $x_1$ is distributed with zero mean and unit variance, and there are many objects where the transformed $\Delta m_{15}$ is between 0.9 and 1.0. To avoid making an overly stringent cut (that would exclude SN~Ia that are routinely found on cosmological Hubble diagrams), we investigate the color offset distribution for objects with $-2 < x_1 < +1$ in the middle panels of Fig.~\ref{fig:cdfs}. For all three samples, we do not see significant differences compared to the full cosmological $x_1$ distribution. 
    
Because the distribution of $x_1$ is markedly different between Foundation and the other datasets (see Figure ~\ref{fig:x1_c}), we have also investigated whether this could be playing a role in the lack of velocity color offset seen in Foundation. We resampled the W09/FK11 and CSP datasets to mimic an $x_1$ distribution based on Foundation (weighting objects based on the frequency of their $x_1$ value in the Foundation sample). We repeated this procedure 50 times; the median color offset of the W09 objects modified to have a Foundation-like $x_1$ distribution was $0.031$ mag, while for a combined W09/FK11+CSP sample it was $0.018$ mag. Thus we conclude that the difference in $x_1$ distribution between Foundation and other datasets does not explain the lack of a velocity color offset seen in the Foundation objects.

Similarly we can test if the difference in host-galaxy masses between Foundation the W09/FK11 and CSP could be playing a role in the color offset. In the bottom panel of Fig.~\ref{fig:cdfs}, we plot the CDFs for each sample with cuts applied such that there are only SNe with host-galaxy stellar masses $M_* > 10^{9.5} M_{\odot}$. Note that applying this cut requires a measured host-galaxy stellar mass, so these panels include only a subset of the full data seen in the top panel of Fig.~\ref{fig:cdfs}. There are subtle differences in each CDF, but the larger trends that were seen in the previous two cases persist, and our results are not sensitive to varying the host-galaxy stellar mass cut at $\log(M_*/M_\odot) > 9.0$, 9.5, or 10.0.
    
\section{Discussion and Conclusions}\label{sec: Discussion}

By combining data from Foundation, CSP, and W09/FK11, thus creating a larger sample than has previously been used to study correlations of this kind, we find that there is 2$\sigma$ evidence for a color offset, $\Delta c = 0.017 \pm 0.007$, between high- and normal-velocity SNe Ia based on SALT2 light-curve fits. The color offset is weakest (undetected) in the Foundation sample that is fundamentally different than the other two, comprising supernovae found from wide-field, untargeted surveys, unlike the higher-mass host galaxies targeted in the discovery surveys that yielded the W09/FK11 and CSP supernovae. The tail of low-mass host galaxies seen in the Foundation SN~Ia sample is in good accord with other untargeted surveys like PTF \citep{Pan2020}, but it does not appear that host-galaxy differences can entirely explain our results.
    
Restricting to the W09/FK11 sample, as seen in Fig.~\ref{fig:W09 Comp Fits}, we find a color offset with lower significance using a SALT2 analysis than from an analysis based on $B-V$ and $\Delta m_{15}$. This may be partly due to the difference between the color parameters in the two approaches, particularly over the restricted color range we analyze, where both intrinsic color variations and dust reddening play significant roles in the observed color. It may also be possible that, through its training, SALT2 has "learned" about velocity-dependent effects on the light curves, and the fit parameters are already accounting for these, partially mitigating the color offset \citep[though see also][]{Pierel2020}.
    
Nonetheless, even in a SALT2 based analysis, we see differences in the strength of both a velocity-dependent color offset and a velocity-dependent Hubble-residual step between subsamples, stronger in W09/FK11 and CSP than Foundation. Both Foundation and CSP also seem to show a lower fraction of high-velocity SN~Ia than W09/FK11. While the explosion velocity is clearly correlated with SN host-galaxy stellar mass (Fig.~\ref{fig:velHostmass}), restricting the samples to a similar range of host masses does not remove the color offset differences between the samples. Host-galaxy stellar mass is thus a confounding, not explanatory, factor in the analysis. If the velocity-based differences are due to sample selection (and not just ``unlucky'' small number statistics), the Foundation objects must be probing a different population of SN~Ia, even in high stellar-mass hosts, than W09/FK11 and CSP. It will be illuminating to repeat a color-offset analysis on upcoming low-redshift SN~Ia data sets; in particular, will results from CSP-II be more similar to CSP-I or Foundation?  
    
Other recent work has further explored the nature of high-velocity SN~Ia populations. \citet{Zhang2020} found that the sample of SNe from \citet{Siebert19} is well described with a bimodal velocity distribution consisting of a sharply peaked, normal velocity population and a more broadly distributed, higher-velocity population. They suggest high-velocity objects may be linked to asymmetric explosions; in that case, our results show that the geometry of the explosion would then also be correlated with environmental factors like the host-galaxy stellar mass.  \citet{Burrow20} also note a distinct high-velocity (what they call ``fast'') population in the CSP data, using a parameter space of maximum \textit{B}-band brightness, \vsi, and the pseudo-equivalent widths of Si II $\lambda6355$ and $\lambda5972$. They also note that these objects are redder than their normal counterparts, ascribing this due to dust reddening based on their location in color-color space and their preferential location in dusty or central regions of their hosts. \citet{Foley2012} also show evidence for a connection between SN ejecta velocity and circumstellar or interstellar absorption. Similar to our results, this suggests an intriguing connection between SN~Ia intrinsic properties (i.e., the explosion velocity) and environment.

Going forward, the near-term future of high-precision supernova cosmology, including from flagship surveys like the Vera Rubin Observatory Legacy Survey of Space and Time (LSST) and the Nancy Grace Roman Space Telescope, will be limited by systematic uncertainties. Astrophysical systematics, potentially including the effects of explosion velocity on SN~Ia luminosity and colors, will play an important role \citep{Pierel2020}. Further investigation is needed to determine whether direct spectroscopic measurements of supernova ejecta velocity will be necessary for the highest fidelity results. Given these will be large samples (thousands to hundreds of thousands of supernovae) of distant, thus faint, supernovae, it will be extremely difficult, if not prohibitive, to collect the spectroscopic data required for ejecta velocity measurements.

Our work suggests that perhaps some of the velocity-dependent information is encoded in the photometry, particularly in redder optical rest-frame bands (Figs.~\ref{fig:ri_BV} and \ref{fig:colDiff}). If so, supernovae observed with the Roman Space Telescope may be especially valuable, as its near-infrared data will cover rest-frame \textit{r} and \textit{i} for the majority of SN~Ia it can observe\footnote{One potential worry is that the lowest-redshift SN~Ia may not have sufficient rest-frame blue coverage with Roman alone.} \citep{Hounsell2018}. Velocity-dependent systematics may play a bigger role for Rubin Observatory LSST SN~Ia.

Improvements in light-curve fitting also provide a potential avenue to address velocity-dependent effects. Our results show that SALT2 may already be doing something ``under the hood'' to mitigate velocity effects (Fig.~\ref{fig:W09 Comp Fits}) or identify high-velocity objects (compare the right panels of Figs.~\ref{fig:ri_BV} and \ref{fig:colDiff}) relative to direct analysis of the $B-V$ color at maximum. Even if SALT2 is superseded in the future, it will be important to analyze the relationship between SN~Ia standardized luminosity, any color parameter (or parameters), and the explosion velocity. In fact, we suggest that, along with supernova photometry, host-galaxy information, etc., \vsi\ also be included in the training of future SN~Ia light-curve fitters. The benefit of this will be twofold: first, it will allow better distances to SN~Ia for which the explosion velocity can be spectroscopically measured; but also second, it can be used to unveil differences in light curves as a function of velocity (and host-galaxy properties, light-curve shape and colors) that can be applied to SN~Ia without measured velocities. Moreover, such an approach can go beyond the hints of velocity-dependent information we see in our SALT2 analysis, and remove the somewhat arbitrary bifurcation of our sample into just normal and high-velocity bins.

We are unfortunately lacking both physical understanding and empirical data to know whether the velocity correlations we are seeing at low redshift apply to high-redshift SN~Ia. The perhaps surprising connection between intrinsic properties of the supernova explosions (like ejecta velocity or light-curve shape) and environmental factors (interstellar dust, host-galaxy stellar mass, star-formation rate, or morphology) must presumably be mediated through the progenitor systems and explosion mechanisms of SN~Ia. 

Even as we develop a better picture of SN~Ia progenitor and explosion channels, and thus the underlying population of supernovae, our analysis of differences between subsamples suggests that survey selection effects are not always straightforward to account for, even in the nearby Universe. It may be that the best use of large, future SN~Ia surveys will be, counterintuitively, to provide more \emph{restricted} and homogeneous samples. For instance, SN~Ia in the lowest-mass host galaxies may be more uniform and standardizable than the rest \citep{ScolnicBrout20}; they tend to occupy a rather narrow range of light-curve parameters (Fig.~\ref{fig:x1_c}, lower panels). Confirming the results of \citet{Pan2015} and \citet{Pan2020}, we show that these objects also predominantly have normal ejecta velocities (Fig.~\ref{fig:velHostmass}), leaving little leverage for velocity-dependent effects on luminosity or colors to operate. If these supernovae also come from young stellar populations, they could remain an identifiably homogeneous class at every cosmic epoch. It may be that the biggest cosmological benefit of upcoming huge SN~Ia surveys, like Rubin and Roman, will be to provide large samples of these ``most standard'' of our standardizable candles.

\acknowledgments

We thank Rick Kessler and Lou Strolger for helpful suggestions.

This research at Rutgers University (K.G.D., S.W.J., M.D.) was supported by NASA contract NNG16PJ34C and DOE award DE-SC0011636. M.D.\ is also supported by the Horizon Fellowship at the Johns Hopkins University.

The UCSC team is supported in part by NASA grant NNG17PX03C, NSF grants AST-1518052 and AST-1815935, the Gordon \& Betty Moore Foundation, the Heising-Simons Foundation, and by a fellowship from the David and Lucile Packard Foundation to R.J.F.  M.R.S.\ is supported by the National Science Foundation Graduate Research Fellowship Program Under Grant No.\ 1842400.  D.A.C.\ acknowledges support from the National Science Foundation Graduate Research Fellowship under Grant DGE1339067.

This paper is based in part on observations made with the Southern African Large Telescope (SALT) using the Robert Stobie Spectrograph (RSS), through allocations made to Rutgers University via programs 2015-1-MLT-002, 2016-1-MLT-007, and 2017-1-MLT-002 (PI: S.W.\ Jha).

Based in part on observations obtained at the Southern Astrophysical Research (SOAR) telescope (NOIRLab Prop.\ IDs 2015A-0253, 2015B-0313, 2017A-0306, 2017B-0169, 2018A-0277; PI: R.\ Foley), which is a joint project of the Minist\'erio da Ci\^encia, Tecnologia e Inova\c{c}\~oes do Brasil (MCTI/LNA), the US National Science Foundation's NOIRLab, the University of North Carolina at Chapel Hill (UNC), and Michigan State University (MSU).  

Based in part on observations at Kitt Peak National Observatory at NSF's NOIRLab (NOIRLab Prop.\ IDs 2015A-0253, 2015B-0313, 2017A-0306, 2017B-0169, 2018A-0277; PI: R.\ Foley), which is managed by the Association of Universities for Research in Astronomy (AURA) under a cooperative agreement with the National Science Foundation. The authors are honored to be permitted to conduct astronomical research on Iolkam Du'ag (Kitt Peak), a mountain with particular significance to the Tohono O'odham.

A major upgrade of the Kast spectrograph on the Shane 3 m telescope at Lick Observatory was made possible through generous gifts from the Heising-Simons Foundation as well as William and Marina Kast.  Research at Lick Observatory is partially supported by a generous gift from Google. 

Some of The data presented herein were obtained at the W.\ M.\ Keck Observatory, which is operated as a scientific partnership among the California Institute of Technology, the University of California and the National Aeronautics and Space Administration. The Observatory was made possible by the generous financial support of the W.\ M.\ Keck Foundation.  The authors wish to recognize and acknowledge the very significant cultural role and reverence that the summit of Maunakea has always had within the indigenous Hawaiian community.  We are most fortunate to have the opportunity to conduct observations from this mountain.

\bibliography{foundCorr}
\bibliographystyle{aasjournal}

\appendix

\startlongtable
\begin{deluxetable*}{lcDcDcDcccccDcDcc}
\tablecaption{Supernova velocities, light-curve parameters, and Hubble residuals for the Foundation sample. \label{table: data with cuts}}
\tabletypesize{\scriptsize}
\tablehead{\colhead{Supernova} & \colhead{$z$} & \twocolhead{\vsimax} & \colhead{$\pm$} & \twocolhead{$c$} & \colhead{$\pm$}  & \twocolhead{$x_1$} & \colhead{$\pm$} & \colhead{$t_0$} & \colhead{$\pm$} & \colhead{$m_B$} & \colhead{$\pm$}  & \twocolhead{$\mu_{\rm res}$} & \colhead{$\pm$} & \twocolhead{Host mass} & \colhead{$\pm$} & \colhead{Figures} \\[-8pt]
\colhead{} & \colhead{} & \multicolumn{3}{c}{(10$^3$ km s$^{-1}$)} &  \multicolumn{3}{c}{}  &  \multicolumn{3}{c}{} & \colhead{(MJD)} & \colhead{} & \multicolumn{2}{c}{(mag)} & \multicolumn{3}{c}{(mag)} & \multicolumn{3}{c}{log ($M_*/M_\odot$)} & \colhead{}
}
\decimals
\startdata
2016afk & $0.046$ & $-12.62$ & $0.25$ & $0.035$ & $0.035$ & $-0.27$ & $0.20$ & $57440.77$ & $0.25$ & $17.48$ & $0.04$ & $0.12$ & $0.15$ & $9.77$ & $0.16$ & 2,4,6,7,8,9a,9b,9c \\ 
2016coj & $0.005$ & $-11.95$ & $0.22$ & $-0.011$ & $0.032$ & $-1.51$ & $0.07$ & $57548.16$ & $0.09$ & $12.84$ & $0.04$ & ... & ... & $10.83$ & $0.16$ & 6,7 \\ 
2016cor & $0.050$ & $-9.63$ & $0.23$ & $0.144$ & $0.034$ & $-0.57$ & $0.28$ & $57541.73$ & $0.24$ & $18.10$ & $0.04$ & $0.20$ & $0.15$ & $10.29$ & $0.16$ & 2,4,6,7,8,9a,9b,9c \\ 
2016cvv & $0.045$ & $-12.26$ & $0.57$ & $0.067$ & $0.038$ & $1.37$ & $0.29$ & $57558.34$ & $0.62$ & $17.21$ & $0.05$ & $-0.00$ & $0.17$ & $11.10$ & $0.16$ & 2,4,6,7,8,9a,9c \\ 
2016esh & $0.045$ & $-10.45$ & $0.22$ & $-0.097$ & $0.038$ & $-1.32$ & $0.25$ & $57613.29$ & $0.20$ & $16.83$ & $0.04$ & $-0.25$ & $0.16$ & $9.94$ & $0.16$ & 2,4,6,7,8,9a,9b,9c \\ 
2016glp & $0.085$ & $-16.77$ & $0.24$ & $0.133$ & $0.041$ & $0.96$ & $0.35$ & $57658.74$ & $0.22$ & $18.86$ & $0.06$ & $0.00$ & $0.15$ & $10.90$ & $0.16$ & 2,4,6,7,8,9a,9b,9c \\ 
2016gmg & $0.049$ & $-10.93$ & $0.23$ & $-0.046$ & $0.041$ & $-1.76$ & $0.26$ & $57662.14$ & $0.22$ & $17.52$ & $0.05$ & $0.03$ & $0.16$ & $11.40$ & $0.16$ & 2,4,6,7,8,9a,9b,9c \\ 
2016gsu & $0.076$ & $-10.04$ & $0.23$ & $-0.114$ & $0.045$ & $-1.88$ & $0.54$ & $57671.15$ & $0.47$ & $18.30$ & $0.04$ & $0.13$ & $0.18$ & $10.37$ & $0.16$ & 2,4,6,7,8,9a,9b,9c \\ 
2016hhv & $0.062$ & $-11.42$ & $0.24$ & $-0.175$ & $0.056$ & $-0.76$ & $0.45$ & $57691.77$ & $1.83$ & $17.60$ & $0.11$ & $0.03$ & $0.18$ & $11.50$ & $0.16$ & 2,4,6,7,8,9a,9b,9c \\ 
2016htn & $0.053$ & $-10.32$ & $0.23$ & $0.080$ & $0.035$ & $-0.18$ & $0.32$ & $57701.31$ & $0.26$ & $17.78$ & $0.04$ & $-0.02$ & $0.16$ & $10.74$ & $0.16$ & 2,4,6,7,8,9a,9b,9c \\ 
2016ixf & $0.066$ & $-10.84$ & $0.22$ & $-0.107$ & $0.039$ & $0.76$ & $0.33$ & $57742.63$ & $0.71$ & $17.79$ & $0.05$ & $0.19$ & $0.17$ & $7.87$ & $0.16$ & 2,4,6,7,8,9a,9b \\ 
2017cfc & $0.024$ & $-10.04$ & $0.24$ & $-0.008$ & $0.030$ & $-0.90$ & $0.08$ & $57837.52$ & $0.10$ & $15.89$ & $0.03$ & $-0.01$ & $0.17$ & $10.65$ & $0.16$ & 2,4,6,7,8,9a,9b,9c \\ 
2017cii & $0.033$ & $-10.16$ & $0.22$ & $-0.103$ & $0.033$ & $0.34$ & $0.14$ & $57837.44$ & $0.54$ & $16.28$ & $0.04$ & $0.09$ & $0.16$ & $9.78$ & $0.16$ & 2,4,6,7,8,9a,9b,9c \\ 
2017ciy & $0.038$ & $-9.59$ & $0.23$ & $-0.012$ & $0.035$ & $1.20$ & $0.19$ & $57838.64$ & $0.55$ & $17.09$ & $0.04$ & $0.39$ & $0.16$ & $10.21$ & $0.16$ & 2,4,6,7,8,9a,9c \\ 
2017cjv & $0.060$ & $-11.07$ & $0.23$ & $-0.102$ & $0.037$ & $-1.27$ & $0.29$ & $57839.93$ & $0.54$ & $17.69$ & $0.04$ & $0.02$ & $0.17$ & $10.81$ & $0.16$ & 2,4,6,7,8,9a,9b,9c \\ 
2017ckx & $0.027$ & $-9.86$ & $0.23$ & $-0.027$ & $0.038$ & $2.10$ & $0.59$ & $57846.02$ & $1.01$ & $16.02$ & $0.04$ & $0.28$ & $0.21$ & $10.47$ & $0.16$ & 2,4,6,7,8,9a,9c \\ 
2017cpu & $0.054$ & $-11.45$ & $0.23$ & $-0.053$ & $0.035$ & $0.74$ & $0.31$ & $57847.69$ & $0.56$ & $17.11$ & $0.04$ & $-0.25$ & $0.16$ & $10.67$ & $0.16$ & 2,4,6,7,8,9a,9b,9c \\ 
2017hn & $0.024$ & $-11.01$ & $0.22$ & $-0.014$ & $0.033$ & $0.56$ & $0.18$ & $57768.25$ & $0.32$ & $15.64$ & $0.04$ & $-0.04$ & $0.17$ & $10.65$ & $0.16$ & 2,4,6,7,8,9a,9b,9c \\ 
2017mf & $0.026$ & $-10.59$ & $0.22$ & $-0.043$ & $0.031$ & $-0.03$ & $0.11$ & $57779.00$ & $0.23$ & $15.99$ & $0.03$ & $0.17$ & $0.17$ & $10.80$ & $0.16$ & 2,4,6,7,8,9a,9b,9c \\ 
2017oz & $0.056$ & $-13.72$ & $0.22$ & $-0.022$ & $0.035$ & $-0.10$ & $0.28$ & $57788.03$ & $0.27$ & $17.51$ & $0.04$ & $-0.08$ & $0.16$ & $10.04$ & $0.16$ & 2,4,6,7,8,9a,9b,9c \\ 
2017po & $0.032$ & $-12.72$ & $0.23$ & $-0.069$ & $0.032$ & $0.22$ & $0.22$ & $57784.33$ & $0.60$ & $16.13$ & $0.04$ & $-0.07$ & $0.16$ & $9.39$ & $0.16$ & 2,4,6,7,8,9a,9b \\ 
2017yk & $0.046$ & $-10.45$ & $0.26$ & ... & ... & ... & ... & ... & ... & ... & ... & ... & ... & $10.62$ & $0.16$ & 6 \\ 
2017zd & $0.029$ & $-10.64$ & $0.25$ & $-0.043$ & $0.031$ & $0.24$ & $0.09$ & $57793.10$ & $0.10$ & $15.89$ & $0.04$ & $-0.21$ & $0.16$ & $10.11$ & $0.16$ & 2,4,6,7,8,9a,9b,9c \\ 
ASASSN-15fa & $0.027$ & $-10.24$ & $0.22$ & ... & ... & ... & ... & ... & ... & ... & ... & ... & ... & $10.86$ & $0.16$ & 6 \\ 
ASASSN-15go & $0.019$ & $-13.08$ & $0.22$ & $0.235$ & $0.045$ & $2.80$ & $0.67$ & $57122.39$ & $0.05$ & $15.86$ & $0.05$ & ... & ... & ... & ... & 2,4 \\ 
ASASSN-15il & $0.023$ & $-10.83$ & $0.22$ & $-0.118$ & $0.032$ & $1.30$ & $0.10$ & $57161.31$ & $0.16$ & $15.16$ & $0.05$ & $-0.09$ & $0.17$ & $10.13$ & $0.16$ & 2,4,6,7,8 \\ 
ASASSN-15mg & $0.043$ & $-10.65$ & $0.23$ & $-0.032$ & $0.035$ & $-1.55$ & $0.15$ & $57218.11$ & $0.13$ & $17.22$ & $0.04$ & $0.03$ & $0.16$ & $10.13$ & $0.16$ & 2,4,6,7,8 \\ 
ASASSN-15np & $0.038$ & $-9.29$ & $0.26$ & $0.059$ & $0.034$ & $-1.07$ & $0.21$ & $57241.66$ & $0.53$ & $17.05$ & $0.04$ & $-0.15$ & $0.16$ & $10.46$ & $0.16$ & 2,4,6,7,8 \\ 
ASASSN-15nr & $0.023$ & $-10.59$ & $0.23$ & $0.041$ & $0.033$ & $1.25$ & $0.16$ & $57250.08$ & $0.12$ & $15.57$ & $0.04$ & $-0.12$ & $0.17$ & $10.27$ & $0.16$ & 2,4,6,7,8 \\ 
ASASSN-15od & $0.018$ & $-11.93$ & $0.23$ & $-0.086$ & $0.033$ & $-1.20$ & $0.12$ & $57256.58$ & $0.25$ & $15.04$ & $0.04$ & $0.02$ & $0.19$ & $9.72$ & $0.16$ & 2,4,6,7,8 \\ 
ASASSN-15pn & $0.038$ & $-10.94$ & $0.24$ & $0.011$ & $0.033$ & $0.42$ & $0.16$ & $57280.42$ & $0.23$ & $16.94$ & $0.04$ & $0.03$ & $0.16$ & $9.48$ & $0.16$ & 2,4,6,7,8 \\ 
ASASSN-15pr & $0.033$ & $-11.61$ & $0.24$ & $-0.048$ & $0.042$ & $-0.70$ & $0.24$ & $57282.95$ & $0.94$ & $16.64$ & $0.06$ & $0.18$ & $0.17$ & $9.88$ & $0.16$ & 2,4,6,7,8 \\ 
ASASSN-15sf & $0.025$ & $-9.87$ & $0.22$ & $-0.069$ & $0.031$ & $0.79$ & $0.10$ & $57333.39$ & $0.13$ & $15.68$ & $0.03$ & $0.12$ & $0.17$ & $9.37$ & $0.16$ & 2,3b,4,6,7,8 \\ 
ASASSN-15ss & $0.036$ & $-10.61$ & $0.22$ & $-0.111$ & $0.038$ & $0.79$ & $0.23$ & $57338.15$ & $0.79$ & $16.32$ & $0.06$ & $-0.03$ & $0.16$ & $10.84$ & $0.16$ & 2,4,6,7,8 \\ 
ASASSN-15tg & $0.036$ & $-11.50$ & $0.26$ & $-0.041$ & $0.033$ & $1.72$ & $0.20$ & $57361.08$ & $0.39$ & $16.25$ & $0.03$ & $-0.19$ & $0.16$ & $10.29$ & $0.16$ & 2,4,6,7,8 \\ 
ASASSN-15uu & $0.027$ & $-10.70$ & $0.26$ & ... & ... & ... & ... & ... & ... & ... & ... & ... & ... & $9.74$ & $0.16$ & 6 \\ 
ASASSN-15uv & $0.020$ & $-12.35$ & $0.22$ & $0.267$ & $0.030$ & $0.30$ & $0.10$ & $57392.46$ & $0.14$ & $16.32$ & $0.04$ & $0.13$ & $0.18$ & ... & ... & 2,4 \\ 
ASASSN-15uw & $0.031$ & $-12.33$ & $0.22$ & $0.008$ & $0.031$ & $-0.46$ & $0.10$ & $57392.99$ & $0.13$ & $16.57$ & $0.03$ & $0.14$ & $0.16$ & $10.42$ & $0.16$ & 2,4,6,7,8 \\ 
ASASSN-16ad & $0.016$ & $-12.35$ & $0.22$ & ... & ... & ... & ... & ... & ... & ... & ... & ... & ... & $9.46$ & $0.16$ & 6 \\ 
ASASSN-16aj & $0.031$ & $-11.62$ & $0.22$ & $-0.018$ & $0.033$ & $-1.76$ & $0.14$ & $57401.97$ & $0.37$ & $16.41$ & $0.04$ & $-0.10$ & $0.16$ & $10.82$ & $0.16$ & 2,4,6,7,8 \\ 
ASASSN-16ay & $0.028$ & $-11.16$ & $0.25$ & $0.209$ & $0.033$ & $1.04$ & $0.14$ & $57416.48$ & $0.45$ & $16.76$ & $0.04$ & $0.09$ & $0.17$ & $10.18$ & $0.16$ & 2,4,6,7,8 \\ 
ASASSN-16bq & $0.025$ & $-12.58$ & $0.25$ & $-0.052$ & $0.032$ & $-1.69$ & $0.06$ & $57437.87$ & $0.18$ & $16.04$ & $0.04$ & $0.09$ & $0.17$ & $10.84$ & $0.16$ & 2,6,7,8 \\ 
ASASSN-16ch & $0.027$ & $-11.04$ & $0.28$ & $0.308$ & $0.032$ & $-0.41$ & $0.07$ & $57454.84$ & $0.14$ & $17.13$ & $0.04$ & ... & ... & $10.01$ & $0.16$ & 4,6,7 \\ 
ASASSN-16ct & $0.042$ & $-9.82$ & $0.24$ & $-0.008$ & $0.032$ & $1.54$ & $0.17$ & $57459.81$ & $0.44$ & $16.80$ & $0.04$ & $-0.11$ & $0.16$ & $9.32$ & $0.16$ & 2,4,6,7,8 \\ 
ASASSN-16dw & $0.035$ & $-10.65$ & $0.22$ & $0.072$ & $0.032$ & $-0.11$ & $0.12$ & $57482.49$ & $0.27$ & $16.71$ & $0.04$ & $-0.23$ & $0.16$ & $10.28$ & $0.16$ & 2,4,6,7,8 \\ 
ASASSN-16fo & $0.029$ & $-15.07$ & $0.22$ & $-0.113$ & $0.030$ & $0.54$ & $0.14$ & $57540.13$ & $0.29$ & $15.83$ & $0.03$ & $0.00$ & $0.16$ & $10.23$ & $0.16$ & 2,4,6,7,8 \\ 
ASASSN-16fs & $0.029$ & $-12.23$ & $0.22$ & $-0.024$ & $0.034$ & $-1.42$ & $0.12$ & $57542.50$ & $0.30$ & $16.14$ & $0.04$ & $-0.20$ & $0.17$ & $10.97$ & $0.16$ & 2,4,6,7,8 \\ 
ASASSN-16hr & $0.031$ & $-16.83$ & $0.94$ & ... & ... & ... & ... & ... & ... & ... & ... & ... & ... & $10.04$ & $0.16$ & 6 \\ 
ASASSN-16jf & $0.011$ & $-10.83$ & $0.24$ & $-0.030$ & $0.043$ & $-2.26$ & $0.18$ & $57630.05$ & $0.22$ & $14.56$ & $0.05$ & $0.19$ & $0.25$ & ... & ... & 2,4 \\ 
ASASSN-16la & $0.015$ & $-9.93$ & $0.22$ & $-0.102$ & $0.032$ & $0.84$ & $0.09$ & $57671.75$ & $0.09$ & $14.24$ & $0.05$ & $-0.14$ & $0.20$ & $9.14$ & $0.16$ & 2,4,6,7,8 \\ 
ASASSN-16lg & $0.021$ & $-10.30$ & $0.23$ & $-0.039$ & $0.034$ & $-1.32$ & $0.15$ & $57674.57$ & $0.17$ & $15.21$ & $0.04$ & $-0.37$ & $0.18$ & $10.57$ & $0.16$ & 2,4,6,7,8 \\ 
ASASSN-17aj & $0.021$ & $-10.36$ & $0.22$ & $0.001$ & $0.030$ & $0.19$ & $0.10$ & $57773.09$ & $0.10$ & $15.53$ & $0.03$ & $0.01$ & $0.17$ & $10.48$ & $0.16$ & 2,4,6,7,8 \\ 
ASASSN-17at & $0.025$ & $-11.76$ & $0.22$ & $-0.015$ & $0.033$ & $-1.55$ & $0.21$ & $57777.02$ & $0.04$ & $16.16$ & $0.04$ & $0.08$ & $0.17$ & $10.31$ & $0.16$ & 2,4,6,7,8 \\ 
ASASSN-17eb & $0.049$ & $-10.90$ & $0.22$ & $0.051$ & $0.036$ & $-1.52$ & $0.18$ & $57836.89$ & $0.50$ & $17.52$ & $0.05$ & $-0.21$ & $0.16$ & $11.07$ & $0.16$ & 2,4,6,7,8 \\ 
ATLAS16agv & $0.048$ & $-10.10$ & $0.23$ & $-0.088$ & $0.032$ & $1.37$ & $0.18$ & $57469.94$ & $0.31$ & $16.84$ & $0.03$ & $-0.09$ & $0.15$ & $5.00$ & $0.16$ & 2,4,6,7,8,9a \\ 
ATLAS16bwu & $0.072$ & $-9.47$ & $0.25$ & $0.044$ & $0.037$ & $0.08$ & $0.28$ & $57609.32$ & $0.47$ & $18.22$ & $0.04$ & $-0.03$ & $0.16$ & $10.01$ & $0.16$ & 2,4,6,7,8,9a,9b,9c \\ 
ATLAS17ajn & $0.029$ & $-10.31$ & $0.22$ & $-0.099$ & $0.038$ & $-2.02$ & $0.11$ & $57776.45$ & $0.44$ & $16.27$ & $0.06$ & $0.10$ & $0.17$ & $10.62$ & $0.16$ & 2,4,6,7,8,9a,9c \\ 
ATLAS17axb & $0.032$ & $-11.53$ & $0.22$ & $-0.081$ & $0.033$ & $-0.16$ & $0.12$ & $57796.67$ & $0.18$ & $16.10$ & $0.04$ & $-0.10$ & $0.16$ & $9.84$ & $0.16$ & 2,4,6,7,8,9a,9b,9c \\ 
CSS160129 & $0.068$ & $-11.87$ & $0.22$ & $-0.149$ & $0.035$ & $0.60$ & $0.17$ & $57416.44$ & $0.60$ & $17.52$ & $0.04$ & $-0.06$ & $0.16$ & $8.16$ & $0.16$ & 2,4,6,7,8,9a,9b \\ 
Gaia16bba & $0.030$ & $-10.06$ & $0.22$ & ... & ... & ... & ... & ... & ... & ... & ... & ... & ... & $8.18$ & $0.16$ & 6 \\ 
PS15ahs & $0.026$ & $-9.61$ & $0.22$ & $-0.072$ & $0.030$ & $0.18$ & $0.07$ & $57156.71$ & $0.08$ & $15.73$ & $0.03$ & $0.03$ & $0.17$ & $9.82$ & $0.16$ & 2,4,6,7,8,9a,9b,9c \\ 
PS15aii & $0.047$ & $-10.48$ & $0.23$ & $-0.051$ & $0.031$ & $0.44$ & $0.12$ & $57158.90$ & $0.13$ & $17.12$ & $0.03$ & $0.05$ & $0.15$ & $10.90$ & $0.16$ & 2,4,6,7,8,9a,9b,9c \\ 
PS15bwh & $0.073$ & $-11.32$ & $0.26$ & $-0.075$ & $0.036$ & $0.70$ & $0.19$ & $57279.02$ & $0.13$ & $17.67$ & $0.04$ & $-0.22$ & $0.16$ & $5.00$ & $0.16$ & 2,4,6,7,8,9a,9b \\ 
PS15bzz & $0.080$ & $-11.08$ & $0.24$ & $-0.018$ & $0.031$ & $-1.00$ & $0.25$ & $57284.45$ & $0.61$ & $18.34$ & $0.04$ & $-0.09$ & $0.15$ & $9.59$ & $0.16$ & 2,4,6,7,8,9a,9b,9c \\ 
PS15coh & $0.019$ & $-10.78$ & $0.22$ & ... & ... & ... & ... & ... & ... & ... & ... & ... & ... & $8.63$ & $0.16$ & 6 \\ 
PS15cze & $0.039$ & $-11.45$ & $0.23$ & $0.031$ & $0.037$ & $-1.53$ & $0.23$ & $57364.32$ & $0.16$ & $17.57$ & $0.05$ & $0.30$ & $0.16$ & $10.81$ & $0.16$ & 2,4,6,7,8,9a,9b,9c \\ 
PS16bby & $0.053$ & $-10.54$ & $0.22$ & $0.004$ & $0.037$ & $-0.86$ & $0.25$ & $57466.90$ & $0.21$ & $17.68$ & $0.05$ & $0.01$ & $0.16$ & $9.93$ & $0.16$ & 2,4,6,7,8,9a,9b,9c \\ 
PS16bnz & $0.063$ & $-12.82$ & $0.35$ & $-0.128$ & $0.034$ & $-1.12$ & $0.21$ & $57487.48$ & $0.15$ & $17.67$ & $0.04$ & $-0.04$ & $0.16$ & $11.66$ & $0.16$ & 2,4,6,7,8,9a,9b,9c \\ 
PS16cqa & $0.044$ & $-10.91$ & $0.25$ & $0.174$ & $0.056$ & $-2.77$ & $0.36$ & $57545.26$ & $0.81$ & $18.23$ & $0.08$ & ... & ... & ... & ... & 2,4,9a \\ 
PS16cvc & $0.030$ & $-11.32$ & $0.22$ & ... & ... & ... & ... & ... & ... & ... & ... & ... & ... & $9.76$ & $0.16$ & 6 \\ 
PS16dnp & $0.051$ & $-8.82$ & $0.25$ & $-0.081$ & $0.034$ & $0.96$ & $0.24$ & $57597.83$ & $0.34$ & $17.28$ & $0.04$ & $0.15$ & $0.16$ & $8.39$ & $0.16$ & 2,4,6,7,8,9a,9b \\ 
PS16em & $0.070$ & $-10.61$ & $0.23$ & $0.206$ & $0.087$ & $-1.01$ & $0.19$ & $57386.50$ & $1.06$ & $18.80$ & $0.13$ & ... & ... & $10.62$ & $0.16$ & 2,4,6,7,9a,9b,9c \\ 
PS16evk & $0.054$ & $-9.77$ & $0.28$ & $-0.020$ & $0.039$ & $1.45$ & $0.32$ & $57694.72$ & $0.73$ & $17.39$ & $0.05$ & $0.02$ & $0.17$ & $10.25$ & $0.16$ & 2,4,6,7,8,9a,9c \\ 
PS16fa & $0.046$ & $-10.17$ & $0.23$ & $-0.004$ & $0.031$ & $-0.64$ & $0.09$ & $57400.33$ & $0.18$ & $17.42$ & $0.03$ & $0.10$ & $0.15$ & $10.76$ & $0.16$ & 2,4,6,7,8,9a,9b,9c \\ 
PS16fbb & $0.052$ & $-10.07$ & $0.22$ & $-0.119$ & $0.035$ & $0.70$ & $0.23$ & $57719.61$ & $0.21$ & $17.26$ & $0.04$ & $0.14$ & $0.16$ & $9.38$ & $0.16$ & 2,4,6,7,8,9a,9b \\ 
PS16n & $0.053$ & $-10.14$ & $0.23$ & $-0.055$ & $0.039$ & $-0.31$ & $0.18$ & $57392.52$ & $0.17$ & $17.41$ & $0.05$ & $-0.00$ & $0.16$ & $10.33$ & $0.16$ & 2,4,6,7,8,9a,9b,9c \\ 
PS17bii & $0.073$ & $-10.59$ & $0.24$ & $-0.093$ & $0.043$ & $-0.95$ & $0.37$ & $57811.77$ & $0.36$ & $18.08$ & $0.05$ & $0.00$ & $0.17$ & $10.69$ & $0.16$ & 2,4,6,7,8,9a,9b,9c \\ 
PSNJ0153424 & $0.026$ & $-10.51$ & $0.22$ & $0.072$ & $0.031$ & $-0.58$ & $0.07$ & $57393.66$ & $0.09$ & $16.21$ & $0.04$ & $-0.03$ & $0.17$ & $10.66$ & $0.16$ & 2,4,6,7,8,9a,9b,9c \\ 
PTSS-16efw & $0.036$ & $-10.40$ & $0.28$ & $-0.118$ & $0.037$ & $-0.09$ & $0.15$ & $57509.43$ & $0.70$ & $16.71$ & $0.07$ & $0.26$ & $0.16$ & $10.74$ & $0.16$ & 2,4,6,7,8,9a,9b,9c \\ 

\enddata
\tablecomments{Uncertainties for the host-galaxy mass measurements in subsequent tables are based on the standard deviation of individual mass measurements from various sources. Objects with only a single mass measurement are given the median uncertainty of the sample where there are two or more measurements (see section \ref{sec: mass data}). The masses for the Foundation host-galaxies are taken only from \citet{Jones2018}; thus their uncertainties are uniform. The last column of this and two subsequent tables show which objects are used in which Figures in this paper.}
\end{deluxetable*}

\startlongtable
\begin{deluxetable*}{lcDcDcDcccccDcc}
\tablecaption{Supernova velocities and light-curve parameters for the CSP sample. \label{table: csp data with cuts}}
\tabletypesize{\scriptsize}
\tablehead{\colhead{Supernova} & \colhead{$z$} & \twocolhead{\vsimax} & \colhead{$\pm$} & \twocolhead{$c$} & \colhead{$\pm$}  & \twocolhead{$x_1$} & \colhead{$\pm$} & \colhead{$t_0$} & \colhead{$\pm$} & \colhead{$m_B$} & \colhead{$\pm$} & \twocolhead{Host mass} & \colhead{$\pm$} & \colhead{Figures} \\[-8pt]
\colhead{} & \colhead{} & \multicolumn{3}{c}{(10$^3$ km s$^{-1}$)} &  \multicolumn{3}{c}{}  &  \multicolumn{3}{c}{} & \colhead{(MJD)} & \colhead{} & \multicolumn{2}{c}{(mag)} & \multicolumn{3}{c}{log ($M_*/M_\odot$)} & \colhead{}
}
\decimals
\startdata
2004dt & $0.020$ & $-15.93$ & $0.22$ & $-0.041$ & $0.025$ & $-0.30$ & $0.03$ & $53239.79$ & $0.03$ & $14.95$ & $0.03$ & $10.55$ & $0.16$ & 1,2,4,6,7,9a,9b,9c \\ 
2004ef & $0.031$ & $-12.29$ & $0.23$ & $0.087$ & $0.021$ & $-1.40$ & $0.02$ & $53264.22$ & $0.03$ & $16.61$ & $0.02$ & $10.90$ & $0.14$ & 1,2,3a,3b,4,6,7,9a,9b,9c \\ 
2004eo & $0.015$ & $-10.57$ & $0.22$ & $0.005$ & $0.022$ & $-1.22$ & $0.02$ & $53278.42$ & $0.03$ & $14.84$ & $0.03$ & $11.09$ & $0.14$ & 1,2,3a,3b,4,6,7,9a,9b,9c \\ 
2004ey & $0.016$ & $-11.20$ & $0.40$ & $-0.131$ & $0.020$ & $0.08$ & $0.02$ & $53304.53$ & $0.02$ & $14.46$ & $0.03$ & $10.01$ & $0.14$ & 1,2,3a,3b,4,6,7,9a,9b,9c \\ 
2004gc & $0.032$ & $-10.34$ & $0.49$ & $0.092$ & $0.023$ & $-0.64$ & $0.04$ & $53326.26$ & $0.14$ & $16.55$ & $0.04$ & $10.43$ & $0.16$ & 2,3a,3b,4,6,7,9a,9b,9c \\ 
2004gs & $0.028$ & $-11.48$ & $0.24$ & $0.123$ & $0.022$ & $-1.88$ & $0.03$ & $53356.03$ & $0.06$ & $16.90$ & $0.02$ & $10.82$ & $0.16$ & 1,2,3a,3b,4,6,7,9a,9b,9c \\ 
2004gu & $0.046$ & $-10.63$ & $0.37$ & ... & ... & ... & ... & ... & ... & ... & ... & $10.36$ & $0.18$ & 6 \\ 
2005A & $0.018$ & $-14.05$ & $0.24$ & $0.981$ & $0.022$ & $-0.42$ & $0.03$ & $53380.00$ & $0.06$ & $17.93$ & $0.02$ & $11.27$ & $0.16$ & 3a,3b,6,7 \\ 
2005ag & $0.079$ & $-11.56$ & $0.43$ & $-0.045$ & $0.023$ & $0.27$ & $0.05$ & $53414.84$ & $0.09$ & $18.19$ & $0.02$ & $11.46$ & $0.16$ & 2,3a,3b,4,6,7,9a,9b,9c \\ 
2005al & $0.012$ & $-10.74$ & $0.25$ & $-0.129$ & $0.023$ & $-1.31$ & $0.02$ & $53430.68$ & $0.01$ & $14.61$ & $0.02$ & $10.86$ & $0.16$ & 2,3a,3b,4,6,7,9a,9b,9c \\ 
2005am & $0.008$ & $-12.16$ & $0.23$ & $0.007$ & $0.023$ & $-1.86$ & $0.03$ & $53436.18$ & $0.10$ & $13.40$ & $0.02$ & $11.17$ & $0.36$ & 3a,3b,4,6,7 \\ 
2005be & $0.035$ & $-9.32$ & $0.62$ & ... & ... & ... & ... & ... & ... & ... & ... & $10.53$ & $0.20$ & 6 \\ 
2005bg & $0.023$ & $-10.79$ & $0.36$ & $-0.025$ & $0.022$ & $0.39$ & $0.06$ & $53469.84$ & $0.12$ & $15.60$ & $0.02$ & $10.24$ & $0.16$ & 2,3a,3b,4,6,7,9a,9b,9c \\ 
2005bl & $0.024$ & $-9.93$ & $0.24$ & ... & ... & ... & ... & ... & ... & ... & ... & $11.17$ & $0.16$ & 6 \\ 
2005bo & $0.014$ & $-11.39$ & $0.22$ & $0.226$ & $0.023$ & $-1.11$ & $0.12$ & $53478.77$ & $0.15$ & $15.40$ & $0.02$ & $10.43$ & $0.13$ & 1,2,3a,3b,4,9a,9b,9c \\ 
2005el & $0.015$ & $-10.80$ & $0.22$ & $-0.149$ & $0.024$ & $-1.27$ & $0.04$ & $53646.80$ & $0.05$ & $14.61$ & $0.03$ & $10.97$ & $0.16$ & 1,2,3a,3b,4,6,7,9a,9b,9c \\ 
2005eq & $0.029$ & $-10.10$ & $0.25$ & $0.014$ & $0.022$ & $1.45$ & $0.05$ & $53654.77$ & $0.10$ & $16.05$ & $0.03$ & $10.80$ & $0.20$ & 2,3a,3b,4,6,7,9a,9c \\ 
2005hc & $0.045$ & $-11.66$ & $0.23$ & $-0.028$ & $0.021$ & $0.78$ & $0.04$ & $53667.65$ & $0.05$ & $17.05$ & $0.02$ & $10.65$ & $0.18$ & 1,2,3a,3b,4,6,7,9a,9b,9c \\ 
2005hj & $0.058$ & $-10.54$ & $0.23$ & $0.002$ & $0.023$ & $1.59$ & $0.10$ & $53674.41$ & $0.16$ & $17.48$ & $0.02$ & $9.62$ & $0.13$ & 2,3a,3b,4,6,7,9a,9c \\ 
2005iq & $0.033$ & $-10.87$ & $0.23$ & $-0.095$ & $0.022$ & $-1.19$ & $0.04$ & $53687.77$ & $0.04$ & $16.51$ & $0.02$ & $10.62$ & $0.22$ & 1,2,3a,3b,4,6,7,9a,9b,9c \\ 
2005ir & $0.076$ & $-13.41$ & $0.24$ & $0.002$ & $0.023$ & $0.67$ & $0.12$ & $53685.34$ & $0.15$ & $18.17$ & $0.02$ & $10.23$ & $0.13$ & 2,3a,3b,4,6,7,9a,9b,9c \\ 
2005kc & $0.014$ & $-10.63$ & $0.23$ & $0.179$ & $0.022$ & $-0.76$ & $0.04$ & $53697.75$ & $0.03$ & $15.31$ & $0.03$ & $10.95$ & $0.13$ & 1,2,3a,3b,4,6,7,9a,9b,9c \\ 
2005ke & $0.005$ & $-9.81$ & $0.24$ & ... & ... & ... & ... & ... & ... & ... & ... & $10.82$ & $0.15$ & 6 \\ 
2005ki & $0.019$ & $-11.28$ & $0.25$ & $-0.081$ & $0.022$ & $-1.48$ & $0.03$ & $53705.27$ & $0.06$ & $15.30$ & $0.02$ & $11.08$ & $0.16$ & 1,2,3a,3b,4,6,7,9a,9b,9c \\ 
2005ku & $0.045$ & $-12.21$ & $0.73$ & $0.093$ & $0.023$ & $-0.19$ & $0.14$ & $53698.99$ & $0.22$ & $17.33$ & $0.02$ & $10.00$ & $0.20$ & 2,3a,3b,4,6,7,9a,9b,9c \\ 
2005lu & $0.032$ & $-10.43$ & $0.42$ & ... & ... & ... & ... & ... & ... & ... & ... & $10.35$ & $0.26$ & 6 \\ 
2005M & $0.022$ & $-8.79$ & $0.23$ & $0.010$ & $0.021$ & $1.24$ & $0.03$ & $53406.25$ & $0.03$ & $15.68$ & $0.02$ & $10.28$ & $0.13$ & 2,3a,3b,4,6,7,9a,9c \\ 
2005mc & $0.025$ & $-11.28$ & $0.24$ & $0.194$ & $0.025$ & $-1.99$ & $0.04$ & $53730.74$ & $0.07$ & $16.74$ & $0.03$ & $10.95$ & $0.14$ & 2,3a,3b,4,6,7,9a,9b,9c \\ 
2005na & $0.027$ & $-10.74$ & $0.23$ & $-0.085$ & $0.021$ & $-0.42$ & $0.03$ & $53740.69$ & $0.06$ & $15.70$ & $0.02$ & $11.05$ & $0.12$ & 1,2,3a,3b,4,6,7,9a,9b,9c \\ 
2005W & $0.009$ & $-10.62$ & $0.22$ & $0.122$ & $0.021$ & $-0.43$ & $0.05$ & $53412.25$ & $0.03$ & $13.94$ & $0.02$ & $10.64$ & $0.16$ & 2,3a,3b,4,6,7,9a,9b,9c \\ 
2006ax & $0.017$ & $-10.26$ & $0.23$ & $-0.098$ & $0.020$ & $0.16$ & $0.02$ & $53827.52$ & $0.02$ & $14.76$ & $0.02$ & ... & ... & 1,2,3a,3b,4,9a,9b \\ 
2006bd & $0.026$ & $-11.21$ & $0.43$ & $0.784$ & $0.050$ & $-2.40$ & $0.26$ & $53825.86$ & $0.07$ & $18.93$ & $0.06$ & $11.06$ & $0.16$ & 3a,3b,6,7 \\ 
2006bh & $0.011$ & $-15.48$ & $0.22$ & $-0.062$ & $0.023$ & $-1.67$ & $0.04$ & $53833.46$ & $0.07$ & $14.08$ & $0.02$ & $10.87$ & $0.13$ & 2,3a,3b,4,6,7,9a,9b,9c \\ 
2006br & $0.025$ & $-14.17$ & $0.49$ & $0.851$ & $0.031$ & $-0.72$ & $0.12$ & $53851.48$ & $0.48$ & $18.73$ & $0.05$ & $10.95$ & $0.12$ & 3a,3b,6,7 \\ 
2006D & $0.009$ & $-10.62$ & $0.22$ & $-0.012$ & $0.023$ & $-1.59$ & $0.03$ & $53757.59$ & $0.05$ & $13.87$ & $0.02$ & $10.03$ & $0.16$ & 1,2,3a,3b,4,6,7,9a,9b,9c \\ 
2006ef & $0.017$ & $-12.15$ & $0.30$ & $-0.015$ & $0.026$ & $-1.52$ & $0.07$ & $53969.20$ & $0.10$ & $15.28$ & $0.03$ & $10.64$ & $0.13$ & 1,2,3a,3b,6,7,9a,9b,9c \\ 
2006ej & $0.019$ & $-12.46$ & $0.30$ & $-0.016$ & $0.026$ & $-1.53$ & $0.05$ & $53976.09$ & $0.20$ & $15.46$ & $0.03$ & $10.91$ & $0.15$ & 1,2,3a,3b,6,7,9a,9b,9c \\ 
2006eq & $0.049$ & $-11.74$ & $0.63$ & $0.090$ & $0.028$ & $-2.09$ & $0.07$ & $53974.21$ & $0.01$ & $17.95$ & $0.03$ & $10.40$ & $0.21$ & 2,3a,3b,4,6,7,9a,9c \\ 
2006et & $0.022$ & $-10.12$ & $0.31$ & $0.139$ & $0.021$ & $0.76$ & $0.03$ & $53994.23$ & $0.04$ & $15.72$ & $0.02$ & $11.01$ & $0.14$ & 2,3a,3b,4,6,7,9a,9b,9c \\ 
2006ev & $0.029$ & $-10.70$ & $0.23$ & $0.097$ & $0.024$ & $-1.31$ & $0.07$ & $53989.48$ & $0.25$ & $16.81$ & $0.03$ & $10.98$ & $0.17$ & 2,3a,3b,4,6,7,9a,9b,9c \\ 
2006fw & $0.083$ & $-10.82$ & $0.42$ & $0.018$ & $0.025$ & $-0.72$ & $0.11$ & $54003.66$ & $0.14$ & $18.68$ & $0.02$ & $10.05$ & $0.14$ & 2,3a,3b,4,6,7,9a,9b,9c \\ 
2006gj & $0.028$ & $-11.34$ & $0.25$ & ... & ... & ... & ... & ... & ... & ... & ... & $11.39$ & $0.59$ & 6 \\ 
2006gt & $0.045$ & $-10.39$ & $0.25$ & $0.115$ & $0.023$ & $-2.12$ & $0.07$ & $54002.75$ & $0.12$ & $17.93$ & $0.02$ & $10.20$ & $0.16$ & 2,3a,3b,4,6,7,9a,9c \\ 
2006hx & $0.045$ & $-10.59$ & $0.45$ & $0.042$ & $0.024$ & $-0.14$ & $0.07$ & $54022.04$ & $0.05$ & $17.26$ & $0.02$ & $10.39$ & $0.12$ & 2,3a,3b,4,6,7,9a,9b,9c \\ 
2006is & $0.031$ & $-14.12$ & $0.31$ & $-0.089$ & $0.024$ & $1.94$ & $0.09$ & $54007.95$ & $0.20$ & $15.89$ & $0.03$ & ... & ... & 2,3a,3b,4,9a \\ 
2006kf & $0.021$ & $-11.43$ & $0.23$ & $-0.049$ & $0.026$ & $-2.08$ & $0.05$ & $54041.26$ & $0.07$ & $15.69$ & $0.05$ & $10.97$ & $0.12$ & 1,2,3a,3b,4,6,7,9a,9c \\ 
2006lu & $0.053$ & $-16.59$ & $0.29$ & ... & ... & ... & ... & ... & ... & ... & ... & $10.56$ & $0.56$ & 6 \\ 
2006mr & $0.006$ & $-10.21$ & $0.23$ & ... & ... & ... & ... & ... & ... & ... & ... & $11.35$ & $0.16$ & 6 \\ 
2006ob & $0.058$ & $-8.98$ & $0.68$ & $-0.012$ & $0.025$ & $-2.16$ & $0.10$ & $54063.31$ & $0.12$ & $17.95$ & $0.02$ & $11.43$ & $0.22$ & 1,2,3a,3b,4,6,7,9a,9c \\ 
2006os & $0.032$ & $-11.91$ & $0.66$ & $0.314$ & $0.023$ & $-0.67$ & $0.06$ & $54064.75$ & $0.15$ & $17.38$ & $0.03$ & $11.30$ & $0.31$ & 3a,3b,6,7 \\ 
2006ot & $0.053$ & $-13.90$ & $0.91$ & ... & ... & ... & ... & ... & ... & ... & ... & $11.36$ & $0.16$ & 6 \\ 
2006py & $0.058$ & $-11.40$ & $0.27$ & $0.016$ & $0.023$ & $0.21$ & $0.20$ & $54070.87$ & $0.21$ & $17.64$ & $0.02$ & $9.96$ & $0.13$ & 2,3a,3b,4,6,7,9a,9b,9c \\ 
2006X & $0.003$ & $-13.90$ & $0.22$ & ... & ... & ... & ... & ... & ... & ... & ... & $10.80$ & $0.12$ & 6 \\ 
2007A & $0.018$ & $-10.56$ & $0.22$ & $0.148$ & $0.022$ & $0.59$ & $0.10$ & $54113.27$ & $0.07$ & $15.46$ & $0.03$ & $10.60$ & $0.22$ & 2,3a,3b,4,9a,9b \\ 
2007af & $0.005$ & $-10.99$ & $0.22$ & $-0.014$ & $0.021$ & $-0.57$ & $0.03$ & $54174.52$ & $0.03$ & $12.85$ & $0.02$ & $9.84$ & $0.12$ & 3a,3b,4,6,7 \\ 
2007ai & $0.032$ & $-9.98$ & $0.25$ & $0.160$ & $0.028$ & $0.98$ & $0.09$ & $54173.58$ & $0.17$ & $16.80$ & $0.06$ & $11.08$ & $0.16$ & 2,3a,3b,4,6,7,9a,9b,9c \\ 
2007al & $0.012$ & $-9.85$ & $0.26$ & ... & ... & ... & ... & ... & ... & ... & ... & $9.62$ & $0.16$ & 6 \\ 
2007as & $0.018$ & $-12.96$ & $0.66$ & $0.050$ & $0.023$ & $-1.00$ & $0.04$ & $54181.67$ & $0.09$ & $15.23$ & $0.03$ & $10.41$ & $0.16$ & 2,3a,3b,4,6,7,9a,9b,9c \\ 
2007ax & $0.007$ & $-9.77$ & $0.30$ & $0.505$ & $0.027$ & $-2.00$ & $0.06$ & $54187.67$ & $0.09$ & $15.95$ & $0.03$ & $10.53$ & $0.16$ & 3b,6,7 \\ 
2007ba & $0.038$ & $-10.61$ & $0.31$ & $0.272$ & $0.024$ & $-2.26$ & $0.06$ & $54197.10$ & $0.07$ & $17.52$ & $0.02$ & $11.06$ & $0.16$ & 2,3a,3b,4,6,7,9a,9c \\ 
2007bc & $0.022$ & $-10.49$ & $0.26$ & $-0.002$ & $0.022$ & $-1.10$ & $0.03$ & $54200.35$ & $0.07$ & $15.58$ & $0.02$ & $10.83$ & $0.14$ & 1,2,3a,3b,4,6,7,9a,9b,9c \\ 
2007bd & $0.032$ & $-12.49$ & $0.23$ & $-0.071$ & $0.023$ & $-1.08$ & $0.04$ & $54206.84$ & $0.05$ & $16.27$ & $0.02$ & $10.79$ & $0.16$ & 1,2,3a,3b,4,6,7,9a,9b,9c \\ 
2007bm & $0.007$ & $-10.82$ & $0.49$ & $0.384$ & $0.023$ & $-0.94$ & $0.04$ & $54224.94$ & $0.08$ & $14.15$ & $0.02$ & $10.42$ & $0.23$ & 3a,3b,6,7 \\ 
2007ca & $0.015$ & $-11.09$ & $0.28$ & $0.234$ & $0.022$ & $0.57$ & $0.03$ & $54227.71$ & $0.04$ & $15.63$ & $0.02$ & $9.75$ & $0.27$ & 2,3a,3b,4,6,7,9a,9b,9c \\ 
2007cg & $0.033$ & $-11.65$ & $0.30$ & ... & ... & ... & ... & ... & ... & ... & ... & $10.63$ & $0.18$ & 6 \\ 
2007hj & $0.014$ & $-12.30$ & $0.22$ & $0.121$ & $0.023$ & $-2.06$ & $0.04$ & $54348.82$ & $0.10$ & $15.33$ & $0.03$ & $10.59$ & $0.16$ & 2,3a,3b,4,6,7,9a,9c \\ 
2007jd & $0.073$ & $-15.02$ & $0.28$ & $0.075$ & $0.024$ & $-0.89$ & $0.07$ & $54361.69$ & $0.01$ & $18.63$ & $0.03$ & $10.05$ & $0.16$ & 2,3a,3b,4,6,7,9a,9b,9c \\ 
2007jg & $0.037$ & $-12.62$ & $0.45$ & $-0.004$ & $0.022$ & $-0.58$ & $0.06$ & $54366.63$ & $0.08$ & $17.01$ & $0.03$ & $9.10$ & $0.16$ & 2,3a,3b,4,6,7,9a,9b \\ 
2007jh & $0.041$ & $-10.68$ & $0.32$ & ... & ... & ... & ... & ... & ... & ... & ... & $11.11$ & $0.12$ & 6 \\ 
2007le & $0.007$ & $-13.31$ & $0.23$ & $0.250$ & $0.021$ & $0.25$ & $0.03$ & $54399.38$ & $0.03$ & $13.61$ & $0.02$ & $10.46$ & $0.16$ & 3a,3b,4,6,7 \\ 
2007N & $0.013$ & $-10.41$ & $0.22$ & $0.844$ & $0.041$ & $-2.96$ & $0.11$ & $54120.36$ & $0.08$ & $17.61$ & $0.05$ & $10.33$ & $0.16$ & 3a,3b,4,6,7 \\ 
2007nq & $0.045$ & $-12.11$ & $0.36$ & $-0.022$ & $0.023$ & $-1.90$ & $0.05$ & $54398.46$ & $0.10$ & $17.15$ & $0.02$ & $11.53$ & $0.29$ & 2,3a,3b,4,6,7,9a,9b,9c \\ 
2007on & $0.006$ & $-11.24$ & $0.22$ & $0.007$ & $0.040$ & $-2.30$ & $0.08$ & $54419.81$ & $0.09$ & $12.80$ & $0.03$ & $10.91$ & $0.16$ & 3a,3b,4,6,7 \\ 
2007S & $0.014$ & $-10.65$ & $0.22$ & $0.377$ & $0.023$ & $1.09$ & $0.05$ & $54144.69$ & $0.05$ & $15.55$ & $0.02$ & $9.93$ & $0.13$ & 3a,3b,4,6,7 \\ 
2007sr & $0.006$ & $-13.59$ & $0.22$ & $0.053$ & $0.022$ & $-0.02$ & $0.03$ & $54449.58$ & $0.09$ & $12.48$ & $0.02$ & $10.45$ & $0.42$ & 3a,3b,4,6,7 \\ 
2007ux & $0.031$ & $-10.74$ & $0.39$ & ... & ... & ... & ... & ... & ... & ... & ... & $10.28$ & $0.12$ & 6 \\ 
2008ar & $0.026$ & $-10.59$ & $0.22$ & $-0.029$ & $0.022$ & $-0.15$ & $0.04$ & $54534.83$ & $0.04$ & $15.95$ & $0.02$ & $10.38$ & $0.13$ & 2,3a,3b,4,6,7,9a,9b,9c \\ 
2008bc & $0.015$ & $-11.55$ & $0.23$ & $-0.090$ & $0.007$ & $0.57$ & $0.02$ & $54550.05$ & $0.03$ & $14.47$ & $0.04$ & $10.14$ & $0.16$ & 2,3a,3b,4,6,7,9a,9b,9c \\ 
2008bf & $0.022$ & $-11.59$ & $0.24$ & $-0.119$ & $0.021$ & $0.37$ & $0.03$ & $54555.11$ & $0.03$ & $15.42$ & $0.02$ & $11.28$ & $0.16$ & 1,2,3a,3b,4,6,7,9a,9b,9c \\ 
2008bq & $0.034$ & $-10.79$ & $0.31$ & $0.045$ & $0.022$ & $0.36$ & $0.04$ & $54563.52$ & $0.08$ & $16.48$ & $0.03$ & $11.29$ & $0.22$ & 2,3a,3b,4,6,7,9a,9b,9c \\ 
2008bz & $0.060$ & $-11.43$ & $0.24$ & $-0.102$ & $0.023$ & $-0.50$ & $0.08$ & $54579.36$ & $0.16$ & $17.69$ & $0.02$ & $10.88$ & $0.19$ & 2,3a,3b,4,6,7,9a,9b,9c \\ 
2008C & $0.017$ & $-10.87$ & $0.22$ & $0.120$ & $0.023$ & $-0.74$ & $0.07$ & $54467.52$ & $0.16$ & $15.42$ & $0.03$ & $10.50$ & $0.14$ & 2,3a,3b,4,6,7,9a,9b,9c \\ 
2008cd & $0.007$ & $-9.32$ & $0.23$ & ... & ... & ... & ... & ... & ... & ... & ... & $10.65$ & $0.16$ & 6 \\ 
2008cf & $0.046$ & $-10.27$ & $0.33$ & $-0.105$ & $0.029$ & $1.36$ & $0.39$ & $54595.09$ & $0.69$ & $16.77$ & $0.03$ & ... & ... & 2,3a,3b,4,9a \\ 
2008fp & $0.006$ & $-10.99$ & $0.22$ & $0.358$ & $0.022$ & $0.29$ & $0.03$ & $54730.76$ & $0.01$ & $13.61$ & $0.04$ & $10.22$ & $0.16$ & 3a,3b,6,7 \\ 
2008fr & $0.039$ & $-10.96$ & $0.22$ & $-0.086$ & $0.028$ & $0.55$ & $0.11$ & $54732.62$ & $0.32$ & $16.56$ & $0.03$ & $8.70$ & $0.23$ & 2,3a,3b,4,6,7,9a,9b \\ 
2008gl & $0.034$ & $-11.99$ & $0.57$ & $0.002$ & $0.022$ & $-1.38$ & $0.03$ & $54768.21$ & $0.06$ & $16.55$ & $0.02$ & $11.12$ & $0.12$ & 2,3a,3b,4,6,7,9a,9b,9c \\ 
2008gp & $0.033$ & $-9.79$ & $0.22$ & $-0.115$ & $0.022$ & $-0.14$ & $0.03$ & $54779.27$ & $0.03$ & $16.14$ & $0.03$ & $10.89$ & $0.12$ & 2,3a,3b,4,6,7,9a,9b,9c \\ 
2008hj & $0.038$ & $-12.79$ & $0.23$ & $-0.044$ & $0.022$ & $0.19$ & $0.04$ & $54801.88$ & $0.04$ & $16.51$ & $0.02$ & $10.43$ & $0.12$ & 2,3a,3b,4,6,7,9a,9b,9c \\ 
2008hu & $0.050$ & $-12.34$ & $0.71$ & $-0.017$ & $0.024$ & $-1.74$ & $0.06$ & $54806.29$ & $0.13$ & $17.60$ & $0.03$ & $11.42$ & $0.16$ & 2,3a,3b,4,6,7,9a,9b,9c \\ 
2008hv & $0.013$ & $-10.93$ & $0.23$ & $-0.074$ & $0.022$ & $-1.22$ & $0.03$ & $54817.02$ & $0.03$ & $14.48$ & $0.02$ & $10.80$ & $0.17$ & 2,3a,3b,4,6,7,9a,9b,9c \\ 
2008ia & $0.022$ & $-11.18$ & $0.22$ & $-0.064$ & $0.025$ & $-1.24$ & $0.04$ & $54812.97$ & $0.07$ & $15.64$ & $0.04$ & $11.00$ & $0.16$ & 2,3a,3b,4,6,7,9a,9b,9c \\ 
2008R & $0.013$ & $-9.82$ & $0.22$ & $0.044$ & $0.023$ & $-2.30$ & $0.05$ & $54494.07$ & $0.08$ & $14.98$ & $0.03$ & $11.11$ & $0.16$ & 2,3a,3b,4,6,7,9a,9c \\ 
2009aa & $0.027$ & $-10.55$ & $0.22$ & $-0.047$ & $0.021$ & $-0.78$ & $0.03$ & $54878.44$ & $0.03$ & $16.04$ & $0.02$ & ... & ... & 2,3a,3b,4,9a,9b \\ 
2009ab & $0.011$ & $-10.56$ & $0.22$ & $-0.011$ & $0.024$ & $-0.99$ & $0.05$ & $54883.38$ & $0.04$ & $14.44$ & $0.04$ & $10.62$ & $0.16$ & 2,3a,3b,4,6,7,9a,9b,9c \\ 
2009ad & $0.028$ & $-10.39$ & $0.22$ & $-0.035$ & $0.022$ & $0.19$ & $0.03$ & $54886.61$ & $0.03$ & $15.91$ & $0.03$ & $10.79$ & $0.20$ & 2,3a,3b,4,6,7,9a,9b,9c \\ 
2009ag & $0.009$ & $-10.26$ & $0.29$ & $0.087$ & $0.024$ & $-0.22$ & $0.03$ & $54889.65$ & $0.06$ & $14.26$ & $0.05$ & $10.59$ & $0.16$ & 2,3a,3b,4,6,7,9a,9b,9c \\ 
2009D & $0.025$ & $-9.96$ & $0.22$ & $-0.061$ & $0.021$ & $0.65$ & $0.03$ & $54841.41$ & $0.04$ & $15.46$ & $0.02$ & $10.44$ & $0.19$ & 2,3a,3b,4,6,7,9a,9b,9c \\ 
2009F & $0.013$ & $-13.98$ & $0.22$ & $0.415$ & $0.020$ & $-2.59$ & $0.06$ & $54842.20$ & $0.06$ & $16.69$ & $0.03$ & $10.76$ & $0.16$ & 3a,3b,6,7 \\ 
2009Y & $0.009$ & $-9.85$ & $0.22$ & $0.079$ & $0.027$ & $0.38$ & $0.05$ & $54876.16$ & $0.06$ & $13.72$ & $0.04$ & $11.02$ & $0.16$ & 2,3a,3b,4,6,7,9a,9b,9c \\ 

\enddata
\end{deluxetable*}

\startlongtable
\begin{deluxetable*}{lcDcDcDcccccDcc}
\tablecaption{Supernova velocities and light-curve parameters for the W09/FK11 sample. \label{table: w09 data with cuts}}
\tabletypesize{\scriptsize}
\tablehead{\colhead{Supernova} & \colhead{$z$} & \twocolhead{\vsimax} & \colhead{$\pm$} & \twocolhead{$c$} & \colhead{$\pm$}  & \twocolhead{$x_1$} & \colhead{$\pm$} & \colhead{$t_0$} & \colhead{$\pm$} & \colhead{$m_B$} & \colhead{$\pm$} & \twocolhead{Host mass} & \colhead{$\pm$} & \colhead{Figures} \\[-8pt]
\colhead{} & \colhead{} & \multicolumn{3}{c}{(10$^3$ km s$^{-1}$)} &  \multicolumn{3}{c}{}  &  \multicolumn{3}{c}{} & \colhead{(MJD)} & \colhead{} & \multicolumn{2}{c}{(mag)} & \multicolumn{3}{c}{log ($M_*/M_\odot$)} & \colhead{}
}
\decimals
\startdata
1994ae & $0.005$ & $-10.98$ & $0.23$ & $0.084$ & $0.025$ & $0.42$ & $0.06$ & $49685.33$ & $0.05$ & $12.87$ & $0.03$ & $9.90$ & $0.24$ & 3b,6,7 \\ 
1995E & $0.012$ & $-10.54$ & $0.22$ & $0.764$ & $0.028$ & $-0.51$ & $0.12$ & $49775.18$ & $0.24$ & $16.51$ & $0.04$ & $10.66$ & $0.16$ & 3b \\ 
1996bo & $0.016$ & $-12.25$ & $0.24$ & $0.329$ & $0.026$ & $-1.01$ & $0.07$ & $50387.11$ & $0.08$ & $15.63$ & $0.03$ & $10.37$ & $0.16$ & 3b,6,7 \\ 
1996X & $0.007$ & $-11.17$ & $0.22$ & $-0.035$ & $0.026$ & $-1.08$ & $0.07$ & $50191.32$ & $0.11$ & $12.76$ & $0.03$ & $10.85$ & $0.16$ & 1,3b,6,7 \\ 
1997bp & $0.008$ & $<-11.8$ & ... & $0.172$ & $0.025$ & $0.42$ & $0.10$ & $50549.40$ & $0.21$ & $13.70$ & $0.03$ & $10.24$ & $0.16$ & 2,9a,9b,9c \\ 
1997do & $0.010$ & $<-11.8$ & ... & $0.099$ & $0.026$ & $0.39$ & $0.11$ & $50766.36$ & $0.09$ & $14.15$ & $0.03$ & $9.30$ & $0.16$ & 1,2,9a,9b \\ 
1997dt & $0.008$ & $-11.13$ & $0.22$ & $0.486$ & $0.030$ & $-0.58$ & $0.14$ & $50785.87$ & $0.11$ & $15.23$ & $0.04$ & ... & ... & 3b \\ 
1997E & $0.014$ & $-12.01$ & $0.23$ & $0.038$ & $0.027$ & $-1.75$ & $0.09$ & $50468.03$ & $0.10$ & $14.92$ & $0.04$ & $9.99$ & $1.53$ & 1,2,3b,9a,9b \\ 
1998dh & $0.009$ & $-12.04$ & $0.22$ & $0.080$ & $0.027$ & $-0.75$ & $0.08$ & $51029.67$ & $0.07$ & $13.65$ & $0.03$ & $10.66$ & $0.16$ & 2,3b,6,7,9a,9b,9c \\ 
1998ec & $0.020$ & $-12.88$ & $0.22$ & $0.135$ & $0.031$ & $0.09$ & $0.23$ & $51088.90$ & $0.22$ & $15.92$ & $0.04$ & $10.57$ & $0.16$ & 2,3b,9a,9b,9c \\ 
1998ef & $0.017$ & $-13.26$ & $0.22$ & $-0.088$ & $0.028$ & $-1.19$ & $0.11$ & $51113.71$ & $0.10$ & $14.59$ & $0.03$ & $10.52$ & $0.16$ & 1,2,3b,6,7,9a,9b,9c \\ 
1998eg & $0.024$ & $-10.12$ & $0.23$ & $0.021$ & $0.028$ & $-0.50$ & $0.22$ & $51110.75$ & $0.02$ & $15.94$ & $0.04$ & $11.32$ & $0.16$ & 1,2,3b,9a,9b,9c \\ 
1999cc & $0.032$ & $-12.03$ & $0.24$ & $-0.017$ & $0.030$ & $-1.62$ & $0.12$ & $51315.54$ & $0.21$ & $16.54$ & $0.03$ & $10.97$ & $0.12$ & 1,2,3b,6,7,9a,9b,9c \\ 
1999cp & $0.010$ & $>-11.8$ & ... & $-0.015$ & $0.026$ & $-0.02$ & $0.03$ & $51363.28$ & $0.03$ & $13.71$ & $0.03$ & $9.48$ & $0.16$ & 1,2,9a,9b \\ 
1999dg & $0.023$ & $>-11.8$ & ... & $-0.072$ & $0.030$ & $-1.78$ & $0.13$ & $51392.84$ & $0.23$ & $15.69$ & $0.03$ & ... & ... & 1,2,9a,9b \\ 
1999dk & $0.014$ & $<-11.8$ & ... & $0.033$ & $0.025$ & $-0.15$ & $0.05$ & $51414.74$ & $0.05$ & $14.57$ & $0.03$ & $10.20$ & $0.16$ & 1,2,9a,9b,9c \\ 
1999ej & $0.015$ & $>-11.8$ & ... & $-0.014$ & $0.035$ & $-1.52$ & $0.23$ & $51482.50$ & $0.62$ & $15.12$ & $0.05$ & $10.58$ & $0.16$ & 1,2,9a,9b,9c \\ 
1999gd & $0.019$ & $-10.37$ & $0.23$ & $0.391$ & $0.029$ & $-1.06$ & $0.12$ & $51520.34$ & $0.28$ & $16.72$ & $0.03$ & $10.38$ & $0.16$ & 3b \\ 
2000cf & $0.036$ & $-9.95$ & $0.24$ & $-0.028$ & $0.028$ & $-1.02$ & $0.17$ & $51674.24$ & $0.44$ & $16.84$ & $0.03$ & $9.60$ & $0.16$ & 1,2,3b,6,7,9a,9b,9c \\ 
2000cn & $0.023$ & $>-11.8$ & ... & $0.097$ & $0.027$ & $-2.41$ & $0.14$ & $51707.36$ & $0.07$ & $16.32$ & $0.03$ & $10.34$ & $0.16$ & 1,2,9a,9c \\ 
2000cw & $0.029$ & $>-11.8$ & ... & $0.034$ & $0.026$ & $-0.76$ & $0.10$ & $51748.35$ & $0.09$ & $16.48$ & $0.03$ & ... & ... & 1,2,9a,9b \\ 
2000dk & $0.016$ & $>-11.8$ & ... & $-0.015$ & $0.027$ & $-2.04$ & $0.08$ & $51812.42$ & $0.09$ & $15.13$ & $0.03$ & $11.05$ & $0.50$ & 1,2,9a,9c \\ 
2000dm & $0.015$ & $>-11.8$ & ... & $-0.019$ & $0.028$ & $-2.06$ & $0.10$ & $51815.48$ & $0.12$ & $14.91$ & $0.04$ & ... & ... & 1,2,9a \\ 
2000dn & $0.032$ & $>-11.8$ & ... & $-0.064$ & $0.027$ & $-0.35$ & $0.10$ & $51824.62$ & $0.07$ & $16.35$ & $0.03$ & ... & ... & 1,2,9a,9b \\ 
2000dr & $0.018$ & $>-11.8$ & ... & $0.084$ & $0.027$ & $-2.60$ & $0.09$ & $51833.71$ & $0.08$ & $15.74$ & $0.03$ & ... & ... & 1,2,9a \\ 
2000fa & $0.022$ & $<-11.8$ & ... & $0.056$ & $0.027$ & $0.49$ & $0.08$ & $51892.60$ & $0.08$ & $15.67$ & $0.03$ & $9.82$ & $0.16$ & 1,2,9a,9b,9c \\ 
2001bf & $0.015$ & $>-11.8$ & ... & $-0.002$ & $0.026$ & $0.54$ & $0.08$ & $52046.79$ & $0.18$ & $14.54$ & $0.03$ & ... & ... & 1,2,9a,9b \\ 
2001br & $0.021$ & $<-11.8$ & ... & $0.070$ & $0.027$ & $-1.32$ & $0.10$ & $52051.94$ & $0.18$ & $16.01$ & $0.03$ & ... & ... & 1,2,9a,9b \\ 
2001cj & $0.025$ & $>-11.8$ & ... & $-0.082$ & $0.025$ & $0.61$ & $0.08$ & $52065.78$ & $0.10$ & $15.59$ & $0.03$ & ... & ... & 1,2,9a,9b \\ 
2001ck & $0.035$ & $>-11.8$ & ... & $-0.044$ & $0.027$ & $-0.01$ & $0.11$ & $52072.53$ & $0.10$ & $16.48$ & $0.03$ & ... & ... & 1,2,9a,9b \\ 
2001cp & $0.022$ & $>-11.8$ & ... & $-0.000$ & $0.027$ & $1.00$ & $0.07$ & $52088.48$ & $0.06$ & $15.48$ & $0.04$ & ... & ... & 1,2,9a,9b \\ 
2001da & $0.016$ & $-11.53$ & $0.22$ & $0.091$ & $0.034$ & $-0.55$ & $0.72$ & $52107.52$ & $1.08$ & $15.25$ & $0.05$ & $10.34$ & $0.12$ & 2,6,7,9a,9b,9c \\ 
2001dl & $0.020$ & $>-11.8$ & ... & $0.278$ & $0.026$ & $0.20$ & $0.07$ & $52130.60$ & $0.06$ & $16.62$ & $0.03$ & ... & ... & 2,9a,9b \\ 
2001en & $0.015$ & $-13.32$ & $0.22$ & $0.022$ & $0.027$ & $-0.97$ & $0.05$ & $52193.54$ & $0.06$ & $14.87$ & $0.03$ & $10.62$ & $0.38$ & 1,2,3b,6,7,9a,9b,9c \\ 
2001ep & $0.013$ & $-10.80$ & $0.23$ & $0.074$ & $0.026$ & $-0.95$ & $0.06$ & $52200.17$ & $0.09$ & $14.68$ & $0.03$ & $10.37$ & $0.16$ & 1,2,3b,6,7,9a,9b,9c \\ 
2001fe & $0.014$ & $-11.21$ & $0.22$ & $-0.041$ & $0.027$ & $0.75$ & $0.09$ & $52229.30$ & $0.24$ & $14.44$ & $0.03$ & $10.26$ & $0.12$ & 1,2,3b,9a,9b,9c \\ 
2001fh & $0.012$ & $>-11.8$ & ... & $-0.089$ & $0.042$ & $-1.97$ & $0.10$ & $52224.47$ & $0.10$ & $14.43$ & $0.13$ & ... & ... & 1,2,9a,9b \\ 
2002aw & $0.026$ & $>-11.8$ & ... & $0.052$ & $0.026$ & $-0.30$ & $0.09$ & $52325.56$ & $0.05$ & $15.98$ & $0.03$ & ... & ... & 1,2,9a,9b \\ 
2002bf & $0.025$ & $-15.52$ & $0.22$ & $-0.096$ & $0.089$ & $-2.28$ & $0.30$ & $52338.32$ & $0.91$ & $15.80$ & $0.15$ & $10.68$ & $0.13$ & 1,2,9a,9c \\ 
2002bo & $0.005$ & $-13.11$ & $0.22$ & $0.337$ & $0.026$ & $-0.51$ & $0.07$ & $52357.30$ & $0.04$ & $13.68$ & $0.03$ & $10.83$ & $0.16$ & 3b,6,7 \\ 
2002cr & $0.010$ & $-10.05$ & $0.23$ & $-0.021$ & $0.026$ & $-0.37$ & $0.06$ & $52408.93$ & $0.04$ & $13.98$ & $0.03$ & $9.48$ & $0.16$ & 1,2,3b,6,7,9a,9b \\ 
2002cs & $0.016$ & $<-11.8$ & ... & $0.122$ & $0.026$ & $0.60$ & $0.04$ & $52409.72$ & $0.04$ & $14.96$ & $0.04$ & ... & ... & 1,2,9a,9b \\ 
2002cu & $0.023$ & $<-11.8$ & ... & $0.041$ & $0.026$ & $-1.84$ & $0.06$ & $52416.16$ & $0.05$ & $15.90$ & $0.03$ & ... & ... & 1,2,9a,9b \\ 
2002de & $0.028$ & $-11.44$ & $0.25$ & $0.112$ & $0.027$ & $0.47$ & $0.32$ & $52434.04$ & $0.16$ & $16.42$ & $0.03$ & $10.74$ & $0.14$ & 1,2,3b,6,7,9a,9b,9c \\ 
2002dj & $0.009$ & $-13.69$ & $0.22$ & $0.071$ & $0.026$ & $0.09$ & $0.13$ & $52451.41$ & $0.10$ & $13.76$ & $0.03$ & $11.34$ & $0.16$ & 2,3b,6,7,9a,9b,9c \\ 
2002dp & $0.010$ & $-11.09$ & $0.23$ & $0.091$ & $0.028$ & $-0.12$ & $0.22$ & $52451.19$ & $0.15$ & $14.37$ & $0.03$ & $10.67$ & $0.29$ & 1,2,3b,6,7,9a,9b,9c \\ 
2002eb & $0.026$ & $>-11.8$ & ... & $-0.053$ & $0.024$ & $0.77$ & $0.04$ & $52494.22$ & $0.04$ & $15.77$ & $0.03$ & ... & ... & 1,2,9a,9b \\ 
2002ef & $0.023$ & $<-11.8$ & ... & $0.295$ & $0.025$ & $-0.49$ & $0.08$ & $52490.94$ & $0.08$ & $16.42$ & $0.03$ & ... & ... & 2,9a,9b \\ 
2002el & $0.028$ & $>-11.8$ & ... & $-0.038$ & $0.026$ & $-1.29$ & $0.07$ & $52508.16$ & $0.05$ & $15.92$ & $0.03$ & ... & ... & 1,2,9a,9b \\ 
2002er & $0.009$ & $>-11.8$ & ... & $0.099$ & $0.026$ & $-0.95$ & $0.03$ & $52524.28$ & $0.03$ & $13.99$ & $0.04$ & $10.56$ & $0.16$ & 2,9a,9b,9c \\ 
2002fk & $0.007$ & $-10.01$ & $0.24$ & $-0.099$ & $0.026$ & $0.09$ & $0.07$ & $52547.99$ & $0.16$ & $12.91$ & $0.03$ & $10.16$ & $0.25$ & 3b,6,7 \\ 
2002ha & $0.013$ & $-11.29$ & $0.24$ & $-0.059$ & $0.028$ & $-1.29$ & $0.07$ & $52581.33$ & $0.11$ & $14.51$ & $0.03$ & $11.04$ & $0.13$ & 1,2,3b,6,7,9a,9b,9c \\ 
2002he & $0.025$ & $-12.62$ & $0.22$ & $-0.023$ & $0.032$ & $-1.65$ & $0.17$ & $52586.19$ & $0.11$ & $16.03$ & $0.04$ & $10.99$ & $0.17$ & 1,2,3b,6,7,9a,9b,9c \\ 
2002hu & $0.038$ & $-9.88$ & $0.28$ & $-0.089$ & $0.026$ & $0.53$ & $0.09$ & $52592.52$ & $0.09$ & $16.41$ & $0.03$ & $10.18$ & $0.15$ & 1,2,3b,6,7,9a,9b,9c \\ 
2002hw & $0.016$ & $-11.13$ & $0.23$ & $0.400$ & $0.007$ & $-1.89$ & $0.16$ & $52595.79$ & $0.11$ & $16.49$ & $0.02$ & $10.38$ & $0.16$ & 3b \\ 
2003cq & $0.034$ & $-11.85$ & $0.25$ & $0.114$ & $0.049$ & $-0.65$ & $0.14$ & $52739.57$ & $0.16$ & $16.94$ & $0.06$ & $11.39$ & $0.16$ & 1,2,9a,9b,9c \\ 
2003gt & $0.015$ & $>-11.8$ & ... & $0.034$ & $0.025$ & $-0.17$ & $0.03$ & $52861.77$ & $0.03$ & $14.73$ & $0.03$ & ... & ... & 1,2,9a,9b \\ 
2003he & $0.024$ & $>-11.8$ & ... & $0.060$ & $0.025$ & $0.37$ & $0.05$ & $52876.45$ & $0.06$ & $15.99$ & $0.03$ & ... & ... & 1,2,9a,9b \\ 
2003W & $0.021$ & $<-11.8$ & ... & $0.118$ & $0.025$ & $0.08$ & $0.06$ & $52680.53$ & $0.06$ & $15.66$ & $0.03$ & $10.55$ & $0.16$ & 1,2,9a,9b,9c \\ 
2004as & $0.032$ & $-11.84$ & $0.25$ & $0.023$ & $0.028$ & $0.23$ & $0.11$ & $53086.29$ & $0.14$ & $16.69$ & $0.03$ & $9.24$ & $0.17$ & 1,2,3b,6,7,9a,9b \\ 
2004at & $0.024$ & $>-11.8$ & ... & $-0.081$ & $0.025$ & $-0.04$ & $0.05$ & $53091.97$ & $0.04$ & $15.45$ & $0.03$ & ... & ... & 1,2,9a,9b \\ 
2004bw & $0.022$ & $>-11.8$ & ... & $-0.052$ & $0.027$ & $-1.18$ & $0.06$ & $53162.78$ & $0.08$ & $15.60$ & $0.04$ & ... & ... & 1,2,9a,9b \\ 
2004dt & $0.020$ & $-15.93$ & $0.22$ & $-0.041$ & $0.025$ & $-0.30$ & $0.03$ & $53239.79$ & $0.03$ & $14.95$ & $0.03$ & $10.55$ & $0.16$ & 1,2,4,6,7,9a,9b,9c \\ 
2004ef & $0.031$ & $-12.29$ & $0.23$ & $0.087$ & $0.021$ & $-1.40$ & $0.02$ & $53264.22$ & $0.03$ & $16.61$ & $0.02$ & $10.90$ & $0.14$ & 1,2,3a,3b,4,6,7,9a,9b,9c \\ 
2004eo & $0.015$ & $-10.57$ & $0.22$ & $0.005$ & $0.022$ & $-1.22$ & $0.02$ & $53278.42$ & $0.03$ & $14.84$ & $0.03$ & $11.09$ & $0.14$ & 1,2,3a,3b,4,6,7,9a,9b,9c \\ 
2004ey & $0.016$ & $-11.20$ & $0.40$ & $-0.131$ & $0.020$ & $0.08$ & $0.02$ & $53304.53$ & $0.02$ & $14.46$ & $0.03$ & $10.01$ & $0.14$ & 1,2,3a,3b,4,6,7,9a,9b,9c \\ 
2004fz & $0.016$ & $>-11.8$ & ... & $-0.022$ & $0.025$ & $-1.34$ & $0.06$ & $53333.52$ & $0.04$ & $14.67$ & $0.03$ & ... & ... & 1,2,9a,9b \\ 
2004gs & $0.028$ & $-11.48$ & $0.24$ & $0.123$ & $0.022$ & $-1.88$ & $0.03$ & $53356.03$ & $0.06$ & $16.90$ & $0.02$ & $10.82$ & $0.16$ & 1,2,3a,3b,4,6,7,9a,9b,9c \\ 
2004S & $0.010$ & $>-11.8$ & ... & $0.032$ & $0.026$ & $-0.16$ & $0.05$ & $53040.22$ & $0.18$ & $13.97$ & $0.04$ & ... & ... & 1,2,9a,9b \\ 
2005A & $0.018$ & $-14.05$ & $0.24$ & $0.981$ & $0.022$ & $-0.42$ & $0.03$ & $53380.00$ & $0.06$ & $17.93$ & $0.02$ & $11.27$ & $0.16$ & 3a,3b,6,7 \\ 
2005am & $0.008$ & $-12.16$ & $0.23$ & $0.007$ & $0.023$ & $-1.86$ & $0.03$ & $53436.18$ & $0.10$ & $13.40$ & $0.02$ & $11.17$ & $0.36$ & 3a,3b,4,6,7 \\ 
2005bo & $0.014$ & $-11.39$ & $0.22$ & $0.226$ & $0.023$ & $-1.11$ & $0.12$ & $53478.77$ & $0.15$ & $15.40$ & $0.02$ & $10.43$ & $0.13$ & 1,2,3a,3b,4,9a,9b,9c \\ 
2005cf & $0.007$ & $-10.34$ & $0.22$ & $-0.012$ & $0.026$ & $-0.02$ & $0.05$ & $53534.23$ & $0.04$ & $13.06$ & $0.03$ & $9.66$ & $0.16$ & 3b,6,7 \\ 
2005de & $0.015$ & $>-11.8$ & ... & $0.049$ & $0.025$ & $-0.51$ & $0.03$ & $53598.34$ & $0.03$ & $15.17$ & $0.03$ & ... & ... & 1,2,9a,9b \\ 
2005el & $0.015$ & $-10.80$ & $0.22$ & $-0.149$ & $0.024$ & $-1.27$ & $0.04$ & $53646.80$ & $0.05$ & $14.61$ & $0.03$ & $10.97$ & $0.16$ & 1,2,3a,3b,4,6,7,9a,9b,9c \\ 
2005hc & $0.045$ & $-11.66$ & $0.23$ & $-0.028$ & $0.021$ & $0.78$ & $0.04$ & $53667.65$ & $0.05$ & $17.05$ & $0.02$ & $10.65$ & $0.18$ & 1,2,3a,3b,4,6,7,9a,9b,9c \\ 
2005iq & $0.033$ & $-10.87$ & $0.23$ & $-0.095$ & $0.022$ & $-1.19$ & $0.04$ & $53687.77$ & $0.04$ & $16.51$ & $0.02$ & $10.62$ & $0.22$ & 1,2,3a,3b,4,6,7,9a,9b,9c \\ 
2005kc & $0.014$ & $-10.63$ & $0.23$ & $0.179$ & $0.022$ & $-0.76$ & $0.04$ & $53697.75$ & $0.03$ & $15.31$ & $0.03$ & $10.95$ & $0.13$ & 1,2,3a,3b,4,6,7,9a,9b,9c \\ 
2005ki & $0.019$ & $-11.28$ & $0.25$ & $-0.081$ & $0.022$ & $-1.48$ & $0.03$ & $53705.27$ & $0.06$ & $15.30$ & $0.02$ & $11.08$ & $0.16$ & 1,2,3a,3b,4,6,7,9a,9b,9c \\ 
2005lz & $0.044$ & $>-11.8$ & ... & $0.077$ & $0.029$ & $-1.19$ & $0.12$ & $53736.09$ & $0.14$ & $17.40$ & $0.03$ & $10.78$ & $0.54$ & 1,2,9a,9b,9c \\ 
2005ms & $0.026$ & $>-11.8$ & ... & $-0.034$ & $0.027$ & $0.48$ & $0.08$ & $53744.21$ & $0.06$ & $15.91$ & $0.03$ & $10.50$ & $0.22$ & 1,2,9a,9b,9c \\ 
2005na & $0.027$ & $-10.74$ & $0.23$ & $-0.085$ & $0.021$ & $-0.42$ & $0.03$ & $53740.69$ & $0.06$ & $15.70$ & $0.02$ & $11.05$ & $0.12$ & 1,2,3a,3b,4,6,7,9a,9b,9c \\ 
2006ac & $0.024$ & $-13.85$ & $0.22$ & $0.024$ & $0.027$ & $-1.01$ & $0.06$ & $53781.83$ & $0.08$ & $15.89$ & $0.03$ & $10.99$ & $0.13$ & 1,2,3b,6,7,9a,9b,9c \\ 
2006ax & $0.017$ & $-10.26$ & $0.23$ & $-0.098$ & $0.020$ & $0.16$ & $0.02$ & $53827.52$ & $0.02$ & $14.76$ & $0.02$ & ... & ... & 1,2,3a,3b,4,9a,9b \\ 
2006az & $0.031$ & $-10.72$ & $0.26$ & $-0.088$ & $0.026$ & $-1.39$ & $0.04$ & $53826.91$ & $0.08$ & $16.22$ & $0.03$ & $11.28$ & $0.16$ & 1,2,3b,6,7,9a,9b,9c \\ 
2006bq & $0.021$ & $-13.78$ & $0.22$ & $0.047$ & $0.027$ & $-1.63$ & $0.06$ & $53848.26$ & $0.14$ & $15.96$ & $0.03$ & $10.78$ & $0.12$ & 1,2,3b,9a,9b,9c \\ 
2006br & $0.025$ & $-14.17$ & $0.49$ & $0.851$ & $0.031$ & $-0.72$ & $0.12$ & $53851.48$ & $0.48$ & $18.73$ & $0.05$ & $10.95$ & $0.12$ & 3a,3b,6,7 \\ 
2006bt & $0.032$ & $-11.30$ & $0.23$ & ... & ... & ... & ... & ... & ... & ... & ... & $11.09$ & $0.16$ & 6 \\ 
2006cp & $0.023$ & $-13.48$ & $0.24$ & $0.077$ & $0.027$ & $0.34$ & $0.08$ & $53897.74$ & $0.07$ & $15.74$ & $0.03$ & $10.03$ & $0.19$ & 1,2,3b,6,7,9a,9b,9c \\ 
2006D & $0.009$ & $-10.62$ & $0.22$ & $-0.012$ & $0.023$ & $-1.59$ & $0.03$ & $53757.59$ & $0.05$ & $13.87$ & $0.02$ & $10.03$ & $0.16$ & 1,2,3a,3b,4,6,7,9a,9b,9c \\ 
2006dm & $0.021$ & $>-11.8$ & ... & $0.011$ & $0.027$ & $-1.86$ & $0.07$ & $53928.28$ & $0.07$ & $15.80$ & $0.03$ & ... & ... & 1,2,9a,9b \\ 
2006ef & $0.017$ & $-12.15$ & $0.30$ & $-0.015$ & $0.026$ & $-1.52$ & $0.07$ & $53969.20$ & $0.10$ & $15.28$ & $0.03$ & $10.64$ & $0.13$ & 1,2,3a,3b,6,7,9a,9b,9c \\ 
2006ej & $0.019$ & $-12.46$ & $0.30$ & $-0.016$ & $0.026$ & $-1.53$ & $0.05$ & $53976.09$ & $0.20$ & $15.46$ & $0.03$ & $10.91$ & $0.15$ & 1,2,3a,3b,6,7,9a,9b,9c \\ 
2006en & $0.032$ & $>-11.8$ & ... & $0.036$ & $0.028$ & $-0.34$ & $0.08$ & $53973.02$ & $0.05$ & $16.55$ & $0.03$ & $10.84$ & $0.19$ & 1,2,9a,9b,9c \\ 
2006gj & $0.028$ & $-11.34$ & $0.25$ & ... & ... & ... & ... & ... & ... & ... & ... & $11.39$ & $0.59$ & 6 \\ 
2006gr & $0.033$ & $-11.85$ & $0.32$ & $0.089$ & $0.027$ & $0.84$ & $0.07$ & $54013.38$ & $0.06$ & $16.75$ & $0.03$ & $11.01$ & $0.16$ & 1,2,3b,6,7,9a,9b,9c \\ 
2006hb & $0.015$ & $-9.88$ & $0.22$ & ... & ... & ... & ... & ... & ... & ... & ... & $10.72$ & $0.24$ & 6 \\ 
2006kf & $0.021$ & $-11.43$ & $0.23$ & $-0.049$ & $0.026$ & $-2.08$ & $0.05$ & $54041.26$ & $0.07$ & $15.69$ & $0.05$ & $10.97$ & $0.12$ & 1,2,3a,3b,4,6,7,9a,9c \\ 
2006le & $0.017$ & $<-11.8$ & ... & $-0.050$ & $0.034$ & $0.90$ & $0.04$ & $54048.52$ & $0.04$ & $14.79$ & $0.08$ & $10.19$ & $0.16$ & 1,2,9a,9b,9c \\ 
2006lf & $0.013$ & $>-11.8$ & ... & $-0.157$ & $0.049$ & $-1.44$ & $0.06$ & $54045.34$ & $0.06$ & $14.00$ & $0.16$ & ... & ... & 2,9a,9b \\ 
2006N & $0.014$ & $-11.32$ & $0.23$ & $-0.037$ & $0.027$ & $-2.06$ & $0.06$ & $53760.79$ & $0.15$ & $14.88$ & $0.03$ & $10.64$ & $0.17$ & 1,2,3b,6,7,9a,9c \\ 
2006ob & $0.058$ & $-8.98$ & $0.68$ & $-0.012$ & $0.025$ & $-2.16$ & $0.10$ & $54063.31$ & $0.12$ & $17.95$ & $0.02$ & $11.43$ & $0.22$ & 1,2,3a,3b,4,6,7,9a,9c \\ 
2006os & $0.032$ & $-11.91$ & $0.66$ & $0.314$ & $0.023$ & $-0.67$ & $0.06$ & $54064.75$ & $0.15$ & $17.38$ & $0.03$ & $11.30$ & $0.31$ & 3a,3b,6,7 \\ 
2006S & $0.033$ & $>-11.8$ & ... & $0.041$ & $0.026$ & $0.91$ & $0.07$ & $53770.32$ & $0.09$ & $16.59$ & $0.03$ & $10.46$ & $0.16$ & 1,2,9a,9b,9c \\ 
2006sr & $0.023$ & $-12.39$ & $0.23$ & $-0.004$ & $0.028$ & $-1.46$ & $0.10$ & $54092.91$ & $0.11$ & $15.92$ & $0.03$ & $10.74$ & $0.12$ & 1,2,3b,6,7,9a,9b,9c \\ 
2006td & $0.016$ & $-10.81$ & $0.30$ & $0.106$ & $0.027$ & $-1.43$ & $0.10$ & $54099.02$ & $0.23$ & $15.51$ & $0.03$ & $9.23$ & $0.17$ & 1,2,3b,6,7,9a,9b \\ 
2006X & $0.003$ & $-13.90$ & $0.22$ & ... & ... & ... & ... & ... & ... & ... & ... & $10.80$ & $0.12$ & 6 \\ 
2007af & $0.005$ & $-10.99$ & $0.22$ & $-0.014$ & $0.021$ & $-0.57$ & $0.03$ & $54174.52$ & $0.03$ & $12.85$ & $0.02$ & $9.84$ & $0.12$ & 3a,3b,4,6,7 \\ 
2007bc & $0.022$ & $-10.49$ & $0.26$ & $-0.002$ & $0.022$ & $-1.10$ & $0.03$ & $54200.35$ & $0.07$ & $15.58$ & $0.02$ & $10.83$ & $0.14$ & 1,2,3a,3b,4,6,7,9a,9b,9c \\ 
2007bd & $0.032$ & $-12.49$ & $0.23$ & $-0.071$ & $0.023$ & $-1.08$ & $0.04$ & $54206.84$ & $0.05$ & $16.27$ & $0.02$ & $10.79$ & $0.16$ & 1,2,3a,3b,4,6,7,9a,9b,9c \\ 
2007bm & $0.007$ & $-10.82$ & $0.49$ & $0.384$ & $0.023$ & $-0.94$ & $0.04$ & $54224.94$ & $0.08$ & $14.15$ & $0.02$ & $10.42$ & $0.23$ & 3a,3b,6,7 \\ 
2007ca & $0.015$ & $-11.09$ & $0.28$ & $0.234$ & $0.022$ & $0.57$ & $0.03$ & $54227.71$ & $0.04$ & $15.63$ & $0.02$ & $9.75$ & $0.27$ & 2,3a,3b,4,6,7,9a,9b,9c \\ 
2007ci & $0.019$ & $>-11.8$ & ... & $0.037$ & $0.028$ & $-2.75$ & $0.10$ & $54246.65$ & $0.03$ & $15.64$ & $0.03$ & $11.13$ & $0.16$ & 1,2,9a,9c \\ 
2007co & $0.027$ & $-12.02$ & $0.23$ & $0.074$ & $0.026$ & $-0.17$ & $0.06$ & $54265.16$ & $0.06$ & $16.25$ & $0.03$ & $10.21$ & $0.20$ & 1,2,3b,6,7,9a,9b,9c \\ 
2007F & $0.024$ & $>-11.8$ & ... & $-0.049$ & $0.026$ & $0.56$ & $0.04$ & $54124.03$ & $0.04$ & $15.63$ & $0.03$ & $10.07$ & $0.12$ & 1,2,9a,9b,9c \\ 
2007le & $0.007$ & $-13.31$ & $0.23$ & $0.250$ & $0.021$ & $0.25$ & $0.03$ & $54399.38$ & $0.03$ & $13.61$ & $0.02$ & $10.46$ & $0.16$ & 3a,3b,4,6,7 \\ 
2007O & $0.037$ & $>-11.8$ & ... & $-0.042$ & $0.029$ & $-0.52$ & $0.10$ & $54125.64$ & $0.35$ & $16.55$ & $0.04$ & $10.86$ & $0.17$ & 1,2,9a,9b,9c \\ 
2007qe & $0.024$ & $-14.38$ & $0.23$ & $0.069$ & $0.026$ & $0.74$ & $0.04$ & $54429.83$ & $0.04$ & $15.83$ & $0.03$ & ... & ... & 1,2,3b,9a,9b \\ 
2007sr & $0.006$ & $-13.59$ & $0.22$ & $0.053$ & $0.022$ & $-0.02$ & $0.03$ & $54449.58$ & $0.09$ & $12.48$ & $0.02$ & $10.45$ & $0.42$ & 3a,3b,4,6,7 \\ 
2008bf & $0.022$ & $-11.59$ & $0.24$ & $-0.119$ & $0.021$ & $0.37$ & $0.03$ & $54555.11$ & $0.03$ & $15.42$ & $0.02$ & $11.28$ & $0.16$ & 1,2,3a,3b,4,6,7,9a,9b,9c \\ 

\enddata
\end{deluxetable*}

\startlongtable
\begin{deluxetable*}{lDDcDcDcDcDcDc}
\tablecolumns{21}
\tablecaption{Results of our two methods of obtaining \vsimax for the Foundation objects. \label{table: velocity fits}}
\tabletypesize{\footnotesize}
\tablehead{
\colhead{} & \twocolhead{} & \multicolumn{3}{c}{Gaussian fit} &  \multicolumn{3}{c}{Smoothed minimum} & \multicolumn{6}{c}{Adopted here}\\[-13pt]
\colhead{} & \twocolhead{} & \multicolumn{3}{c}{\hrulefill} &  \multicolumn{3}{c}{\hrulefill} & \multicolumn{6}{c}{\hrulefill} \\[-9pt]
\colhead{Supernova} & \twocolhead{phase} & \twocolhead{\vsi} & \colhead{$\pm$} & \twocolhead{\vsi} & \colhead{$\pm$} & \twocolhead{\vsi} & \colhead{$\pm$}  & \twocolhead{\vsimax} & \colhead{$\pm$}\\[-6pt]
\colhead{} & \twocolhead{(days)} & \multicolumn{3}{c}{(10$^3$ km s$^{-1}$)} &  \multicolumn{3}{c}{(10$^3$ km s$^{-1}$)}  &  \multicolumn{6}{c}{(10$^3$ km s$^{-1}$)}
}
\decimals
\startdata
2016afk & $-5.2$ & $-13.07$ & $0.03$ & $-13.45$ & $0.12$ & $-13.26$ & $0.12$ & $-12.62$ & $0.25$ \\ 
2016coj & $-1.1$ & $-12.06$ & $0.03$ & ... & ... & $-12.06$ & $0.03$ & $-11.95$ & $0.22$ \\ 
2016cor & $-2.6$ & $-9.70$ & $0.07$ & $-9.53$ & $0.06$ & $-9.70$ & $0.07$ & $-9.63$ & $0.23$ \\ 
2016cvv & $+1.8$ & $-12.07$ & $0.52$ & ... & ... & $-12.07$ & $0.52$ & $-12.26$ & $0.57$ \\ 
2016esh & $-6.0$ & $-10.76$ & $0.04$ & $-11.00$ & $0.04$ & $-10.76$ & $0.04$ & $-10.45$ & $0.22$ \\ 
2016glp & $+8.9$ & $-14.51$ & $0.08$ & $-13.55$ & $0.27$ & $-14.51$ & $0.08$ & $-16.77$ & $0.24$ \\ 
2016gmg & $-2.2$ & $-11.07$ & $0.08$ & $-11.00$ & $0.45$ & $-11.07$ & $0.08$ & $-10.93$ & $0.23$ \\ 
2016gsu & $-1.5$ & $-10.10$ & $0.05$ & $-9.83$ & $0.24$ & $-10.10$ & $0.05$ & $-10.04$ & $0.23$ \\ 
2016hhv & $+2.5$ & $-11.13$ & $0.05$ & $-11.29$ & $0.07$ & $-11.21$ & $0.09$ & $-11.42$ & $0.24$ \\ 
2016htn & $-4.7$ & $-10.55$ & $0.08$ & $-10.12$ & $0.16$ & $-10.55$ & $0.08$ & $-10.32$ & $0.23$ \\ 
2016ixf & $+5.9$ & $-10.46$ & $0.02$ & $-10.51$ & $0.08$ & $-10.46$ & $0.02$ & $-10.84$ & $0.22$ \\ 
2017cfc & $+0.4$ & $-9.93$ & $0.02$ & $-10.12$ & $0.08$ & $-10.02$ & $0.08$ & $-10.04$ & $0.24$ \\ 
2017cii & $+0.5$ & $-10.14$ & $0.03$ & $-10.22$ & $0.19$ & $-10.14$ & $0.03$ & $-10.16$ & $0.22$ \\ 
2017ciy & $-1.4$ & $-9.62$ & $0.06$ & $-10.31$ & $0.16$ & $-9.62$ & $0.06$ & $-9.59$ & $0.23$ \\ 
2017cjv & $+1.9$ & $-10.93$ & $0.06$ & ... & ... & $-10.93$ & $0.06$ & $-11.07$ & $0.23$ \\ 
2017ckx & $-5.4$ & $-10.03$ & $0.07$ & $-11.00$ & $0.50$ & $-10.03$ & $0.07$ & $-9.86$ & $0.23$ \\ 
2017cpu & $-1.3$ & $-11.55$ & $0.06$ & $-11.68$ & $0.08$ & $-11.55$ & $0.06$ & $-11.45$ & $0.23$ \\ 
2017hn & $-4.0$ & $-11.29$ & $0.02$ & $-11.19$ & $0.06$ & $-11.29$ & $0.02$ & $-11.01$ & $0.22$ \\ 
2017mf & $-4.1$ & $-10.81$ & $0.02$ & $-10.80$ & $0.05$ & $-10.81$ & $0.02$ & $-10.59$ & $0.22$ \\ 
2017oz & $-3.6$ & $-14.27$ & $0.02$ & $-14.14$ & $0.03$ & $-14.27$ & $0.02$ & $-13.72$ & $0.22$ \\ 
2017po & $-0.3$ & $-12.31$ & $0.01$ & $-12.76$ & $0.05$ & $-12.76$ & $0.05$ & $-12.72$ & $0.23$ \\ 
2017yk & $-4.7$ & $-10.57$ & $0.03$ & $-10.80$ & $0.13$ & $-10.69$ & $0.13$ & $-10.45$ & $0.26$ \\ 
2017zd & $-4.0$ & $-10.65$ & $0.03$ & $-11.10$ & $0.11$ & $-10.87$ & $0.11$ & $-10.64$ & $0.25$ \\ 
ASASSN-15fa & $-2.2$ & $-10.34$ & $0.03$ & $-10.02$ & $0.04$ & $-10.34$ & $0.03$ & $-10.24$ & $0.22$ \\ 
ASASSN-15go & $+0.3$ & $-13.04$ & $0.01$ & $-13.06$ & $0.07$ & $-13.04$ & $0.01$ & $-13.08$ & $0.22$ \\ 
ASASSN-15il & $-2.4$ & $-10.99$ & $0.02$ & $-10.90$ & $0.03$ & $-10.99$ & $0.02$ & $-10.83$ & $0.22$ \\ 
ASASSN-15mg & $-4.9$ & $-10.93$ & $0.07$ & ... & ... & $-10.93$ & $0.07$ & $-10.65$ & $0.23$ \\ 
ASASSN-15np & $-1.5$ & $-9.31$ & $0.14$ & $-9.44$ & $0.12$ & $-9.31$ & $0.14$ & $-9.29$ & $0.26$ \\ 
ASASSN-15nr & $+4.8$ & $-10.32$ & $0.05$ & $-9.63$ & $0.27$ & $-10.32$ & $0.05$ & $-10.59$ & $0.23$ \\ 
ASASSN-15od & $-1.4$ & $-11.65$ & $0.01$ & $-12.07$ & $0.08$ & $-12.07$ & $0.08$ & $-11.93$ & $0.23$ \\ 
ASASSN-15pn & $+4.4$ & $-10.64$ & $0.10$ & ... & ... & $-10.64$ & $0.10$ & $-10.94$ & $0.24$ \\ 
ASASSN-15pr & $+0.9$ & $-11.53$ & $0.09$ & ... & ... & $-11.53$ & $0.09$ & $-11.61$ & $0.24$ \\ 
ASASSN-15sf & $+6.7$ & $-9.65$ & $0.01$ & $-9.73$ & $0.06$ & $-9.65$ & $0.01$ & $-9.87$ & $0.22$ \\ 
ASASSN-15ss & $+1.5$ & $-10.53$ & $0.02$ & ... & ... & $-10.53$ & $0.02$ & $-10.61$ & $0.22$ \\ 
ASASSN-15tg & $-4.2$ & $-11.86$ & $0.14$ & $-9.53$ & $0.55$ & $-11.86$ & $0.14$ & $-11.50$ & $0.26$ \\ 
ASASSN-15uu & $-1.7$ & $-10.43$ & $0.05$ & $-10.80$ & $0.14$ & $-10.80$ & $0.14$ & $-10.70$ & $0.26$ \\ 
ASASSN-15uv & $-5.6$ & $-12.99$ & $0.04$ & ... & ... & $-12.99$ & $0.04$ & $-12.35$ & $0.22$ \\ 
ASASSN-15uw & $-1.7$ & $-12.52$ & $0.03$ & $-12.66$ & $0.09$ & $-12.52$ & $0.03$ & $-12.33$ & $0.22$ \\ 
ASASSN-16ad & $-4.4$ & $-12.85$ & $0.02$ & ... & ... & $-12.85$ & $0.02$ & $-12.35$ & $0.22$ \\ 
ASASSN-16aj & $-0.0$ & $-11.63$ & $0.03$ & $-11.88$ & $0.07$ & $-11.63$ & $0.03$ & $-11.62$ & $0.22$ \\ 
ASASSN-16ay & $+0.8$ & $-10.70$ & $0.03$ & $-11.10$ & $0.12$ & $-11.10$ & $0.12$ & $-11.16$ & $0.25$ \\ 
ASASSN-16bq & $-5.6$ & $-12.79$ & $0.02$ & $-13.25$ & $0.11$ & $-13.25$ & $0.11$ & $-12.58$ & $0.25$ \\ 
ASASSN-16ch & $-5.6$ & $-11.43$ & $0.18$ & $-11.39$ & $0.15$ & $-11.43$ & $0.18$ & $-11.04$ & $0.28$ \\ 
ASASSN-16ct & $-2.4$ & $-9.90$ & $0.11$ & $-9.63$ & $0.07$ & $-9.90$ & $0.11$ & $-9.82$ & $0.24$ \\ 
ASASSN-16dw & $-1.1$ & $-10.72$ & $0.03$ & $-10.80$ & $0.06$ & $-10.72$ & $0.03$ & $-10.65$ & $0.22$ \\ 
ASASSN-16fo & $+6.9$ & $-13.68$ & $0.01$ & $-13.84$ & $0.11$ & $-13.68$ & $0.01$ & $-15.07$ & $0.22$ \\ 
ASASSN-16fs & $+4.0$ & $-11.80$ & $0.01$ & $-11.78$ & $0.05$ & $-11.80$ & $0.01$ & $-12.23$ & $0.22$ \\ 
ASASSN-16hr & $-0.0$ & $-16.84$ & $0.91$ & ... & ... & $-16.84$ & $0.91$ & $-16.83$ & $0.94$ \\ 
ASASSN-16jf & $-5.7$ & $-10.51$ & $0.02$ & $-11.19$ & $0.09$ & $-11.19$ & $0.09$ & $-10.83$ & $0.24$ \\ 
ASASSN-16la & $-2.6$ & $-10.02$ & $0.01$ & $-9.92$ & $0.12$ & $-10.02$ & $0.01$ & $-9.93$ & $0.22$ \\ 
ASASSN-16lg & $-4.5$ & $-10.31$ & $0.03$ & $-10.70$ & $0.06$ & $-10.50$ & $0.07$ & $-10.30$ & $0.23$ \\ 
ASASSN-17aj & $-3.0$ & $-10.51$ & $0.02$ & $-10.51$ & $0.01$ & $-10.51$ & $0.02$ & $-10.36$ & $0.22$ \\ 
ASASSN-17at & $-3.9$ & $-12.13$ & $0.02$ & $-12.27$ & $0.05$ & $-12.13$ & $0.02$ & $-11.76$ & $0.22$ \\ 
ASASSN-17eb & $+0.2$ & $-10.89$ & $0.03$ & $-11.19$ & $0.05$ & $-10.89$ & $0.03$ & $-10.90$ & $0.22$ \\ 
ATLAS16agv & $-3.8$ & $-10.25$ & $0.07$ & $-10.02$ & $0.04$ & $-10.25$ & $0.07$ & $-10.10$ & $0.23$ \\ 
ATLAS16bwu & $-1.6$ & $-9.66$ & $0.10$ & $-9.34$ & $0.06$ & $-9.50$ & $0.12$ & $-9.47$ & $0.25$ \\ 
ATLAS17ajn & $+2.5$ & $-10.19$ & $0.03$ & $-10.41$ & $0.16$ & $-10.19$ & $0.03$ & $-10.31$ & $0.22$ \\ 
ATLAS17axb & $-5.5$ & $-12.01$ & $0.02$ & ... & ... & $-12.01$ & $0.02$ & $-11.53$ & $0.22$ \\ 
CSS160129 & $+0.3$ & $-11.84$ & $0.03$ & $-12.57$ & $0.06$ & $-11.84$ & $0.03$ & $-11.87$ & $0.22$ \\ 
Gaia16bba & $+0.1$ & $-10.05$ & $0.01$ & $-10.02$ & $0.07$ & $-10.05$ & $0.01$ & $-10.06$ & $0.22$ \\ 
PS15ahs & $+4.0$ & $-9.51$ & $0.01$ & $-9.53$ & $0.07$ & $-9.51$ & $0.01$ & $-9.61$ & $0.22$ \\ 
PS15aii & $-0.6$ & $-10.24$ & $0.01$ & $-10.51$ & $0.06$ & $-10.51$ & $0.06$ & $-10.48$ & $0.23$ \\ 
PS15bwh & $+0.4$ & $-11.29$ & $0.13$ & $-11.10$ & $0.12$ & $-11.29$ & $0.13$ & $-11.32$ & $0.26$ \\ 
PS15bzz & $-3.1$ & $-11.22$ & $0.05$ & $-11.39$ & $0.08$ & $-11.30$ & $0.09$ & $-11.08$ & $0.24$ \\ 
PS15cfn & $-3.5$ & $-10.32$ & $0.05$ & ... & ... & $-10.32$ & $0.05$ & $-10.17$ & $0.23$ \\ 
PS15coh & $+8.4$ & $-10.26$ & $0.02$ & ... & ... & $-10.26$ & $0.02$ & $-10.78$ & $0.22$ \\ 
PS15cze & $-5.2$ & $-10.98$ & $0.06$ & $-11.88$ & $0.08$ & $-11.88$ & $0.08$ & $-11.45$ & $0.23$ \\ 
PS16bby & $-5.5$ & $-10.84$ & $0.01$ & ... & ... & $-10.84$ & $0.01$ & $-10.54$ & $0.22$ \\ 
PS16bnz & $-3.3$ & $-13.03$ & $0.22$ & $-13.45$ & $0.16$ & $-13.24$ & $0.27$ & $-12.82$ & $0.35$ \\ 
PS16cqa & $+1.7$ & $-10.17$ & $0.02$ & $-10.80$ & $0.11$ & $-10.80$ & $0.11$ & $-10.91$ & $0.25$ \\ 
PS16cvc & $-2.4$ & $-11.52$ & $0.04$ & $-11.49$ & $0.06$ & $-11.52$ & $0.04$ & $-11.32$ & $0.22$ \\ 
PS16dnp & $+2.9$ & $-9.09$ & $0.08$ & $-8.56$ & $0.10$ & $-8.82$ & $0.13$ & $-8.82$ & $0.25$ \\ 
PS16em & $+8.9$ & $-10.11$ & $0.06$ & ... & ... & $-10.11$ & $0.06$ & $-10.61$ & $0.23$ \\ 
PS16evk & $+0.2$ & $-9.90$ & $0.09$ & $-9.63$ & $0.14$ & $-9.76$ & $0.17$ & $-9.77$ & $0.28$ \\ 
PS16fa & $-2.2$ & $-10.27$ & $0.06$ & $-10.51$ & $0.13$ & $-10.27$ & $0.06$ & $-10.17$ & $0.23$ \\ 
PS16fbb & $-2.4$ & $-10.16$ & $0.03$ & $-10.22$ & $0.06$ & $-10.16$ & $0.03$ & $-10.07$ & $0.22$ \\ 
PS16n & $+8.1$ & $-9.88$ & $0.03$ & $-9.73$ & $0.04$ & $-9.81$ & $0.05$ & $-10.14$ & $0.23$ \\ 
PS17bii & $-3.9$ & $-10.31$ & $0.04$ & $-10.80$ & $0.10$ & $-10.80$ & $0.10$ & $-10.59$ & $0.24$ \\ 
PSNJ0153424 & $-2.6$ & $-10.65$ & $0.01$ & ... & ... & $-10.65$ & $0.01$ & $-10.51$ & $0.22$ \\ 
PTSS-16efw & $+3.1$ & $-10.17$ & $0.03$ & $-10.31$ & $0.17$ & $-10.24$ & $0.17$ & $-10.40$ & $0.28$ \\ 

\cutinhead{Objects with spectra outside nominal phase range}
2016ayg & $-8.1$ & $-10.85$ & $0.07$ & $-10.80$ & $0.27$ & ... & ... & ... & ... \\ 
2016cyt & $-6.3$ & $-11.69$ & $0.09$ & $-12.07$ & $0.08$ & ... & ... & ... & ... \\ 
2016eoa & $-8.5$ & $-14.17$ & $0.02$ & $-14.24$ & $0.04$ & ... & ... & ... & ... \\ 
2016fbk & $-7.3$ & $-11.04$ & $0.08$ & $-10.90$ & $0.09$ & ... & ... & ... & ... \\ 
2016ffh & $-8.4$ & $-17.91$ & $0.04$ & $-19.11$ & $0.38$ & ... & ... & ... & ... \\ 
2016gfr & $-9.1$ & $-13.32$ & $0.11$ & $-12.37$ & $0.22$ & ... & ... & ... & ... \\ 
2016glz & $-9.0$ & $-16.47$ & $0.11$ & $-15.82$ & $0.22$ & ... & ... & ... & ... \\ 
2016gmb & $+12.5$ & $-11.13$ & $0.05$ & $-11.00$ & $0.06$ & ... & ... & ... & ... \\ 
2016gou & $-10.3$ & $-11.28$ & $0.03$ & $-11.39$ & $0.05$ & ... & ... & ... & ... \\ 
2016grz & $-7.3$ & $-10.68$ & $0.29$ & $-10.80$ & $0.06$ & ... & ... & ... & ... \\ 
2016hnk & $+16.4$ & $-9.25$ & $0.04$ & $-9.92$ & $0.09$ & ... & ... & ... & ... \\ 
2016hpx & $-9.1$ & $-7.85$ & $0.04$ & $-13.06$ & $0.11$ & ... & ... & ... & ... \\ 
2016htm & $-10.5$ & $-11.23$ & $0.04$ & $-11.19$ & $0.08$ & ... & ... & ... & ... \\ 
2016W & $-8.8$ & $-11.41$ & $0.05$ & $-11.78$ & $0.11$ & ... & ... & ... & ... \\ 
2017dfq & $-6.1$ & $-11.62$ & $0.02$ & $-11.68$ & $0.05$ & ... & ... & ... & ... \\ 
2017jl & $-9.8$ & $-11.92$ & $0.04$ & $-11.98$ & $0.06$ & ... & ... & ... & ... \\ 
2017lm & $-6.7$ & $-10.81$ & $0.01$ & $-11.10$ & $0.07$ & ... & ... & ... & ... \\ 
2017ms & $-6.1$ & $-11.69$ & $0.02$ & $-11.68$ & $0.03$ & ... & ... & ... & ... \\ 
2017ns & $-7.0$ & $-12.54$ & $0.05$ & $-11.98$ & $0.16$ & ... & ... & ... & ... \\ 
2017wb & $-6.9$ & $-13.70$ & $0.10$ & $-14.54$ & $0.07$ & ... & ... & ... & ... \\ 
ASASSN-15kx & $-7.5$ & $-11.60$ & $0.03$ & $-12.37$ & $0.24$ & ... & ... & ... & ... \\ 
ASASSN-15lg & $-7.3$ & $-12.56$ & $0.02$ & $-12.96$ & $0.09$ & ... & ... & ... & ... \\ 
ASASSN-15mf & $+15.0$ & $-10.61$ & $0.04$ & $-10.02$ & $0.09$ & ... & ... & ... & ... \\ 
ASASSN-16bc & $-7.8$ & $-10.90$ & $0.08$ & $-12.47$ & $0.36$ & ... & ... & ... & ... \\ 
ASASSN-16bc & $-7.8$ & $-10.90$ & $0.08$ & $-12.47$ & $0.36$ & ... & ... & ... & ... \\ 
ASASSN-16dn & $-9.7$ & $-11.35$ & $0.03$ & $-11.39$ & $0.04$ & ... & ... & ... & ... \\ 
ASASSN-16et & $-14.9$ & $-16.14$ & $0.06$ & $-13.25$ & $0.25$ & ... & ... & ... & ... \\ 
ASASSN-16ex & $-12.3$ & $-13.08$ & $0.04$ & $-13.06$ & $0.05$ & ... & ... & ... & ... \\ 
ASASSN-16ip & $-8.2$ & $-12.28$ & $0.02$ & $-12.76$ & $0.04$ & ... & ... & ... & ... \\ 
ASASSN-16oz & $-6.7$ & $-10.76$ & $0.06$ & $-11.10$ & $0.18$ & ... & ... & ... & ... \\ 
ASASSN-17bs & $-6.0$ & $-11.06$ & $0.01$ & $-11.49$ & $0.08$ & ... & ... & ... & ... \\ 
ASASSN-17co & $-6.2$ & $-11.98$ & $0.03$ & $-11.78$ & $0.20$ & ... & ... & ... & ... \\ 
ASASSN-17fd & $-6.3$ & $-9.91$ & $0.73$ & ... & ... & ... & ... & ... & ... \\ 
ATLAS16dpb & $-6.2$ & $-11.81$ & $0.12$ & $-12.37$ & $0.07$ & ... & ... & ... & ... \\ 
ATLAS16dqf & $-10.5$ & ... & ... & $-13.55$ & $0.18$ & ... & ... & ... & ... \\ 
ATLAS16dtf & $-12.6$ & $-11.60$ & $0.04$ & $-12.57$ & $0.40$ & ... & ... & ... & ... \\ 
PS15bif & $-6.7$ & $-11.37$ & $0.03$ & $-11.58$ & $0.09$ & ... & ... & ... & ... \\ 
PS15bjg & $-9.2$ & $-13.03$ & $0.10$ & $-13.25$ & $0.29$ & ... & ... & ... & ... \\ 
PS15bsq & $-7.6$ & $-12.24$ & $0.01$ & $-12.27$ & $0.07$ & ... & ... & ... & ... \\ 
PS15cge & $-10.8$ & $-11.57$ & $0.04$ & $-11.49$ & $0.11$ & ... & ... & ... & ... \\ 
PS16aer & $-7.0$ & $-12.97$ & $0.02$ & $-12.66$ & $0.08$ & ... & ... & ... & ... \\ 
PS17akj & $-7.3$ & $-12.41$ & $0.03$ & $-11.88$ & $0.14$ & ... & ... & ... & ... \\ 

\enddata
\tablecomments{Both the measurement at the phase of the spectrum and the measurement phased to maximum light are shown. Objects with a measurement of the \ion{Si}{2} line, but with an incompatible phase are shown at the bottom of the table.}
\end{deluxetable*}

\startlongtable
\begin{deluxetable*}{llccDDc}
\tablecaption{Properties of Foundation spectral data release}
\tabletypesize{\footnotesize}
\tablehead{\colhead{Supernova} &\colhead{Telescope/Instrument} & \colhead{$z$} & \colhead{Date of Observation} & \twocolhead{$t_0$} & \twocolhead{approx. phase} & \colhead{Wavelength Range}\\[-6pt]
\colhead{} & \colhead{} & \colhead{}  &  \colhead{} & \twocolhead{(MJD)} & \twocolhead{(days)} & \colhead{(\AA)}
}
\decimals
\startdata
2016ac &  FLWO1.5m/FAST & $0.026$ & 2016-02-04 & $57409.1$ & $12.6$ & 3463-7403 \\ 
2016bew &  SALT/RSS & $0.054$ & 2016-03-14 & $57466.9$ & $-5.6$ & 3348-9250 \\ 
2016bln &  KPNO4m/KOSMOS & $0.023$ & 2016-05-06 & $57499.6$ & $14.1$ & 4165-7065 \\ 
2016eoa &  KPNO4m/KOSMOS & $0.021$ & 2016-08-07 & $57615.7$ & $-8.5$ & 4180-7060 \\ 
2016glp &  SOAR/Goodman & $0.085$ & 2016-10-08 & $57659.3$ & $8.9$ & 3700-7100 \\ 
2016glz &  SOAR/Goodman & $0.041$ & 2016-10-30 & $57664.3$ & $25.6$ & 3600-7100 \\ 
2016gmb &  SOAR/Goodman & $0.058$ & 2016-10-08 & $57655.8$ & $12.5$ & 3700-7100 \\ 
2016hnk &  SOAR/Goodman & $0.016$ & 2016-10-30 & $57691.2$ & $-0.2$ & 3600-7100 \\ 
2016hnk &  SALT/RSS & $0.016$ & 2016-10-31 & $57691.2$ & $0.8$ & 3495-9397 \\ 
2016hns &  SOAR/Goodman & $0.037$ & 2016-10-30 & $57697.5$ & $-6.2$ & 3600-7100 \\ 
2017cbv &  SOAR/Goodman & $0.037$ & 2017-03-29 & \nodata & \nodata & 3600-9020 \\ 
2017cfb &  Lick3m/KAST & $0.023$ & 2017-05-25 & $57836.8$ & $59.8$ & 3260-10001 \\ 
2017ckx &  FLWO1.5m/FAST & $0.027$ & 2017-03-29 & $57846.6$ & $-5.4$ & 3475-7416 \\ 
2017cpu &  FLWO1.5m/FAST & $0.054$ & 2017-04-03 & $57847.4$ & $-1.3$ & 3481-7422 \\ 
2017dit &  Lick3m/KAST & $0.019$ & 2017-04-30 & $57881.2$ & $-8.1$ & 3220-10249 \\ 
2017erp &  SALT/RSS & $0.006$ & 2017-06-13 & \nodata & \nodata & 3496-9399 \\ 
2017fbs &  Lick3m/KAST & $0.036$ & 2017-07-03 & $57945.7$ & $-8.4$ & 3240-9999 \\ 
2017fms &  SOAR/Goodman & $0.030$ & 2017-07-18 & $57960.3$ & $-8.1$ & 3600-7100 \\ 
2017fmz &  SOAR/Goodman & $0.028$ & 2017-07-18 & $57953.5$ & $-1.5$ & 3600-7100 \\ 
2017fnz &  Lick3m/KAST & $0.081$ & 2017-07-22 & \nodata & \nodata & 3400-10000 \\ 
2017gav &  KPNO4m/KOSMOS & $0.033$ & 2017-08-15 & $57979.9$ & $0.0$ & 4180-7060 \\ 
2017hju &  KPNO4m/KOSMOS & $0.015$ & 2017-10-17 & $58051.6$ & $-8.5$ & 4180-7060 \\ 
2017hpj &  Lick3m/KAST & $0.037$ & 2017-10-28 & \nodata & \nodata & 3350-9698 \\ 
2017hyx &  SOAR/Goodman & $0.038$ & 2017-11-13 & \nodata & \nodata & 3700-7100 \\ 
2017iez &  Lick3m/KAST & $0.046$ & 2017-11-22 & $58082.8$ & $-3.6$ & 3200-10000 \\ 
2017iye &  Lick3m/KAST & $0.046$ & 2017-12-22 & $58109.5$ & $-0.5$ & 3200-10500 \\ 
2017lm &  SALT/RSS & $0.031$ & 2017-01-19 & $57778.9$ & $-6.7$ & 3494-9397 \\ 
2017oz &  KPNO4m/KOSMOS & $0.056$ & 2017-01-31 & $57788.0$ & $-3.8$ & 4180-7060 \\ 
2017po &  KPNO4m/KOSMOS & $0.032$ & 2017-01-31 & $57784.5$ & $-0.5$ & 4180-7060 \\ 
2017ya &  KPNO4m/KOSMOS & $0.070$ & 2017-01-31 & $57782.7$ & $1.2$ & 4180-7060 \\ 
2017yk &  KPNO4m/KOSMOS & $0.046$ & 2017-01-31 & $57788.9$ & $-4.7$ & 4180-7060 \\ 
2018bie &  SOAR/Goodman & $0.023$ & 2018-05-13 & \nodata & \nodata & 3600-7124 \\ 
2018bs &  Lick3m/KAST & $0.070$ & 2018-01-14 & \nodata & \nodata & 3300-10500 \\ 
2018cnw &  APO3.5m/DIS & $0.024$ & 2018-06-15 & \nodata & \nodata & 3401-8879 \\ 
2018oh &  Lick3m/KAST & $0.011$ & 2018-02-08 & \nodata & \nodata & 3200-9999 \\ 
2018oh &  SOAR/Goodman & $0.011$ & 2018-03-22 & \nodata & \nodata & 3500-9040 \\ 
2018oh &  SOAR/Goodman & $0.011$ & 2018-04-21 & \nodata & \nodata & 3600-9055 \\ 
2018pj &  Lick3m/KAST & $0.055$ & 2018-02-08 & \nodata & \nodata & 3500-10197 \\ 
ASASSN-15bm &  SOAR/Goodman & $0.020$ & 2015-04-10 & $57054.6$ & $66.0$ & 3500-10197 \\ 
ASASSN-15ga &  SOAR/Goodman & $0.007$ & 2015-04-10 & \nodata & \nodata & 3500-7091 \\ 
ASASSN-15go &  SOAR/Goodman & $0.019$ & 2015-04-10 & $57122.7$ & $-0.7$ & 3200-7090 \\ 
ASASSN-15hg &  SOAR/Goodman & $0.030$ & 2015-05-16 & $57131.3$ & $25.9$ & 3600-7050 \\ 
ASASSN-15hy &  SOAR/Goodman & $0.018$ & 2015-05-17 & $57153.2$ & $5.7$ & 3500-7050 \\ 
ASASSN-15il &  SOAR/Goodman & $0.023$ & 2015-05-17 & $57161.4$ & $-2.4$ & 3250-7050 \\ 
ASASSN-15jt &  SOAR/Goodman & $0.023$ & 2015-06-16 & $57169.8$ & $18.8$ & 3900-7050 \\ 
ASASSN-15lg &  SOAR/Goodman & $0.020$ & 2015-06-16 & $57196.4$ & $-7.3$ & 3600-7050 \\ 
ASASSN-15mf &  SOAR/Goodman & $0.026$ & 2015-07-24 & $57212.4$ & $14.2$ & 3500-7050 \\ 
ASASSN-15mg &  Keck/DEIMOS & $0.043$ & 2015-09-19 & $57218.3$ & $63.0$ & 4458-9630 \\ 
ASASSN-15np &  SOAR/Goodman & $0.000$ & 2015-08-21 & $57241.6$ & $13.4$ & 3920-7050 \\ 
ASASSN15nq &  Keck/DEIMOS & $0.030$ & 2015-09-19 & $57247.8$ & $35.2$ & 4442-9630 \\ 
ASASSN-15nr &  SOAR/Goodman & $0.023$ & 2015-08-21 & $57250.1$ & $4.8$ & 3750-7050 \\ 
ASASSN-15od &  SOAR/Goodman & $0.018$ & 2015-08-21 & $57256.6$ & $-1.6$ & 3500-7050 \\ 
ASASSN15oh &  Keck/DEIMOS & $0.000$ & 2015-09-19 & $57255.9$ & $28.1$ & 4447-9630 \\ 
ASASSN-15pm &  SOAR/Goodman & $0.049$ & 2015-10-11 & $57277.9$ & $26.8$ & 3700-7050 \\ 
ASASSN15pn &  Keck/DEIMOS & $0.000$ & 2015-09-19 & $57279.4$ & $4.6$ & 4458-9630 \\ 
ASASSN15pr &  Keck/DEIMOS & $0.000$ & 2015-09-19 & $57283.1$ & $0.9$ & 4442-9630 \\ 
ASASSN-15py &  SOAR/Goodman & $0.000$ & 2015-10-11 & $57289.0$ & $17.0$ & 3700-7050 \\ 
ASASSN-15sb &  SOAR/Goodman & $0.000$ & 2015-11-14 & $57321.2$ & $18.8$ & 3900-7050 \\ 
ASASSN-15sf &  SOAR/Goodman & $0.025$ & 2015-11-14 & $57333.1$ & $6.7$ & 3500-7050 \\ 
ASASSN-15uu &  KPNO4m/KOSMOS & $0.027$ & 2016-01-04 & $57388.8$ & $2.2$ & 4180-7060 \\ 
ASASSN-15uw &  KPNO4m/KOSMOS & $0.031$ & 2016-01-04 & $57392.8$ & $-1.7$ & 4180-7060 \\ 
ASASSN-16dn &  FLWO1.5m/FAST & $0.013$ & 2016-04-02 & $57489.9$ & $-9.7$ & 3477-7419 \\ 
ASASSN-16fo &  KPNO4m/KOSMOS & $0.000$ & 2016-06-08 & $57539.9$ & $7.1$ & 4170-7065 \\ 
ASASSN-16fs &  KPNO4m/KOSMOS & $0.029$ & 2016-06-08 & $57542.9$ & $4.0$ & 4165-7065 \\ 
ASASSN-16hz &  SOAR/Goodman & $0.025$ & 2016-09-07 & $57605.2$ & $32.0$ & 3700-7100 \\ 
ASASSN-16la &  SOAR/Goodman & $0.015$ & 2016-10-08 & $57671.6$ & $-2.6$ & 3700-7100 \\ 
ASASSN-17bs &  KPNO4m/KOSMOS & $0.020$ & 2017-02-01 & $57791.1$ & $-6.0$ & 4180-7060 \\ 
ASASSN-17jq &  KPNO4m/KOSMOS & $0.029$ & 2017-07-26 & \nodata & \nodata & 4180-7060 \\ 
ATLAS16bdg &  SOAR/Goodman & $0.014$ & 2016-07-27 & $57554.8$ & $40.6$ & 3700-7100 \\ 
ATLAS16dpb &  FLWO1.5m/FAST & $0.023$ & 2016-11-02 & $57689.3$ & $4.6$ & 3476-7417 \\ 
ATLAS16dtf &  FLWO1.5m/FAST & $0.013$ & 2016-11-24 & $57710.8$ & $5.1$ & 3478-7419 \\ 
ATLAS16eej &  SOAR/Goodman & $0.049$ & 2017-01-03 & $57755.9$ & $0.1$ & 3600-7100 \\ 
ATLAS17dzg &  Lick3m/KAST & $0.052$ & 2017-04-20 & $57864.3$ & $-1.3$ & 3200-10250 \\ 
CSS160129 &  SOAR/Goodman & $0.068$ & 2016-02-09 & $57416.7$ & $9.7$ & 3700-7050 \\ 
CSS160129 &  KPNO4m/KOSMOS & $0.068$ & 2016-06-08 & $57416.7$ & $122.0$ & 4180-7060 \\ 
CSS160428 &  KPNO4m/KOSMOS & $0.030$ & 2016-06-08 & $57517.6$ & $28.5$ & 4180-7060 \\ 
Gaia16acv &  KPNO4m/KOSMOS & $0.078$ & 2016-04-06 & $57435.1$ & $45.3$ & 4165-7065 \\ 
MASTER151647 &  SOAR/Goodman & $0.056$ & 2016-07-28 & $57582.5$ & $13.8$ & 3700-7100 \\ 
MASTERJ222 &  SOAR/Goodman & $0.024$ & 2015-06-16 & $57183.4$ & $5.5$ & 3500-7050 \\ 
PS15adh &  SOAR/Goodman & $0.103$ & 2015-05-17 & $57136.6$ & $20.3$ & 3600-7050 \\ 
PS15ahs &  KPNO4m/KOSMOS & $0.026$ & 2015-05-19 & $57156.8$ & $4.1$ & 4300-7065 \\ 
PS15aii &  SOAR/Goodman & $0.047$ & 2015-05-17 & $57159.5$ & $-0.5$ & 3500-7050 \\ 
PS15akf &  SOAR/Goodman & $0.059$ & 2015-05-17 & $57164.0$ & $-4.7$ & 3500-7050 \\ 
PS15asb &  SOAR/Goodman & $0.049$ & 2015-07-24 & $57194.6$ & $30.9$ & 3500-7050 \\ 
PS15atx &  SALT/RSS & $0.072$ & 2015-06-20 & $57188.9$ & $3.8$ & 3000-9999 \\ 
PS15bbn &  SOAR/Goodman & $0.037$ & 2015-07-24 & $57213.9$ & $12.6$ & 3500-7050 \\ 
PS15bif &  SOAR/Goodman & $0.079$ & 2015-07-24 & $57234.2$ & $-6.7$ & 3701-7098 \\ 
PS15bjg &  SALT/RSS & $0.069$ & 2015-07-26 & $57238.8$ & $-9.2$ & 3200-10000 \\ 
PS15brr &  SOAR/Goodman & $0.052$ & 2015-10-11 & $57259.3$ & $44.4$ & 3700-7050 \\ 
PS15bsq &  SOAR/Goodman & $0.034$ & 2015-08-21 & $57262.9$ & $-7.6$ & 3500-7050 \\ 
PS15bst &  SOAR/Goodman & $0.088$ & 2015-08-21 & $57258.8$ & $-3.5$ & 3670-7050 \\ 
PS15bwh &  SOAR/Goodman & $0.073$ & 2015-09-14 & $57278.5$ & $0.4$ & 3700-7050 \\ 
PS15bzz &  SALT/RSS & $0.080$ & 2015-09-16 & $57284.4$ & $-3.1$ & 3598-8148 \\ 
PS15cfn &  SOAR/Goodman & $0.111$ & 2015-10-11 & $57309.9$ & $-3.5$ & 3700-7050 \\ 
PS15cge &  SOAR/Goodman & $0.085$ & 2015-10-11 & $57313.5$ & $-6.9$ & 3700-7050 \\ 
PS15cku &  SOAR/Goodman & $0.023$ & 2015-11-14 & $57319.3$ & $20.2$ & 3200-7050 \\ 
PS15mt &  Keck/DEIMOS & $0.071$ & 2015-03-18 & $57105.3$ & $-5.9$ & 4625-9850 \\ 
PS15zn &  SOAR/Goodman & $0.055$ & 2015-04-10 & $57128.0$ & $-5.7$ & 3500-7050 \\ 
PS16aer &  SOAR/Goodman & $0.055$ & 2016-02-09 & $57434.4$ & $-7.0$ & 3700-7100 \\ 
PS16axi &  KPNO4m/KOSMOS & $0.039$ & 2016-04-06 & $57461.4$ & $21.8$ & 4165-7065 \\ 
PS16ayd &  KPNO4m/KOSMOS & $0.054$ & 2016-04-06 & $57460.4$ & $22.3$ & 4165-7064 \\ 
PS16bnz &  KPNO4m/KOSMOS & $0.063$ & 2016-04-06 & $57487.5$ & $-3.3$ & 4165-7060 \\ 
PS16ccn &  KPNO4m/KOSMOS & $0.024$ & 2016-06-08 & $57508.4$ & $37.7$ & 4170-7060 \\ 
PS16cqa &  KPNO4m/KOSMOS & $0.044$ & 2016-06-08 & $57545.3$ & $1.7$ & 4165-7065 \\ 
PS16em &  SALT/RSS & $0.070$ & 2016-01-09 & $57386.5$ & $8.9$ & 2999-9998 \\ 
PS16eqv &  SALT/RSS & $0.085$ & 2016-10-27 & $57683.5$ & $4.2$ & 3349-9249 \\ 
PS16fbb &  SALT/RSS & $0.052$ & 2016-11-25 & $57719.5$ & $-2.4$ & 3348-9251 \\ 
PS16fbb &  FLWO1.5m/FAST & $0.052$ & 2016-12-01 & $57719.5$ & $3.3$ & 3471-7412 \\ 
PS16n &  SOAR/Goodman & $0.053$ & 2016-01-14 & $57392.4$ & $8.2$ & 3600-7116 \\ 
PS17bii &  SALT/RSS & $0.073$ & 2017-02-24 & $57812.1$ & $-3.9$ & 3299-9249 \\ 
PSNJ015342 &  KPNO4m/KOSMOS & $0.026$ & 2016-01-04 & $57393.6$ & $-2.6$ & 4180-7060 \\ 
PSNJ1204 &  SOAR/Goodman & $0.044$ & 2015-06-16 & $57170.3$ & $17.9$ & 3600-7050 \\ 
PSNJ1300 &  SOAR/Goodman & $0.023$ & 2015-06-15 & $57168.8$ & $18.8$ & 3500-7050 \\ 
PSNJ1602 &  SOAR/Goodman & $0.020$ & 2015-07-24 & $57172.4$ & $53.5$ & 3800-7050 \\ 
PSNJ2043 &  SOAR/Goodman & $0.015$ & 2015-07-24 & $57226.0$ & $1.0$ & 3500-7050 \\ 
PSNJ2310 &  KPNO4m/KOSMOS & $0.039$ & 2015-09-15 & $57281.2$ & $-1.2$ & 4201-7060 \\ 
PTSS17niq &  SOAR/Goodman & $0.068$ & 2017-03-29 & $57842.9$ & $-1.8$ & 3600-7100 \\ 

\enddata
\end{deluxetable*}

\begin{figure*}
    \centering
    \includegraphics[height=0.93\textheight]{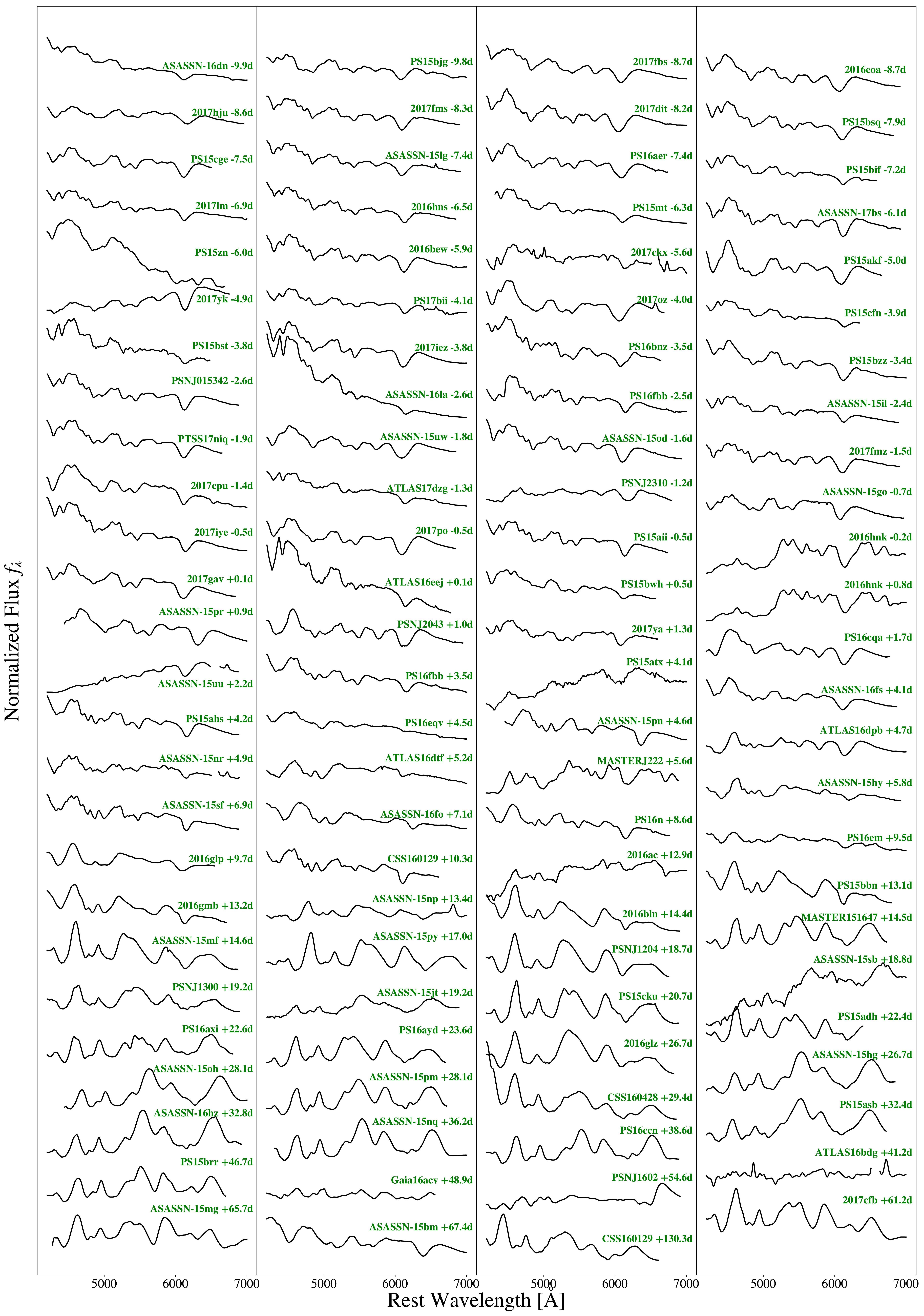}
    \caption{Optical spectra obtained as part of the first data release of the Foundation Supernova Survey. The spectra are ordered according to rest-frame phase. Some host-galaxy lines have been excised here for readability. Machine readable versions of each spectrum are included. A complete archive of these spectra can be found at \url{https://github.com/kdettman/foundVelCorr}.}
    \label{fig:team spec}
\end{figure*}

\end{document}